\documentclass[twocolumn,twocolappendix]{aastex7}
\usepackage{amsmath}     
\usepackage{wasysym} 
\usepackage{tikz}
\usepackage{graphicx}
\usepackage{amssymb}
\usepackage{epstopdf}
\usepackage{mathrsfs}
\usepackage{hyperref}
\usepackage{anyfontsize}
\usepackage{natbib}
\usepackage{color}
\usepackage{lipsum}
\usepackage{diagbox}

\shorttitle{Peak Fallback Rate for partial TDEs}
\shortauthors{Bandopadhyay, Coughlin, \& Nixon}
\begin{document}
\title{The Maximum Gravity Model for partial Tidal Disruption Events: Mass Loss, Peak Fallback Rate and Dependence on Stellar Properties}
\author[0000-0002-5116-844X]{Ananya Bandopadhyay}
\affiliation{Department of Physics, Syracuse University, Syracuse, NY 13210, USA}
\email[show]{abandopa@syr.edu}

\author[0000-0003-3765-6401]{Eric R.~Coughlin}
\affiliation{Department of Physics, Syracuse University, Syracuse, NY 13210, USA}
\email[show]{ecoughli@syr.edu}

\author[0000-0002-2137-4146]{C.~J.~Nixon}
\affiliation{School of Physics and Astronomy, Sir William Henry Bragg Building, Woodhouse Ln., University of Leeds, Leeds LS2 9JT, UK}
\email[show]{c.j.nixon@leeds.ac.uk}

\begin{abstract}
A star entering the tidal sphere of a supermassive black hole (SMBH) can be partially stripped of mass, resulting in a partial tidal disruption event (TDE). Here we develop an analytical model for properties of these events, including the peak fallback rate, $\dot{M}_{\rm peak}$, the time at which the peak occurs, $t_{\rm peak}$, and the amount of mass removed from the star, $\Delta M$, for any star and any pericenter distance associated with the stellar orbit about the black hole. We compare the model predictions to 1276 hydrodynamical simulations of partial TDEs of main-sequence stars by a $10^6 M_\odot$ SMBH. The model yields $t_{\rm peak}$ predictions that are in good agreement (to within tens of percent) with the numerical simulations for any stellar mass and age. The agreement for $\dot{M}_{\rm peak}$ is weaker due to the influence of self-gravity on the debris stream dynamics, which remains dynamically important for partial TDEs; the agreement for $\dot{M}_{\rm peak}$ is, however, to within a factor of $\sim 2-3$ in the majority of cases considered, with larger differences for low-mass stars ($M_\star \lesssim 0.5 M_\odot$) on grazing orbits with small mass loss. We show that partial TDE lightcurves for disruptions caused by $\sim 10^6M_\odot$ SMBHs can span $\sim 20-100$ day peak timescales, whereas grazing encounters of high-mass stars with high-mass SMBHs can yield longer peak timescales ($t\gtrsim 1000$ days), associated with some observed transients. Our model provides a significant step toward an analytical prescription for TDE lightcurves and luminosity functions.
\end{abstract}
\section{Introduction}
The complete or partial tidal disruption of a star by a supermassive black hole (SMBH) is known as a tidal disruption event (TDE; \citealp{hills75,rees88,gezari21}), and produces a stream of disrupted stellar debris. Roughly half of the stripped debris is gravitationally bound to the SMBH and circularizes to form an accretion disk. The accretion process generates a luminous flare, which sheds light on the population of quiescent SMBHs in galactic centers. There are currently $\sim 100$ TDEs detected through time-domain surveys (e.g.,~\citealp{arcavi14,french16,nicholl19,pasham19,wevers19,hinkle21,payne21,vanvelzen21,wevers21,lin22,nicholl22,hammerstein23,wevers23,yao23,guolo24,pasham24,ho25}; see also \citealp{gezari21} and references therein), but the detection rate is expected to grow rapidly in the era of the Vera Rubin Observatory~\citep{ivezic19,bricman20}.

The process of circularization of stellar debris leading to the formation of an accretion disk is not well understood, and has been the subject of extensive numerical investigations (e.g., \citealt{cannizzo90,rosswog09,hayasaki13,Shiokawa15,Bonnerot16,sadowski16,curd19,andalman22,meza25,kubli25}). However, the fallback rate $\dot{M}$, which is the rate of return of bound stellar debris to the SMBH, closely tracks the lightcurve of a TDE, provided the debris rapidly circularizes into an accretion disk and the viscous timescale of the disk is small compared to the fallback time of the debris (\citealt{rees88, cannizzo90}; the results of ~\citealp{mockler19,nicholl22} suggest this is true for most UV/optical TDEs for at least the first few months to a year since disruption; see also \citealt{lodato11}). Thus, significant effort has been dedicated to the modeling of the fallback rate from TDEs, both numerically (e.g., \citealp{lodato09, guillochon13, coughlin15, goicovic19, golightly19,lawsmith19,lawsmith20,ryu20a,jankovic23,fancher25}) and analytically via the ``frozen-in approximation''~\citep{lacy82,bicknell83,lodato09,stone13}; the latter assumes that the entire star moves with the center-of-mass (COM) until it reaches the tidal radius $r_{\rm t} = R_\star (M_\bullet/M_\star)^{1/3}$, which serves as an estimate for the distance at which a star of mass $M_\star$ and radius $R_\star$ is completely destroyed by the tidal field of an SMBH of mass $M_\bullet$ \citep{hills75}. Subsequently, using the energy distribution established at pericenter (which is given by the distribution of Keplerian energies of each fluid element in the star as a function of its position in the SMBH potential), one can estimate, e.g., the bound and unbound mass fractions, the fallback time (the orbital period of a fluid element located at a distance $R_\star$ from the COM), and the mass fallback rate (as calculated explicitly for three polytropic stars in \citealp{lodato09}).

Numerical studies of TDEs showed that the critical pericenter distance for complete disruption depends on stellar structure and is generally different from the canonical tidal radius $r_{\rm t}$. For example, \cite{khoklov93} used an Eulerian code to analyze the disruption of three polytropic stars ($\gamma=3/2, 5/3, 4/3$ polytropes, where the polytropic equation of state is given by $p\propto\rho^\gamma$, with $p$ and $\rho$ being the pressure and density respectively), and found that the critical pericenter distance for complete disruption varies with $\gamma$ and differs from $r_{\rm t}$ by a factor of $\sim a \, few$. \cite{diener97} studied the partial disruption of a $\gamma=5/3$ polytrope by spinning SMBHs, using a combination of numerical and semi-analytic techniques. They showed that the amount of mass lost in a partial TDE, as well as the energy and angular momentum imparted to the star, can be constrained in terms of the trace of the tidal tensor. 

More recent works investigating partial TDEs have focused on the amount of mass stripped $\Delta M$, the fallback rate $\dot{M}(t)$, and the late-time scaling of $\dot{M} \propto t^{n_{\infty}}$. \cite{guillochon13} performed grid-based hydrodynamical simulations of the disruption of two polytropic stars ($\gamma=5/3$ and $\gamma=4/3$) over a range of penetration factors $\beta$ (where $\beta = r_{\rm t}/r_{\rm p}$ with $r_{\rm p}$ the pericenter distance; see also \citealt{mainetti17} for particle-based results over the same parameter space) and generated empirical fits for the amount of mass stripped $\Delta M$ as a function of $\beta$, which were in good agreement with the predictions of \cite{ivanov01}. They also estimated the power-law index of the fallback rate, and found that for some of their partial disruption simulations the power-law index could be steeper than $-5/3$, which they attributed to the gravitational influence of the surviving core. 

\cite{coughlin19} analytically showed that the late-time scaling of the fallback rate from partial TDEs is $\propto t^{-9/4}$, effectively independent of the core mass. They also used a variation of the frozen-in approximation (similar to the method prescribed in ~\citealp{lodato09}) in conjunction with the Lagrangian equation of motion for the disrupted debris stream in the presence of a core, to obtain the fallback rate $\dot{M}$ on to the SMBH. For core mass fractions $M_{\rm c}/(10^{-6}M_\bullet )\lesssim15\%$, the fallback rates closely tracked the power-law scaling for the core-less case ($\propto t^{-5/3}$) before eventually transitioning to a $t^{-9/4}$ scaling, while for larger core masses (more grazing encounters), the $t^{-9/4}$ scaling is reached more rapidly. However, their model of the early-time behavior of the fallback rate does not accurately reproduce the trends seen in numerical simulations of partial TDEs. For example, all of the fallback rates shown in Figure 2 of their paper have a similar peak timescale, whereas numerical simulations show that $t_{\rm peak}$ varies significantly with the penetration factor $\beta$ (e.g., \citealp{guillochon13,mainetti17,golightly19,goicovic19,lawsmith19,ryu20b,nixon21,jankovic23}). 

The asymptotic power-law scaling of the fallback rate $\dot{M}\propto t^{-9/4}$ predicted by \cite{coughlin19} has been recovered in numerical simulations of partial TDEs, e.g., \cite{golightly19,miles20,nixon21,wang21,kremer22,jankovic23,fancher25}. \cite{miles20} studied the partial disruption of a $\gamma=5/3$-polytropic star for a range of penetration factors $\beta$, and found that for $0.70\lesssim\beta\lesssim0.85$, the fallback rate scales as $\propto t^{-5/3}$ at early times, and steepens into a $\propto t^{-9/4}$ decay on a timescale defined as the break timescale, which is a decreasing function of the mass of the surviving core (or equivalently, an increasing function of $\beta$). For more disruptive encounters ($0.70\lesssim\beta\lesssim0.85$), the break timescale can range between $\sim1-100$ years for the $\gamma=5/3-$polytropic star\footnote{For a $\beta=0.9$ orbit, which yields a core mass of $\sim13\%M_\star$ for the $\gamma=5/3$-polytropic star, \cite{miles20} find that the fallback rate tracks the $t^{-5/3}$ scaling at early times, before steepening to a $t^{-2}$ at around $30$ years post-dirsption, before eventually returning back to $\propto t^{-5/3}$ at very late times. This arises from the fact that the core reforms at late times in this case, and the difference in shear along the debris stream results in its reformation in the unbound (from the SMBH) tail \citep{coughlin25}, such that the influence of the surviving core on the dynamics of the bound stream diminishes with time.}. 

While the 1D Lagrangian model developed in \cite{coughlin19} yields excellent agreement with numerical simulations as concerns the late time scaling of the fallback rate, and some previous numerical investigations of partial TDEs (e.g., \citealp{lawsmith19,lawsmith20,ryu20b}) provide empirical fitting formulae for $t_{\rm peak}$, $\dot{M}_{\rm peak}$ and $\Delta M$ as functions of $\beta$, an accurate and analytical prescription (i.e., one that does not rely on ad hoc, empirical fits of numerically obtained results) for the variation of these quantities with $\beta$ does not exist. Here we develop such a prescription, and in particular we extend the maximum gravity (MG) model developed in \cite{coughlin22} -- which postulates that a star is completely destroyed by tides when the tidal field overcomes the maximum self-gravitational field of the star (as opposed to its surface gravity) -- to predict $\Delta M, t_{\rm peak}$ and $\dot{M}_{\rm peak}$ for a partial TDE as a function of $\beta$. In Section~\ref{sec:analytical-model} we describe the analytical model and the predictions for $t_{\rm peak}$, $\dot{M}_{\rm peak}$, and $\Delta M$ as a function of $\beta$ for a given star. In Section~\ref{sec:hydro} we test these predictions against hydrodynamical simulations of the partial disruptions of stars evolved using {\sc mesa} \citep{paxton11,paxton13,paxton15,paxton18} for a range of pericenter distances. We find that the predicted  $t_{\rm peak}$ agrees with the numerical results to within $\sim20\%$ for less evolved stars, and to within $\sim 50\%$ for high-mass and evolved stars. The predictions for $\dot{M}_{\rm peak}$ agree to within a factor of $\sim2-3$ of the numerical results (the model predictions being lower in most cases) for 
most stars and orbital pericenter distances, but worsens for low-mass stars ($M_\star \lesssim 0.5 M_\odot$) on low-$\beta$ orbits; we attribute this disagreement primarily to the neglect of self-gravity in the model, which shifts $t_{\rm peak}$ to earlier times and $\dot{M}_{\rm peak}$ to higher values. We discuss the implications of the model for TDE observations in Section ~\ref{sec:discussion}, and summarize our results in Section~\ref{sec:summary}.

\section{Analytical model}
\label{sec:analytical-model}
\begin{table*}
\begin{center}
\begin{tabular}{|c|>{\centering\arraybackslash}p{10cm}|}
\hline
Mass/Radius variable & Definition \\
\hline
\multicolumn{2}{|c|}{{\bf Quantities relevant for complete disruptions}} \\
\hline
$M_\bullet$ & Mass of the SMBH that tidally disrupts a star. \\
\hline
$M_\star$ & Total mass contained in a star. \\
\hline 
$R_\star$ & Radius of the star (where stellar density equals zero). \\
\hline
$r_{\rm t} \equiv R_\star (M_\bullet/M_\star)^{1/3}$ & Canonical tidal radius, obtained by equating the tidal field to the surface gravity of a star of mass $M_\star$ and radius $R_\star$.\\
\hline
$R_{\rm c}$ &  The ``core radius'': the radius within a star at which its self-gravity is maximized.\\ 
\hline
$M_{\rm c}$ &  Mass contained within $R_{\rm c}$.\\
\hline
$r_{\rm t,c}(R_{\rm c})$ & Distance from the SMBH at which the tidal field equals the maximum self-gravity, and the star is compleley destroyed by tides. \\
\hline
$\beta_{\rm c} \equiv r_{\rm t}/r_{\rm t,c}(R_{\rm c})$ & Critical penetration factor for complete disruption. \\
\hline
\multicolumn{2}{|c|}{{\bf Quantities relevant for partial disruptions}} \\
\hline
$\beta \equiv r_{\rm t}/r_{\rm p}$ & Penetration factor of the orbit. 
\\
\hline
$M(R)$ & Stellar mass contained within some radius $R \leqslant R_\star$. \\
\hline
$r_{\rm t,c} (R) $ & The pericenter distance necessary for the tidal field to exceed $GM(R)/R^2$ (cf.~Equation~\ref{eq:critical_rtc}).\\
\hline
$\Delta M (R)$ & Mass stripped when the tidal field equals the self-gravitational field due to the mass contained interior to  $R$. \\
\hline
\end{tabular}
\caption{\label{tab:table1} Tabulated list of mass and radius variables characterizing complete and partial disruptions using the MG model.}
\end{center}
\end{table*}

The MG model \citep{coughlin22} posits that a star of mass $M_{\star}$ and radius $R_{\star}$ is completely destroyed when the tidal field of an SMBH overcomes its maximum self-gravitational field, which occurs at a radius within the star $R_{\rm c} < R_{\star}$ (the ``core'' radius). This model predicts the critical pericenter distance within which a star must come to be completely destroyed, $r_{\rm t,c}$, as well as $t_{\rm peak}$ and $\dot{M}_{\rm peak}$ for the fallback rate (see Equations 11 and 12 of \citealp{coughlin22}). The $t_{\rm peak}$ and $\dot{M}_{\rm peak}$ predictions from the MG model were tested against hydrodynamical simulations of the disruption of a wide range of stars evolved using {\sc mesa} in \cite{fancher25}, who found excellent agreement with the model for younger (zero age main-sequence; ZAMS), 
and reasonably good agreement (to within $\sim 35-50\%$ of the model prediction) for more chemically evolved stars. 

\begin{figure}
\begin{tikzpicture}
\filldraw[color=orange!50, fill=orange!50, very thick](-2,0) circle (1.6);
\filldraw[even odd rule,inner color=white,outer color=orange!40,color=orange!40](-2,0) circle (1.1);
\draw[color=orange!50](-2,0) circle (1.1);
\filldraw[color=black!100, fill=black!100, very thick](5,0) circle (0.3);
\draw[dashed,thick] (-2,0) -- (5,0);
\draw[thick] (-2,0) -- (-1.3,0.85);
\node[font=\normalsize] at (2,-0.3) {$r_{\rm t,c}(R)$};
\node at (5,-1.2) {SMBH};
\node[font=\normalsize] at (-2,-2) {star ($M_\star,\, R_\star$)};
\node at (-2.5,1.2) {$\Delta M$};
\node at (-1.8,0.55) {$R$};
\end{tikzpicture}
\caption{Schematic showing a star of mass $M_\star$, radius $R_\star$, being partially tidally stripped by an SMBH. The tidal field of the SMBH exceeds the self-gravitational field of the mass contained within some radius $R\leqslant R_\star$ at a pericenter distance $r_{\rm p}=r_{\rm t,c}(R)$, such that the mass exterior to this radius, $\Delta M$, is tidally stripped. When the tidal field of the SMBH exceeds the maximum self gravitational field within the star, it leads to complete disruption of the star, i.e., $\Delta M = M_\star$.  }
\label{fig:schematic}
\end{figure}
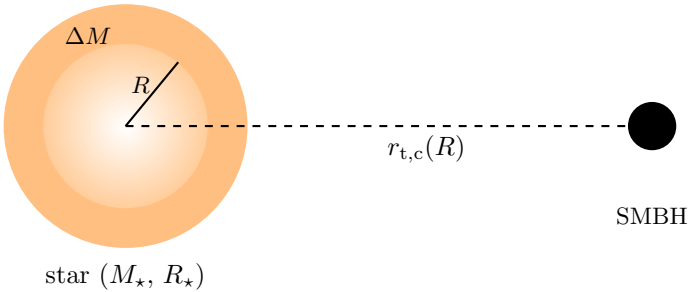

Here we extend the MG model to the partial TDEs, where the star has a pericenter distance $r_{\rm p}$ that is larger than the distance at which it is completely destroyed. Figure~\ref{fig:schematic} illustrates the tidal interaction, depicting the mass stripped on a given orbit, and Table \ref{tab:table1} lists the variables relevant to complete TDEs and those used here to characterize partials. To calculate the amount of mass stripped, we note that the tidal field of the SMBH at a given pericenter distance $r_{\rm p}$ will equal the self-gravitational field appropriate to some inner mass fraction of the star (i.e., the mass contained within some radius $R \leqslant R_\star$, where $R$ is the spherical radius measured from the center of the unperturbed star). Denoting this pericenter distance by $r_{\rm t,c}(R)$, it follows that $r_{\rm t, c}$ satisfies 
\begin{equation}
\label{eq:critical_rtc}
    \frac{4 G M_\bullet R}{r_{\rm t,c}^3(R)} = \frac{G M(R)}{R^2},
\end{equation}
\noindent{}where the tidal field a distance $r_{\rm p}=r_{\rm t,c}$ from the SMBH is given by the left side of Equation~\eqref{eq:critical_rtc} (the factor of $4$ arises from calculating the shear across the stellar diameter, rather than the radius; see Appendix \ref{sec:tidal_field} for further evidence to support this choice) and the self-gravitational field of a star (as a function of radial distance $R$ from its center) is on the right. Given some pericenter distance $r_{\rm p} = r_{\rm t,c} (R)$, we can solve Equation~\eqref{eq:critical_rtc} to obtain the value of $R$ that satisfies the equality. 

The amount of mass stripped, $\Delta M(R)$, would then seemingly and simply be equal to the amount of mass exterior to this radius in the original star, i.e., $\Delta M  = M_\star-M(R)$. However, \citet{coughlin22} argued that the star should be completely destroyed -- and hence $\Delta M = M_{\star}$ -- once the tidal field exceeds the maximum self-gravitational field in the stellar interior, which occurs at some finite radius $R_{\rm c}$. We thus estimate the amount of mass stripped in a partial TDE as
\begin{equation}
    \Delta M (R) = M_\star\left(\frac{M_\star-M(R)}{M_\star-M_{\rm c}}\right), \label{eq:deltaM}
\end{equation}
such that for any value of the spherical radius $R\leqslant R_\star$, there exists a unique value of the pericenter distance $r_{\rm t,c}(R)$ (and hence penetration factor $\beta$) that solves Equation~\eqref{eq:critical_rtc}, implying that $\Delta M$ is a function only of $\beta$. The normalization in Equation \eqref{eq:deltaM} accounts for the fact that the amount of mass stripped is $M_\star$ when the penetration factor $\beta$ equals the critical penetration factor $\beta_{\rm c}$ (as defined in Equation~8 of~\citealt{coughlin22}).

For any given $\beta$, the timescale on which the fallback rate peaks and the peak value itself can be estimated analogously to Equations (11) and (12) of~\cite{coughlin22}:
\begin{equation}
\begin{split}
    &t_{\rm peak}(\beta) = \left(\frac{r_{\rm t,c}^2(R)}{2 R}\right)^{3/2} \frac{\pi}{\sqrt{G M_\bullet}}, \\ 
    &\dot{M}_{\rm peak}(\beta) = \frac{\Delta M(\beta)}{4 t_{\rm peak}}, \label{eq:peak-fbr}
    \end{split}
\end{equation}
where $\Delta M(\beta)$ is the amount of massed stripped (given by Equation~\ref{eq:deltaM}), and $t_{\rm peak}$ depends implicitly on $\beta$ through $r_{\rm t,c}(R)$ (which can be calculated using Equation~\ref{eq:critical_rtc}). In the next section we use this formalism to predict the peak timescale $t_{\rm peak}$ and peak fallback rate $\dot{M}_{\rm peak}$ for a wide range of stellar masses and ages and penetration factors, ranging from $\beta \sim 0.6$ (for which $\Delta M \ll M_{\star}$) to $\beta_{\rm c}$ (for which $\Delta M \simeq M_{\star}$), and compare the predictions of the model to the results of numerical simulations of partial TDEs\footnote{\cite{coughlin22} argued that the partial tidal disruption radius, being the minimum pericenter distance at which the tidal interaction strips off any mass, occurs when $\beta \simeq 0.6$, independent of stellar properties; our results confirm this as a corollary.}. The fallback rates from the hydrodynamical simulations are publicly available on Zenodo \citep{bandopadhyay_2025_17822028}. See \citealt{bandopadhyay24, fancher25} for a comparison of the predictions of the MG model to numerical simulations where the star is completely destroyed.

\section{Hydrodynamical Simulations}
\label{sec:hydro}
\subsection{Simulation Setup}
We simulated the partial disruption of 23 stars of solar metallicity (at zero-age main-sequence; ZAMS) with masses ranging from $0.2-5.0 M_\odot$, at different stages of their main-sequence evolution, using the Smoothed Particle Hydrodynamics code {\sc phantom}~\citep{price18}. We used {\sc mesa} to evolve the stars from their ZAMS stage (when the star has just begun fusing hydrogen in its core) to the terminal age main-sequence (TAMS; when the core hydrogen fraction drops below $0.1\%$). The main-sequence lifetimes of low-mass ($\lesssim 0.9 M_\odot$) stars exceed the age of the Universe. For $M_\star < 0.6 M_\odot$, we performed simulations only for the ZAMS stage, since these stars do not significantly chemically evolve within the age of the Universe. For $M_\star \in \lbrace 0.6,0.7,0.8 \rbrace M_\odot$ we simulated disruptions of the stars at the ZAMS stage and at $14$ Gyr, by which point the star was structurally obviously distinct from its ZAMS state. For $M_\star \geqslant0.9 M_\odot$, we considered three different ages -- ZAMS, MAMS (middle age main sequence, when the core hydrogen fraction drops to $\sim 0.2$), and TAMS. For each star, we simulate the disruption for 22 values of the penetration factor $\beta$, equally spaced between $\beta_{\rm partial} = 0.6$ and $\beta_{\rm c}$, at which the tidal field of the SMBH exceeds the maximum self-gravitational field of the star (see Figure 4 of~\citealp{bandopadhyay24} for the variation of $\beta_{\rm c}$ with stellar mass and age), resulting in a total of $1276$ simulations.

The setup used for the {\sc phantom} simulations is identical to that used in earlier works for {\sc mesa}-evolved stars, e.g., \cite{golightly19,nixon21,bandopadhyay24b,fancher25}.We used $10^6$ SPH particles to model each star and an SMBH mass of $10^6M_\odot$. Each simulation was initialized by placing the star at a distance $5r_{\rm t}$ from the SMBH, with its center-of-mass (COM) on a Keplerian parabolic orbit having a pericenter distance $r_{\rm p} = r_{\rm t}/\beta$. To calculate the fallback rates, the accretion radius of the SMBH was increased to to $3 r_{\rm t}$, and the surviving stellar core was replaced with a sink particle $\sim 2$ days past pericenter passage (except in case of a complete disruption, which for most of the ZAMS stars, occurs at $\beta\lesssim\beta_{\rm c}$). The rate at which particles from the tidally disrupted debris stream return to the accretion radius of the SMBH yields the fallback rate. 
To reduce the numerical noise, we average the fallback rates as described in, e.g.,~\cite{miles20,nixon21}, i.e., they are binned by time at early times, and by particle number at late times. Subsequently, we follow the approach described in \cite{nixon21} and fit each fallback rate by a Pad{\'e} approximant of the form
\begin{figure}
    \advance\leftskip-0.5cm
    \includegraphics[width=0.51\textwidth]{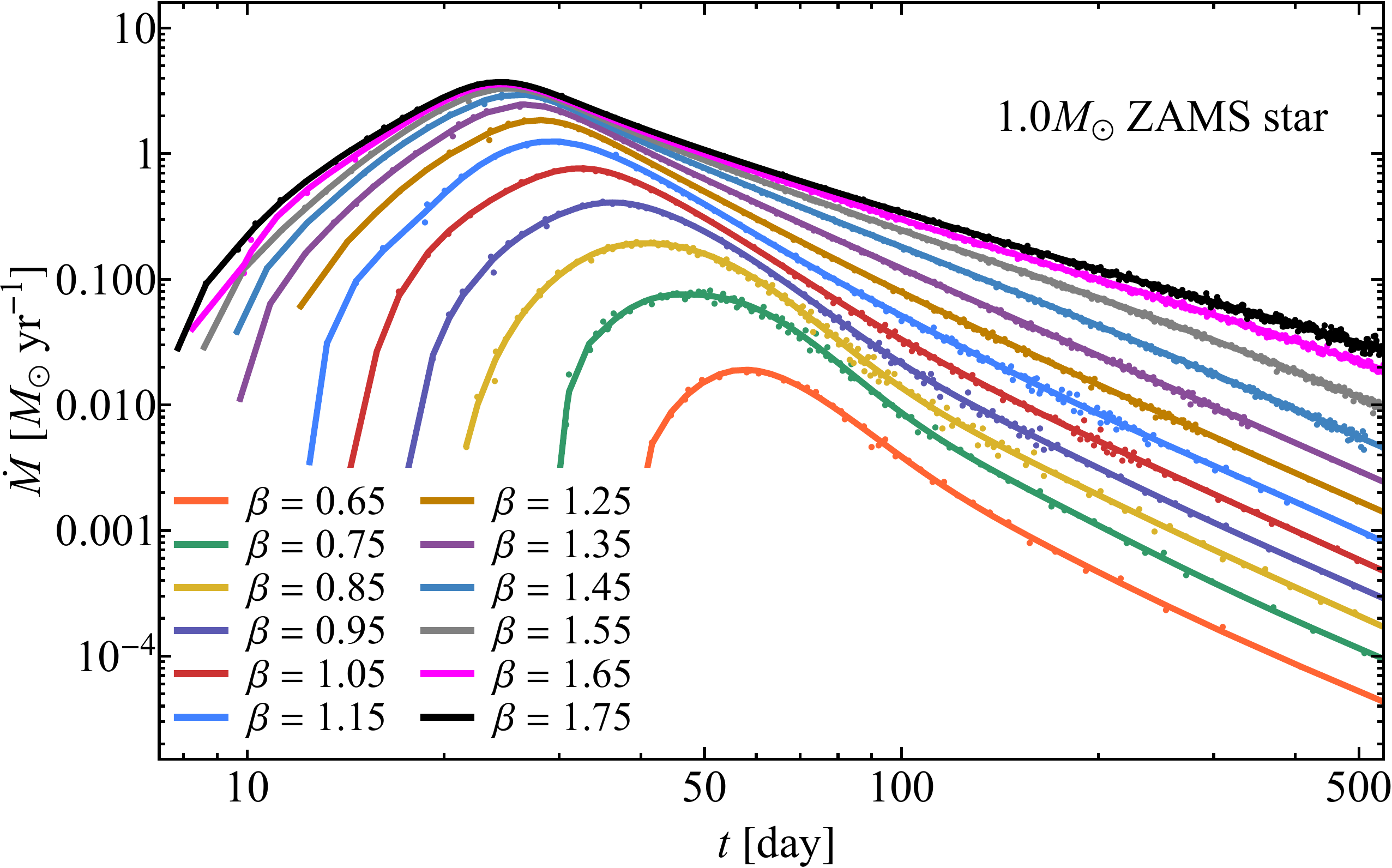} 
    \caption{The numerically obtained fallback rates for a $1.0M_\odot$ ZAMS star (shown by the discrete data points) along with the fits to the functional form given by Equation~\eqref{eq:pade-approx} (solid curves).} \label{fig:pade-approx-eg}
\end{figure}
\begin{equation}
    \dot{M}_{\rm fit} = \frac{a \tilde{t}^{m}}{1+ (a/b) \tilde{t}^{m-n_\infty}}\frac{1+\sum \limits_{i=1}^{N_{\rm max}-1} c_i \tilde{t}^{i}+\tilde{t}^{N_{\rm max}}}{1+\tilde{t}^{N_{\rm max}}}, \label{eq:pade-approx}
\end{equation}
where $\tilde{t}=t/t_{\rm peak}$, and $a,b,m,n_{\infty}$ and $\lbrace c_i\rbrace_{i=1}^{N_{\rm max}-1}$ are constants that are fit by minimizing the $\chi^2$ of the logarithm of the fallback rate\footnote{We note that the parameter $m$, which constrains the early time fallback, is dispensable in the case of fitting numerical data, since a sufficiently large number of $c_i$'s in the ratio of polynomials in Equation~\eqref{eq:pade-approx} can achieve this independent of $m$. However, when the number of coefficients $c_i$ is limited by the physical constraints under consideration (see Appendix \ref{sec:appendix} for an example), we require $m>0$ to restrict the range of $\tilde{t}$ from $0$ to $\infty$. }. Figure~\ref{fig:pade-approx-eg} shows the fits to the fallback data for a $1M_\odot$ ZAMS star obtained using $N_{\rm max}=10$ in Equation~\eqref{eq:pade-approx}. \citet{nixon21} showed that $N_{\rm max}=5$ yields reasonable fits for most fallback rates, but here we require $N_{\rm max}\sim 10$ to accurately reproduce the peak as $\beta$ approaches $\beta_{\rm c}$, and thus choose $N_{\rm max}=10$ to fit the fallback data for all disruptions.

\begin{figure*}
    \includegraphics[width=0.51\textwidth]{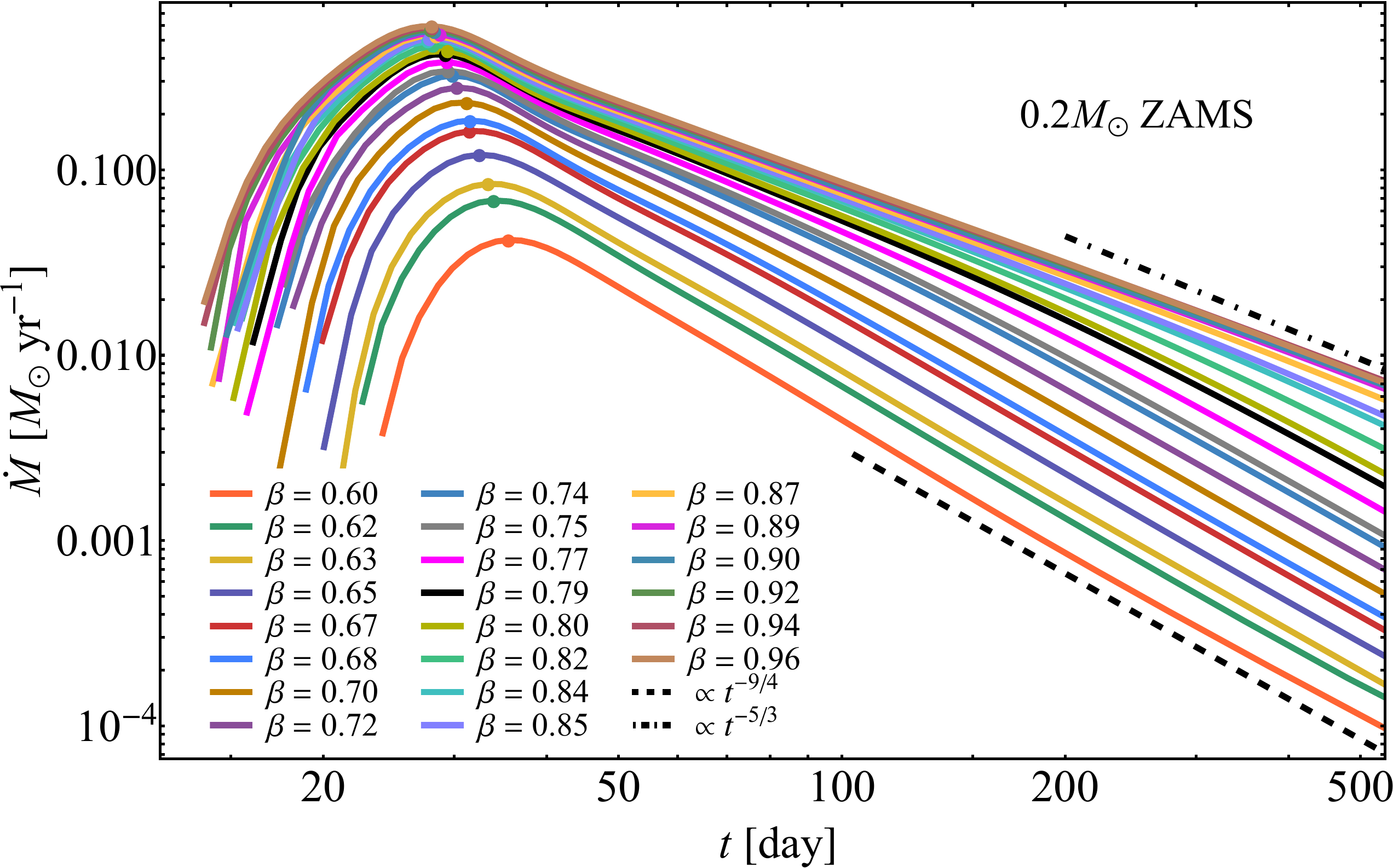}
    \includegraphics[width=0.51\textwidth]{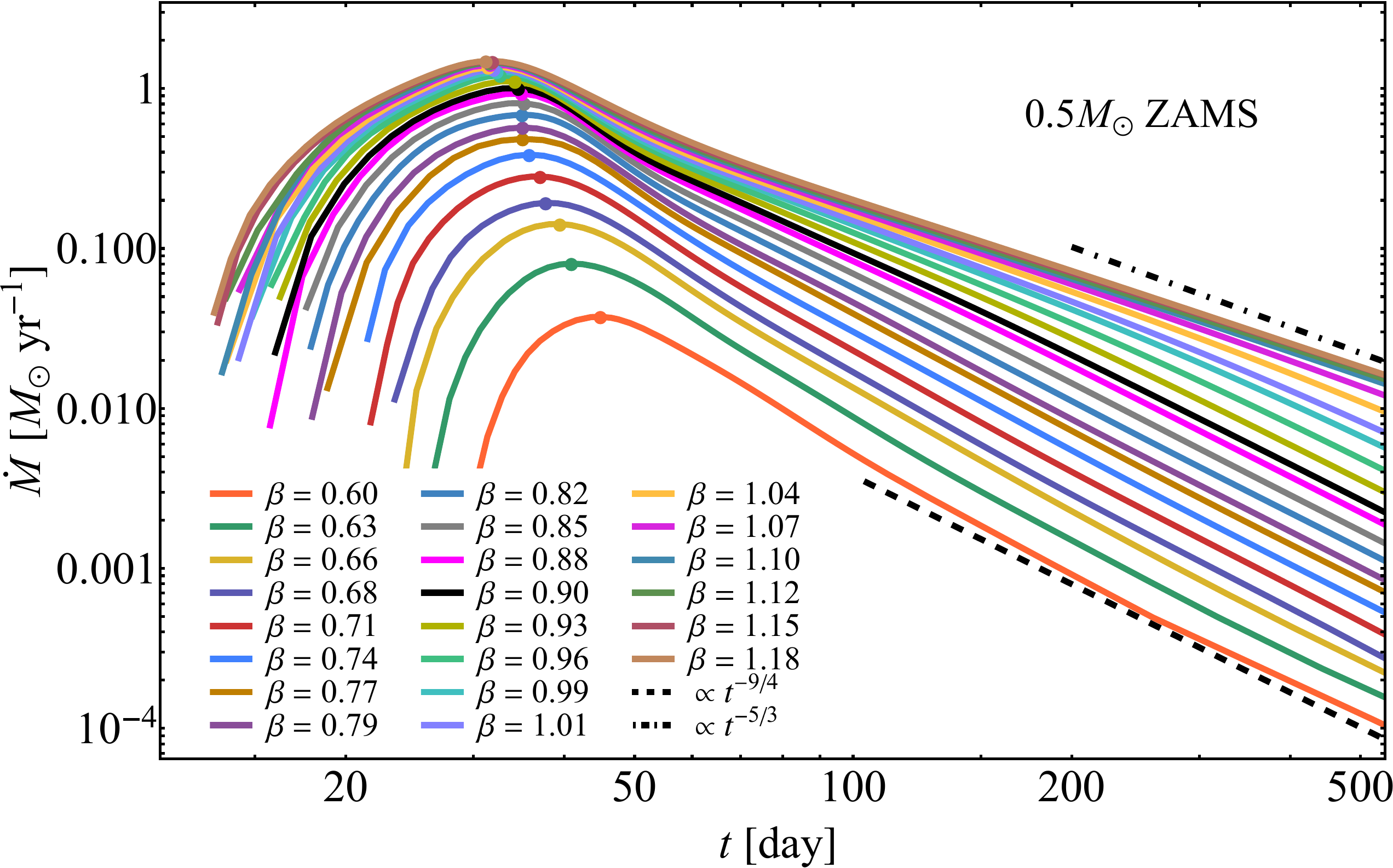} \\
    \includegraphics[width=0.51\textwidth]{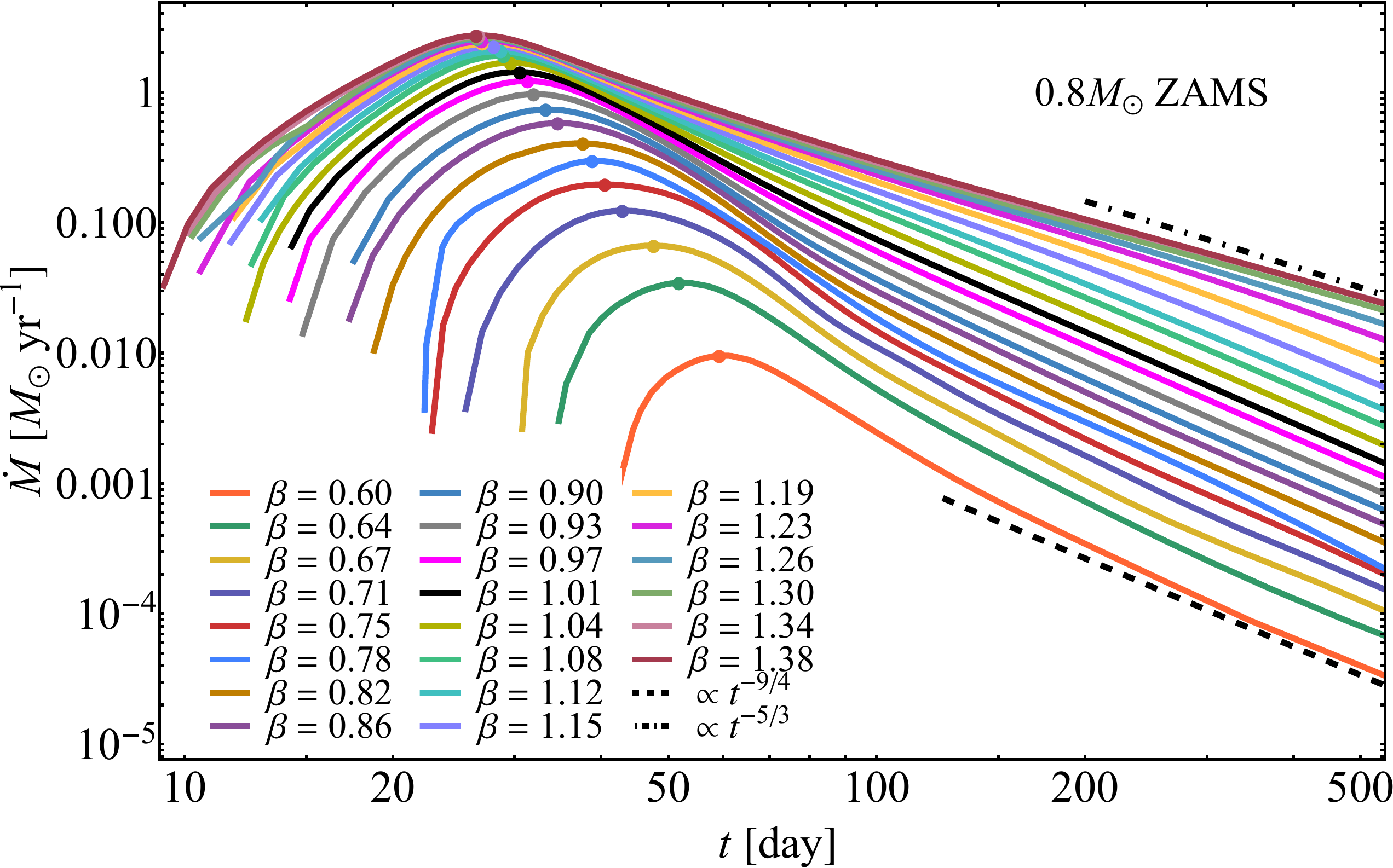}
    \includegraphics[width=0.51\textwidth]{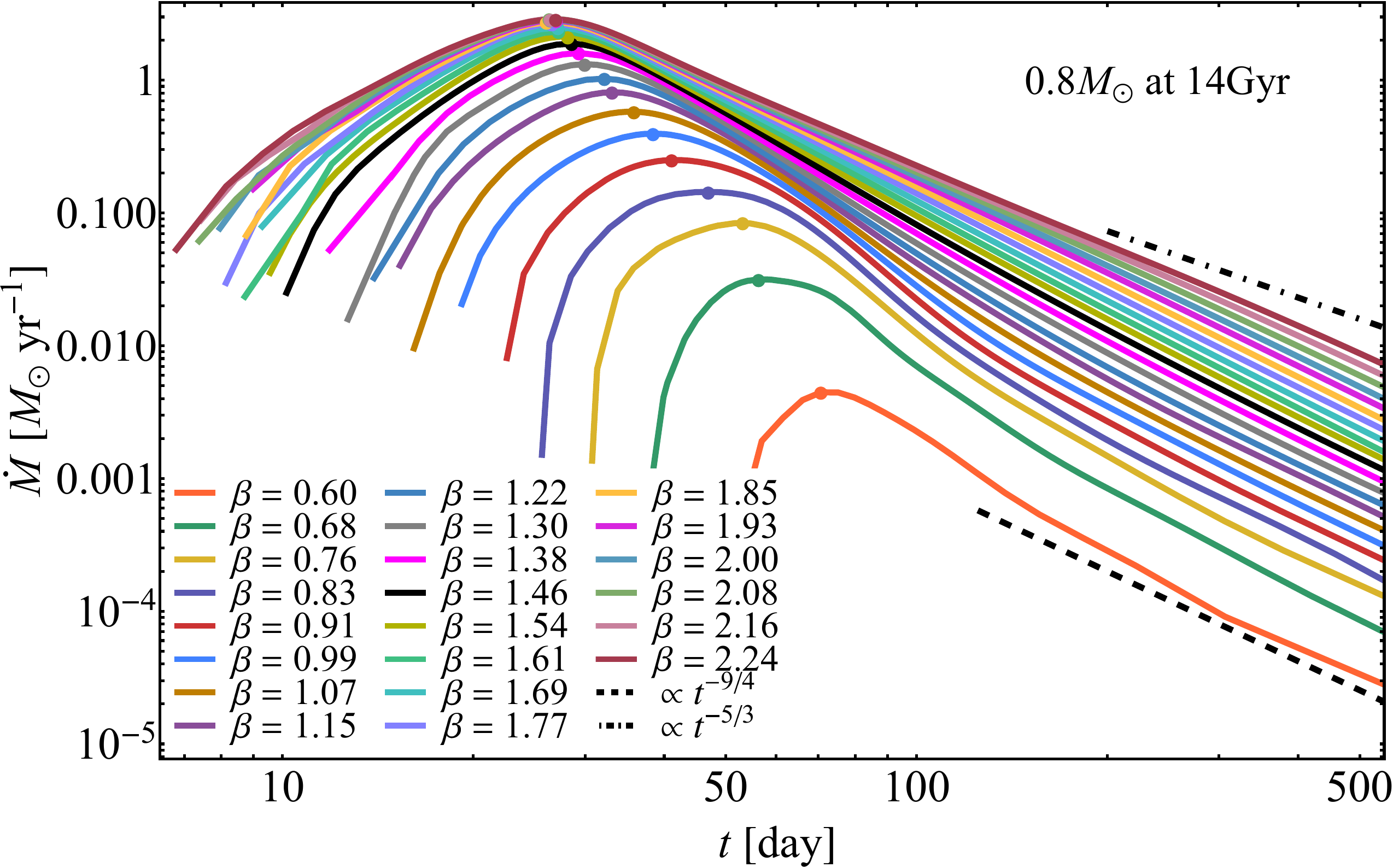}
    \caption{Fallback rates for a $0.2M_\odot$ ZAMS star (top left), a $0.5 M_\odot$ ZAMS star (top right), a $0.8 M_\odot$ ZAMS star (bottom left), and a $0.8M_\odot$ star at $14$Gyr (botom right), with the penetration factor $\beta$ ranging from $0.6$ to $\beta_{\rm c}$ (critical $\beta$ for complete disruption). The fallback rates for the partial disruptions scale as $\propto t^{-9/4}$ at late times. The peak of each fallback rate is indicated with a circle.} \label{fig:low-mass-fallbackrates}
\end{figure*}
\begin{figure*}
    \includegraphics[width=0.51\textwidth]{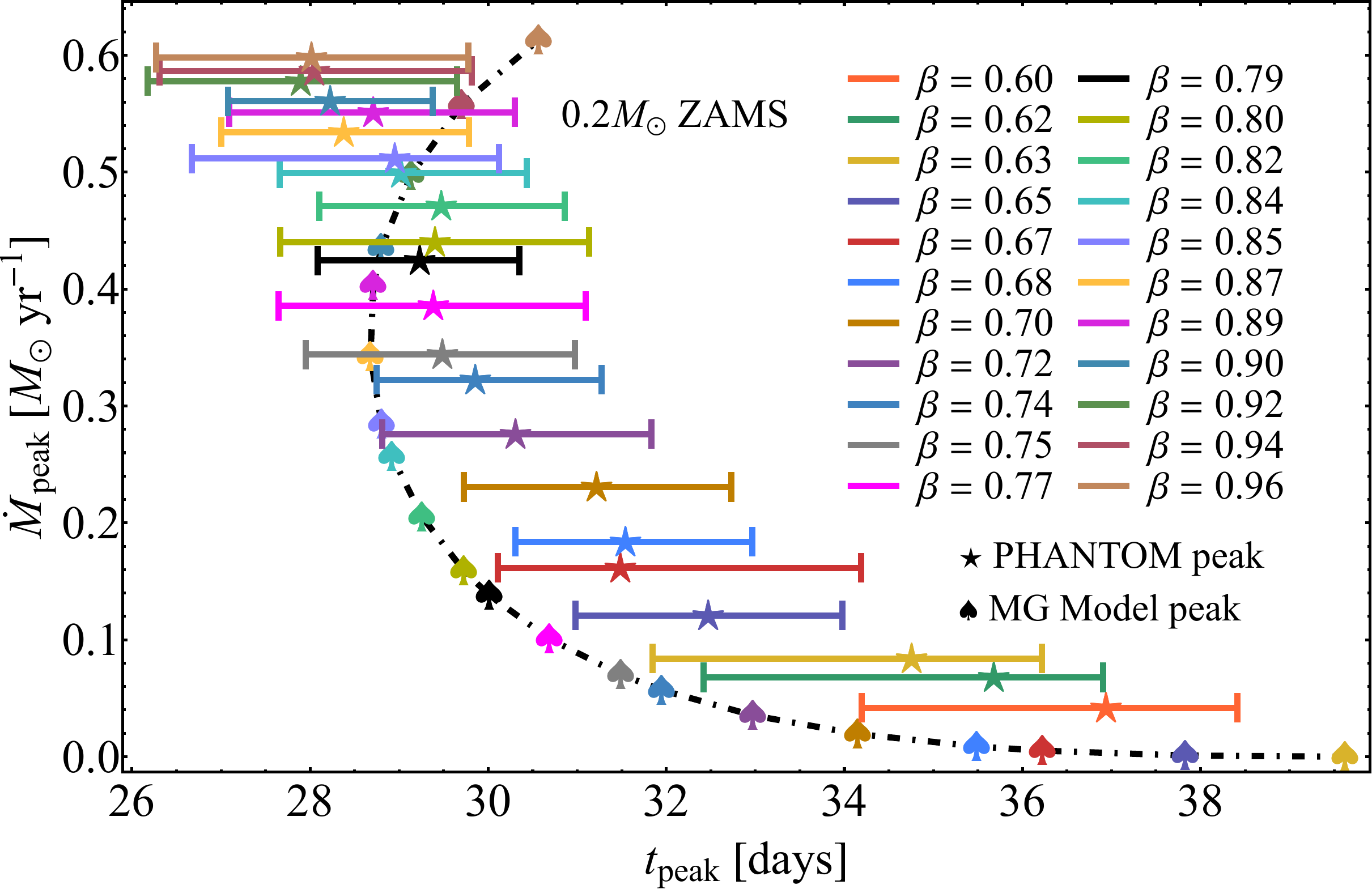}
    \includegraphics[width=0.51\textwidth]{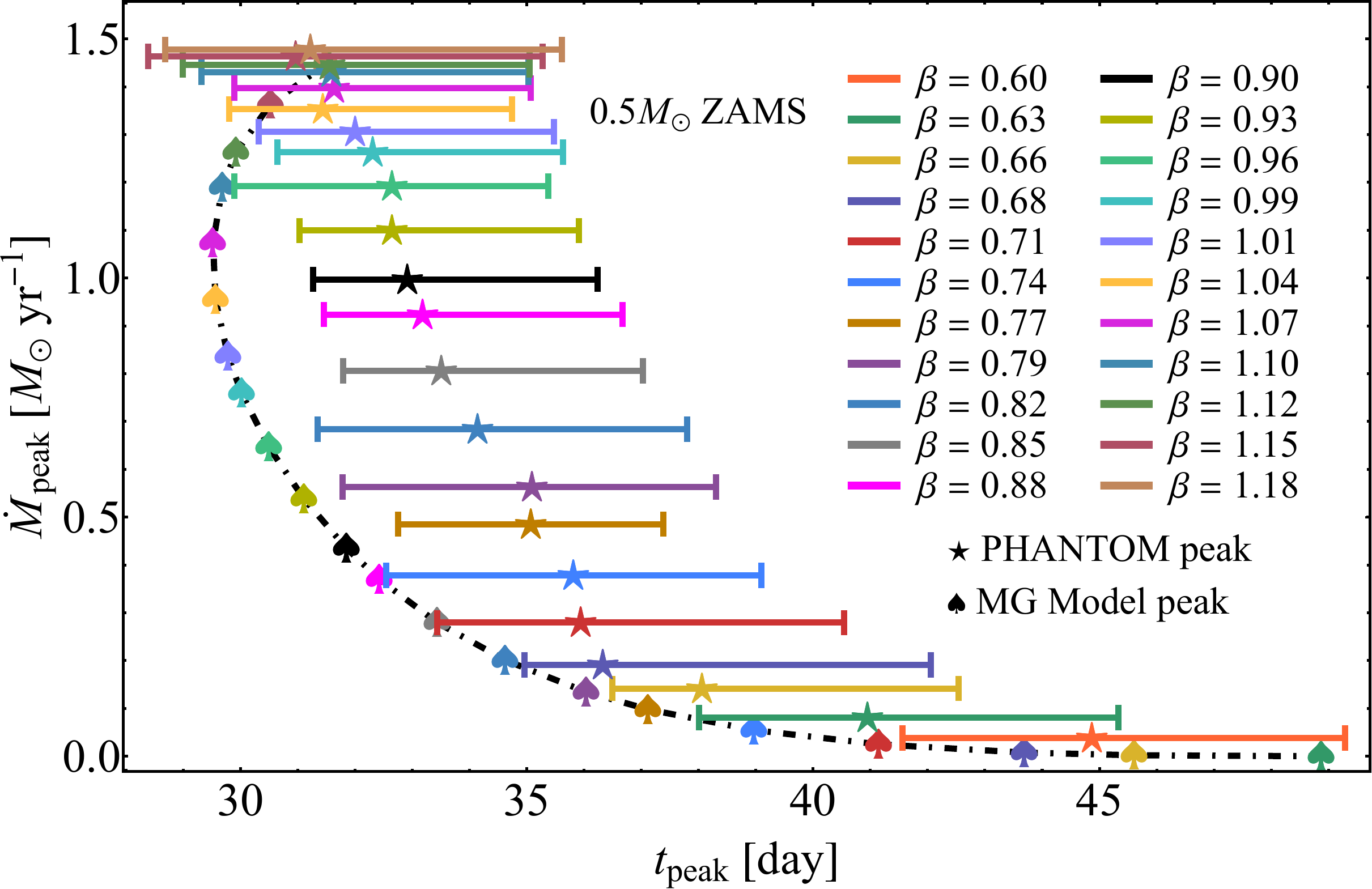} \\
    \includegraphics[width=0.51\textwidth]{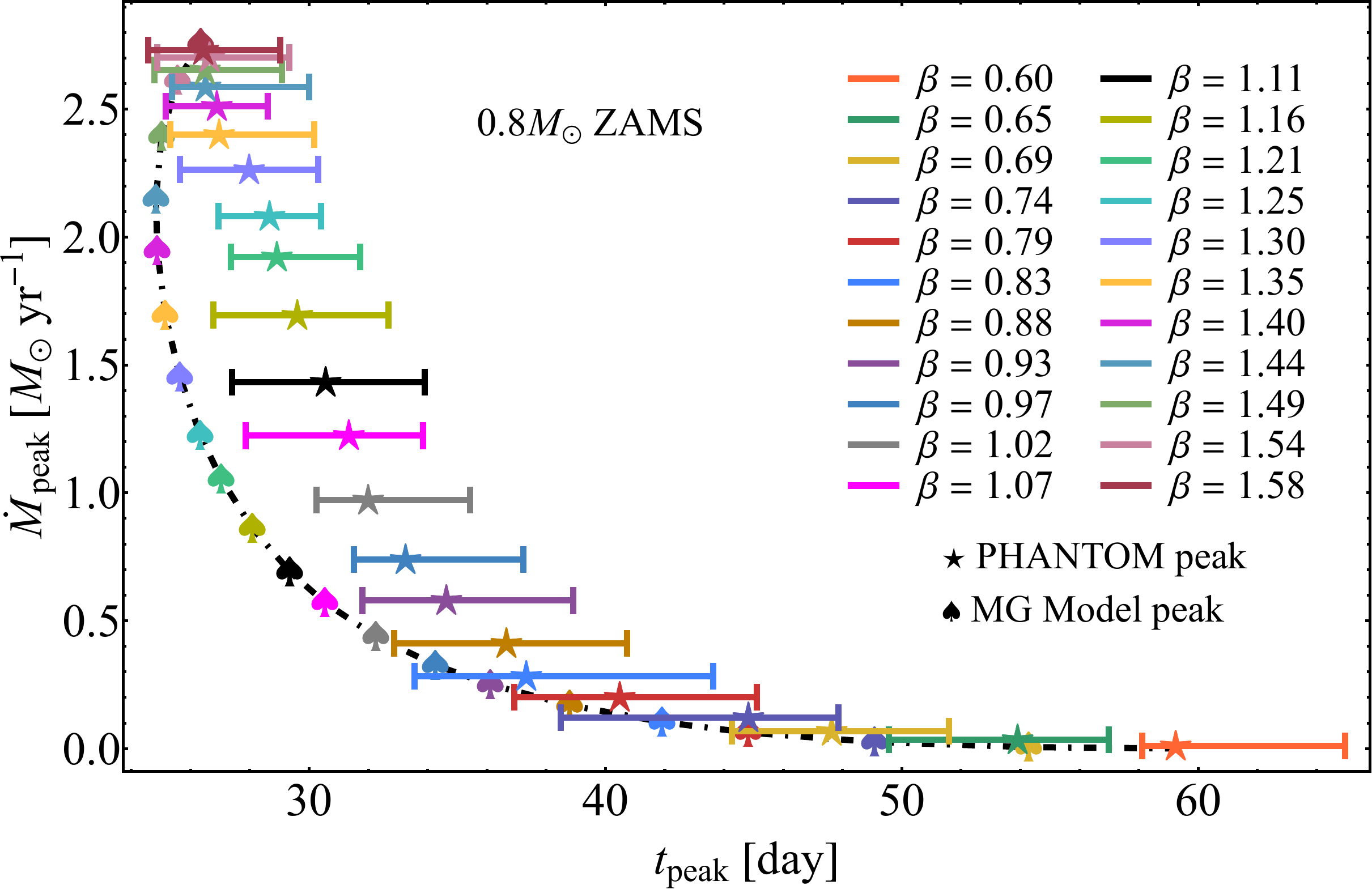}
    \includegraphics[width=0.51\textwidth]{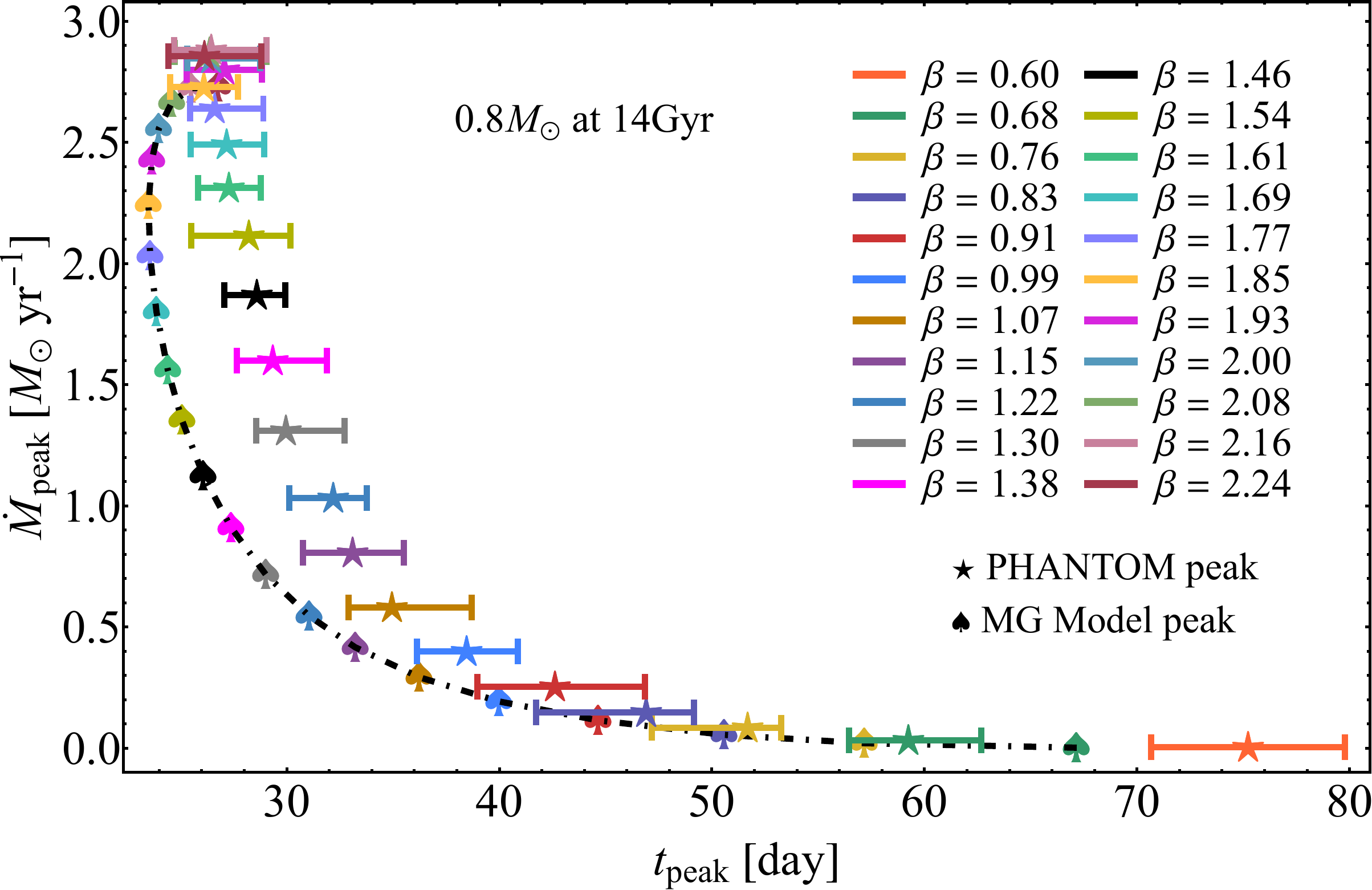}
    \caption{The peak timescale  $t_{\rm peak}$ and peak fallback rate $\dot{M}_{\rm peak}$ for a $0.2M_\odot$ ZAMS star (top left), a $0.5 M_\odot$ ZAMS star (top right), a $0.8 M_\odot$ ZAMS star (bottom left), and a $0.8M_\odot$ star at $14$Gyr (botom right), with the penetration factor $\beta$ ranging from $0.6$ to $\beta_{\rm c}$ (the MG model prediction for the critical penetration factor), as indicated in the legend. With an increase in $\beta$, the peak timescale $t_{\rm peak}$ shifts to earlier times, and the peak fallback rate $\dot{M}_{\rm peak}$ increases. The error bars show the duration for which $\dot{M}>95\%\dot{M}_{\rm peak}$.} \label{fig:low-mass-peak-fbrs}
\end{figure*}

\begin{figure*}
    \includegraphics[width=0.515\textwidth]{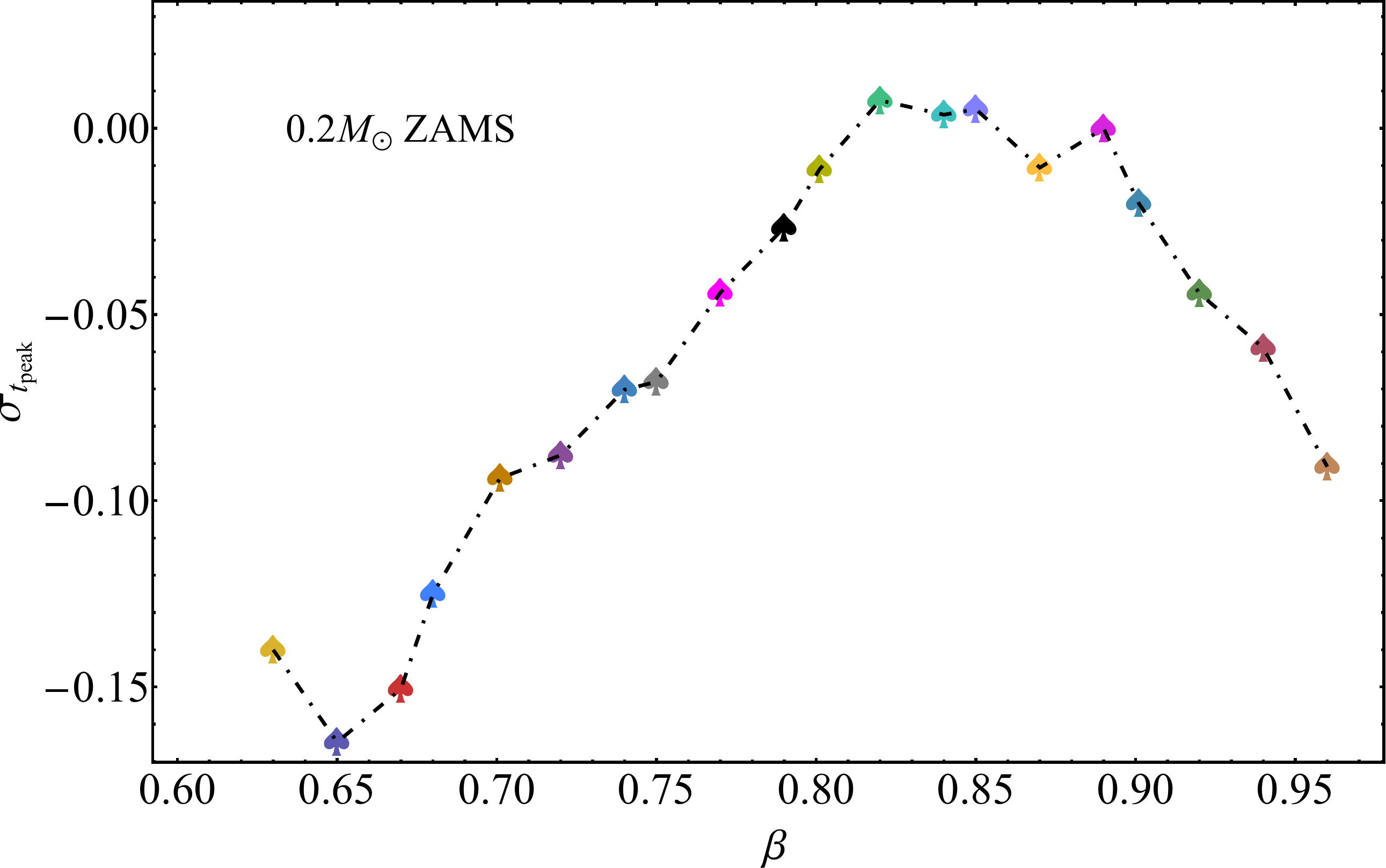}
    \includegraphics[width=0.51\textwidth]{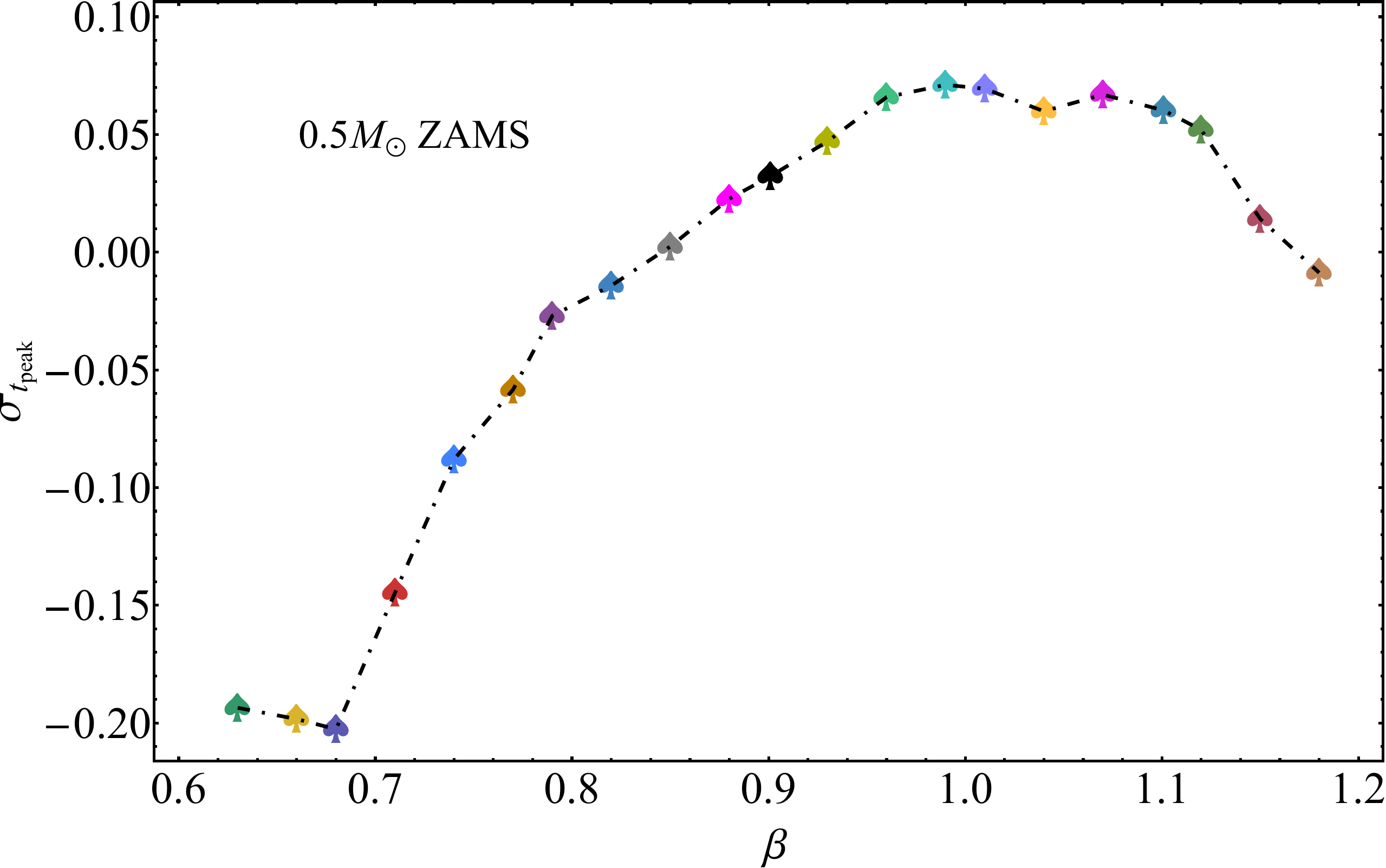} \\
    \includegraphics[width=0.51\textwidth]{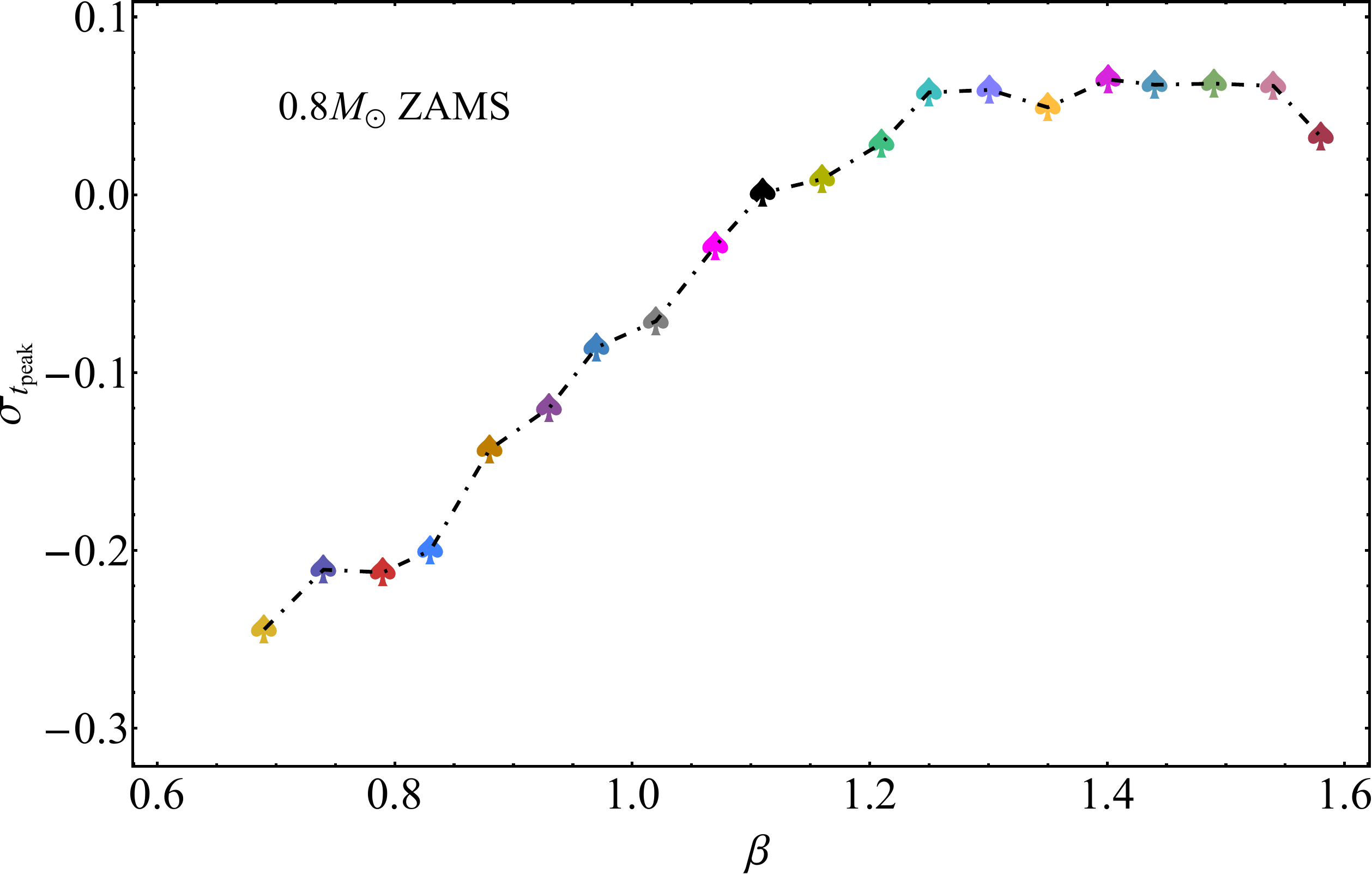}
    \includegraphics[width=0.51\textwidth]{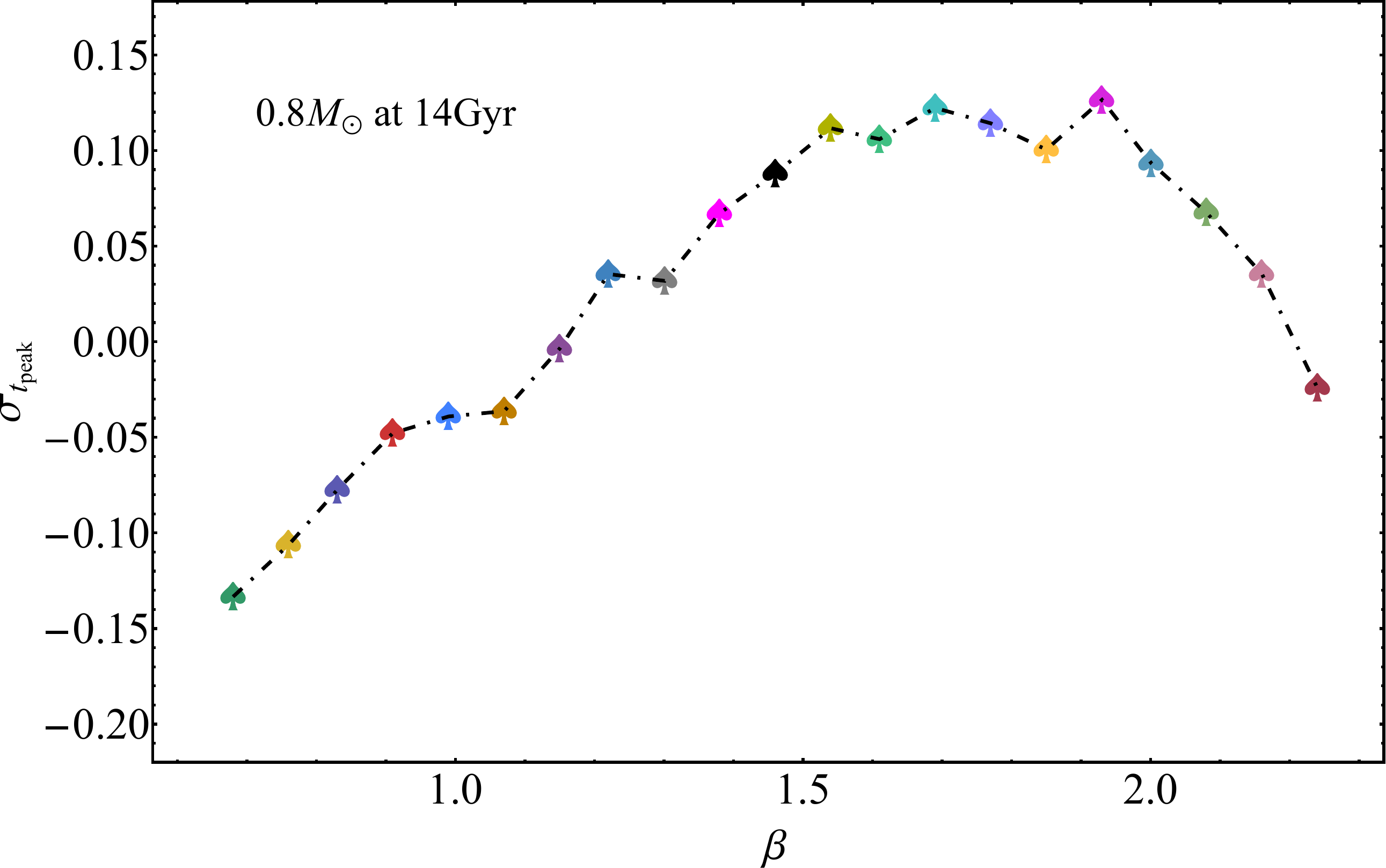}
    \caption{The relative error in  $t_{\rm peak}$, defined as $\sigma_{\rm t_{\rm peak}} \equiv (t_{\rm peak,hydro}-t_{\rm peak,MG})/t_{\rm peak,hydro}$ (where $t_{\rm peak,hydro}$ is the peak timescale measured from the {\sc phantom} simulation and $t_{\rm peak,MG}$ is the model prediction), as a function of $\beta$ for a $0.2M_\odot$ ZAMS star (top left), a $0.5 M_\odot$ ZAMS star (top right), a $0.8 M_\odot$ ZAMS star (bottom left), and a $0.8M_\odot$ star at $14$Gyr (botom right). } \label{fig:low-mass-tpeak-error}
\end{figure*}
\subsection{Mass Fallback Rates}
\subsubsection{Low-mass stars}
The top panels of Figure~\ref{fig:low-mass-fallbackrates} show the fallback rates for the partial disruption of a $0.2M_\odot$ ZAMS star and a $0.5M_\odot$ ZAMS star for $\beta$ given by the legend, from $\beta = 0.6$ to $\beta_{\rm c}$, where $\beta_{\rm c}$ is the MG model prediction for complete disruption. The fallback rates rise and peak on a timescale of $\sim30-50$ days, with the magnitude of the peak growing as a function of $\beta$. The bottom panel shows fallback rates for a $0.8M_\odot$ star at ZAMS (left) and at $14$Gyr (right), by which time its central density has increased by a factor of $\sim 2.5$ (even though it is still on the main sequence), which increases its value of $\beta_{\rm c}$ (relative to the ZAMS star). The fallback rates for the ZAMS star have slightly shorter peak timescales and higher peak values for orbits with comparable $\beta$-values. We also note that all the fallback rates for the $0.8 M_\odot$ star at 14Gyr shown in the right panel of the figure exhibit a late time scaling $\propto t^{-9/4}$, characteristic of partial disruptions~\citep{coughlin19}, indicating that the star does not get completely destroyed at the $\beta_{\rm c}$ predicted by the MG model. 

Figure~\ref{fig:low-mass-peak-fbrs} shows the peak of the fallback rates (i.e., $t_{\rm peak}$ and $\dot{M}_{\rm peak}$) for the same stars and range of pericenter distances as shown in Figure~\ref{fig:low-mass-fallbackrates}, with the $5-$pointed stars (spades) depicting the peak fallback rates from the {\sc phantom} simulations (MG model). To quantify the statistical error in $t_{\rm peak}$ calculated from the hydrodynamical simulations (e.g., due to binning effects in the fallback rates), the {\sc phantom} data points are plotted along with error bars that delineate the time over which the fallback rate is $\geqslant95\%$ of its peak value. The peak timescale ranges between $\sim 28-38$ days for the $0.2M_\odot$ star, and between $\sim 30-45$ days for the $0.5 M_\odot$ star. For the $0.8M_\odot$ star (shown in the left and right panels of the bottom row), the peak timescales range between $\sim20-60$ days for the ZAMS star, and between $\sim 30-75$ days for the star at 14Gyr. 

In Figure~\ref{fig:low-mass-tpeak-error} we plot the relative error in $t_{\rm peak}$, defined as $\sigma_{t_{\rm peak}} \equiv (t_{\rm peak,hydro}-t_{\rm peak,MG})/t_{\rm peak,hydro}$ (where $t_{\rm peak,hydro}$ is the peak timescale obtained from the {\sc phantom} simulation and $t_{\rm peak,MG}$ is the model prediction), as a function of $\beta$ for the four stars discussed above. In all cases the prediction is accurate to within $\sim 30\%$, with the largest discrepancies occurring at the smallest $\beta$ (where the model generally over-predicts $t_{\rm peak}$ relative to the hydrodynamical simulations). The predictions for $\dot{M}_{\rm peak}$ agree to within a factor of $\sim2-3$ of the numerical simulations in most cases, with the exception of the low-$\beta$ orbits, for which the discrepancy in the mass stripped $\Delta M$ is larger (see Section~\ref{sec:critical-beta}). Additionally, the discrepancies for small values of $\beta$ can be attributed to the fact that our model ignores the self-gravity of the debris stream, which acts to shift $t_{\rm peak}$ to earlier times and $\dot{M}_{\rm peak}$ to higher values in the {\sc phantom} simulations. 
\begin{figure*}
    \includegraphics[width=0.495\textwidth]{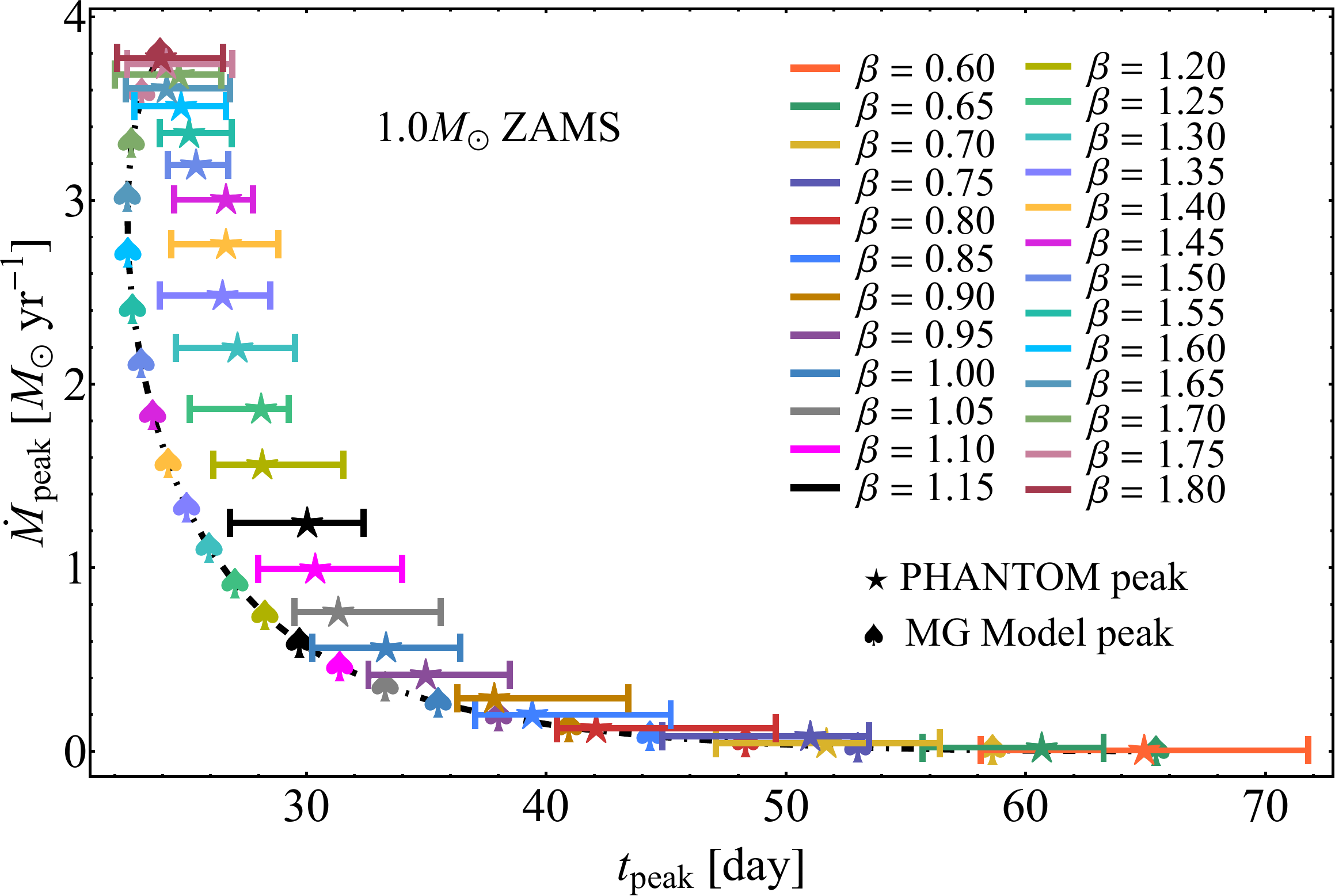}
    \includegraphics[width=0.52\textwidth]{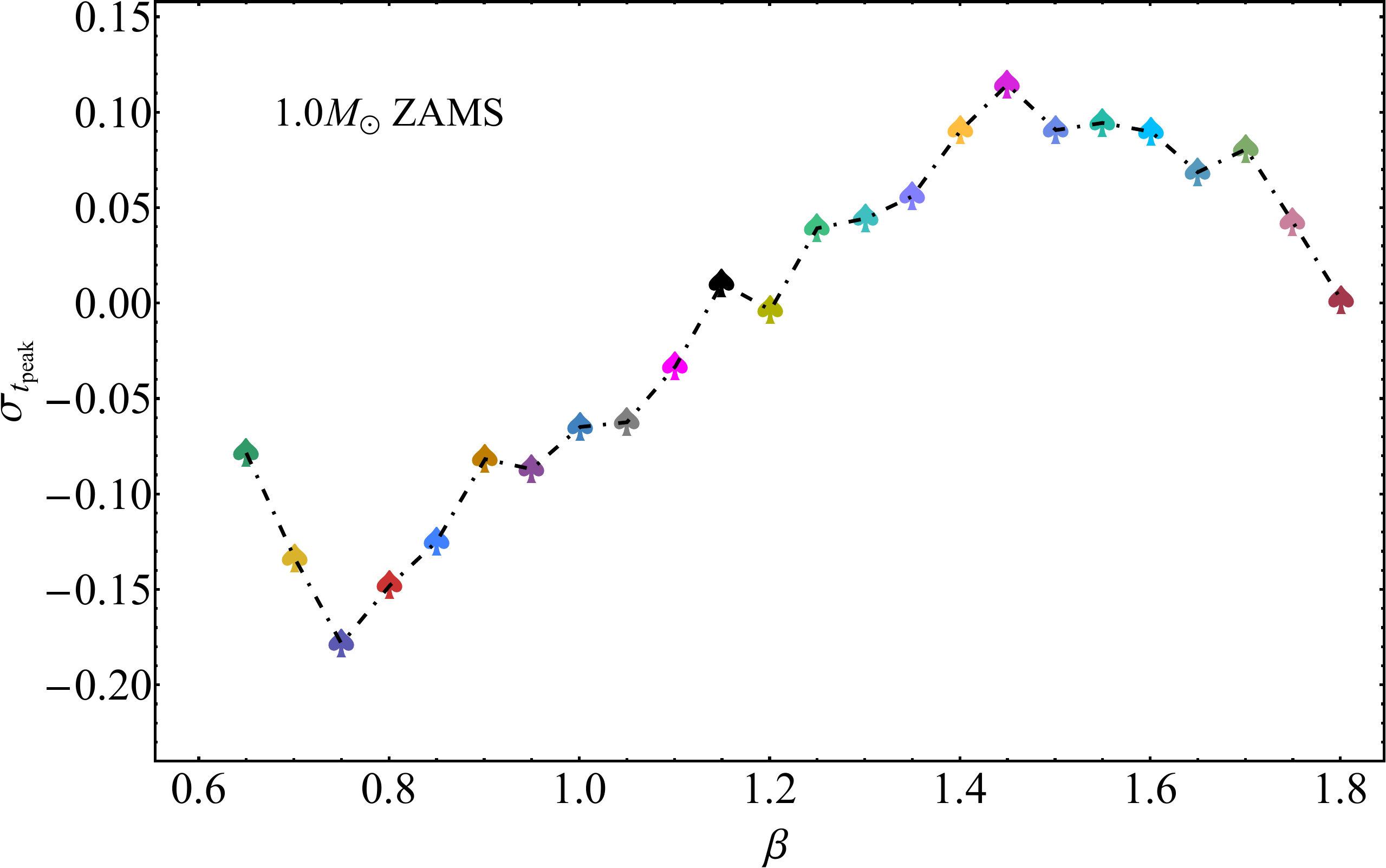}\\
    \includegraphics[width=0.495\textwidth]{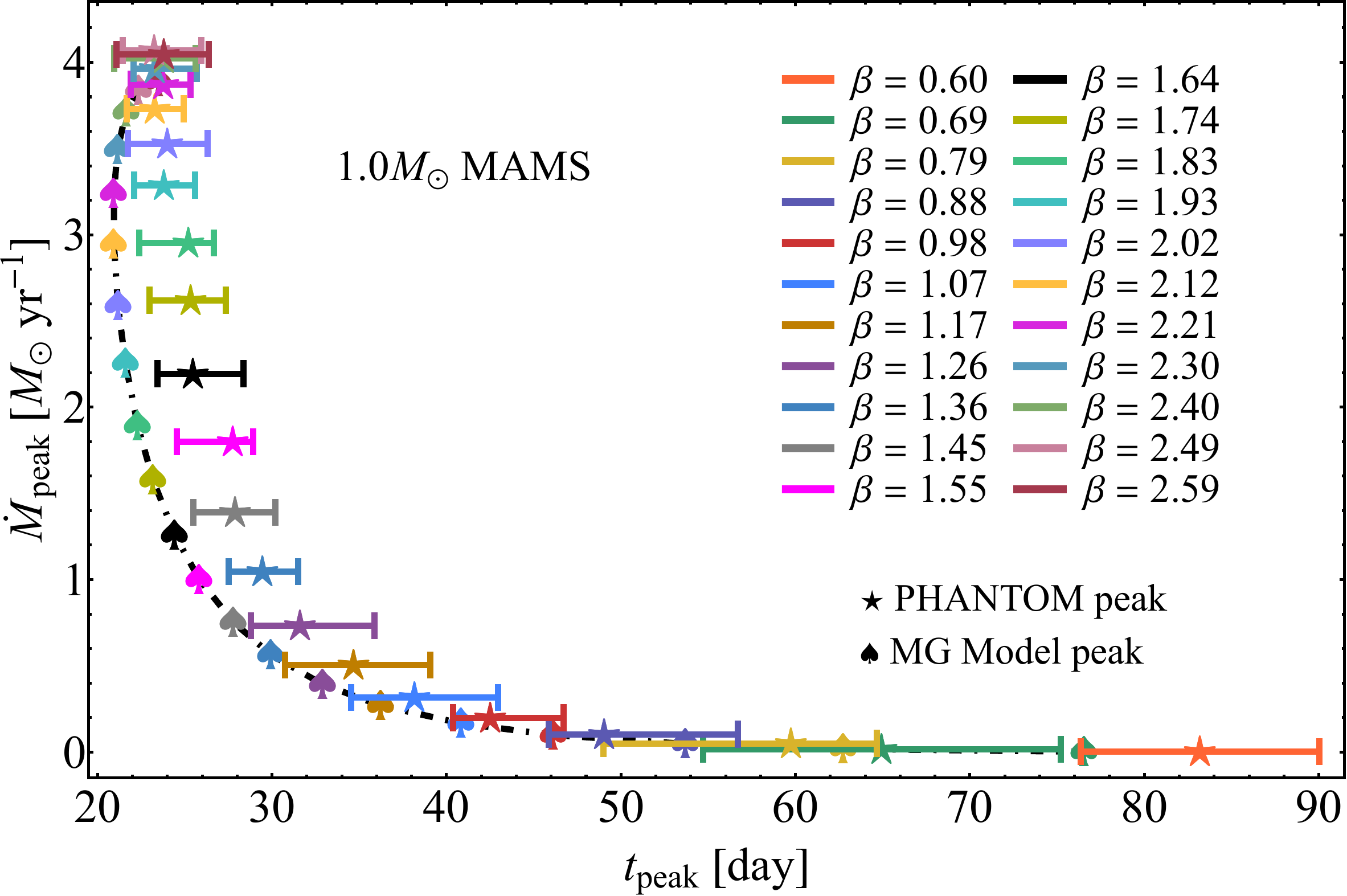} 
    \includegraphics[width=0.52\textwidth]{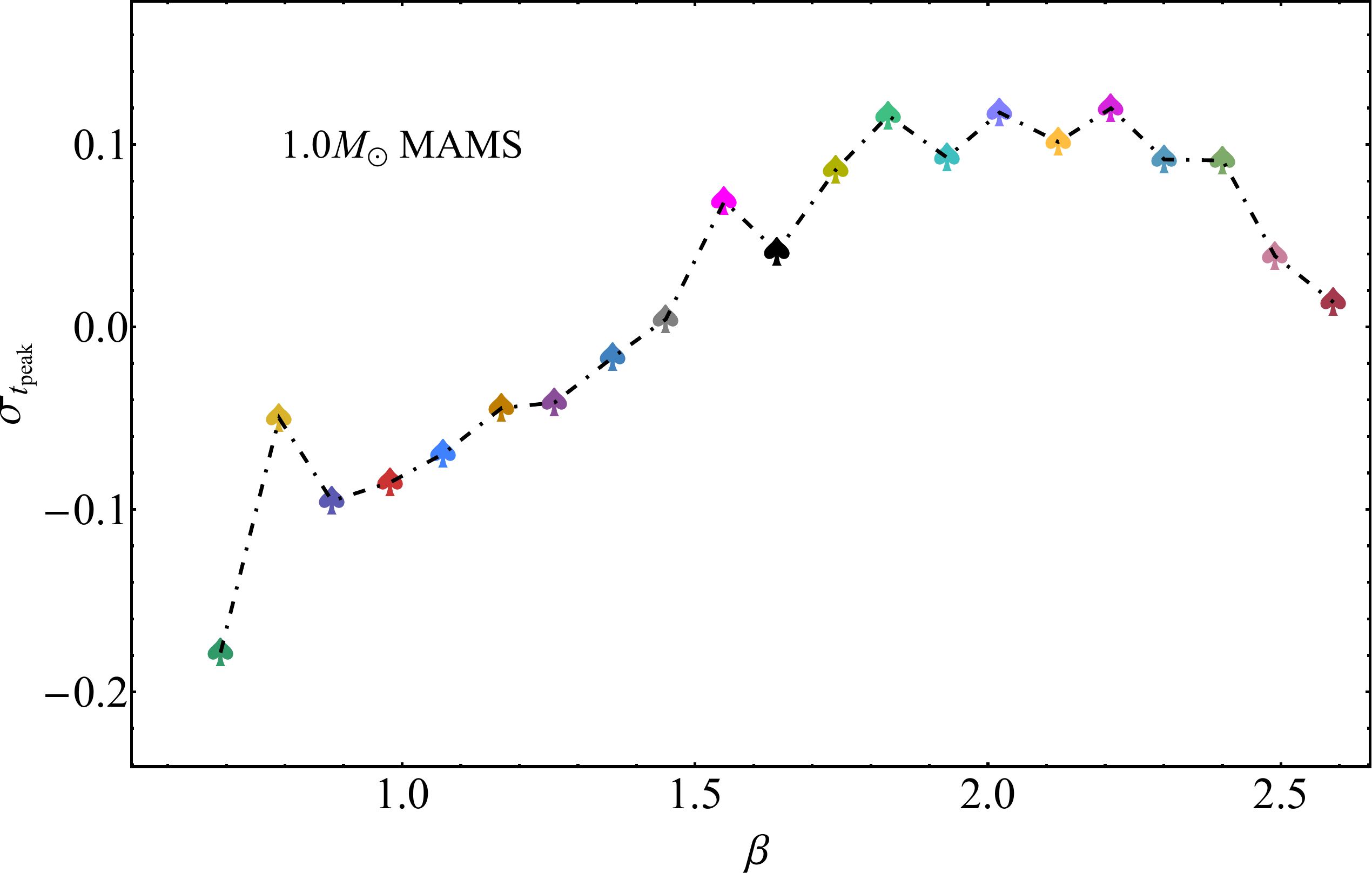}\\
    \includegraphics[width=0.495\textwidth]{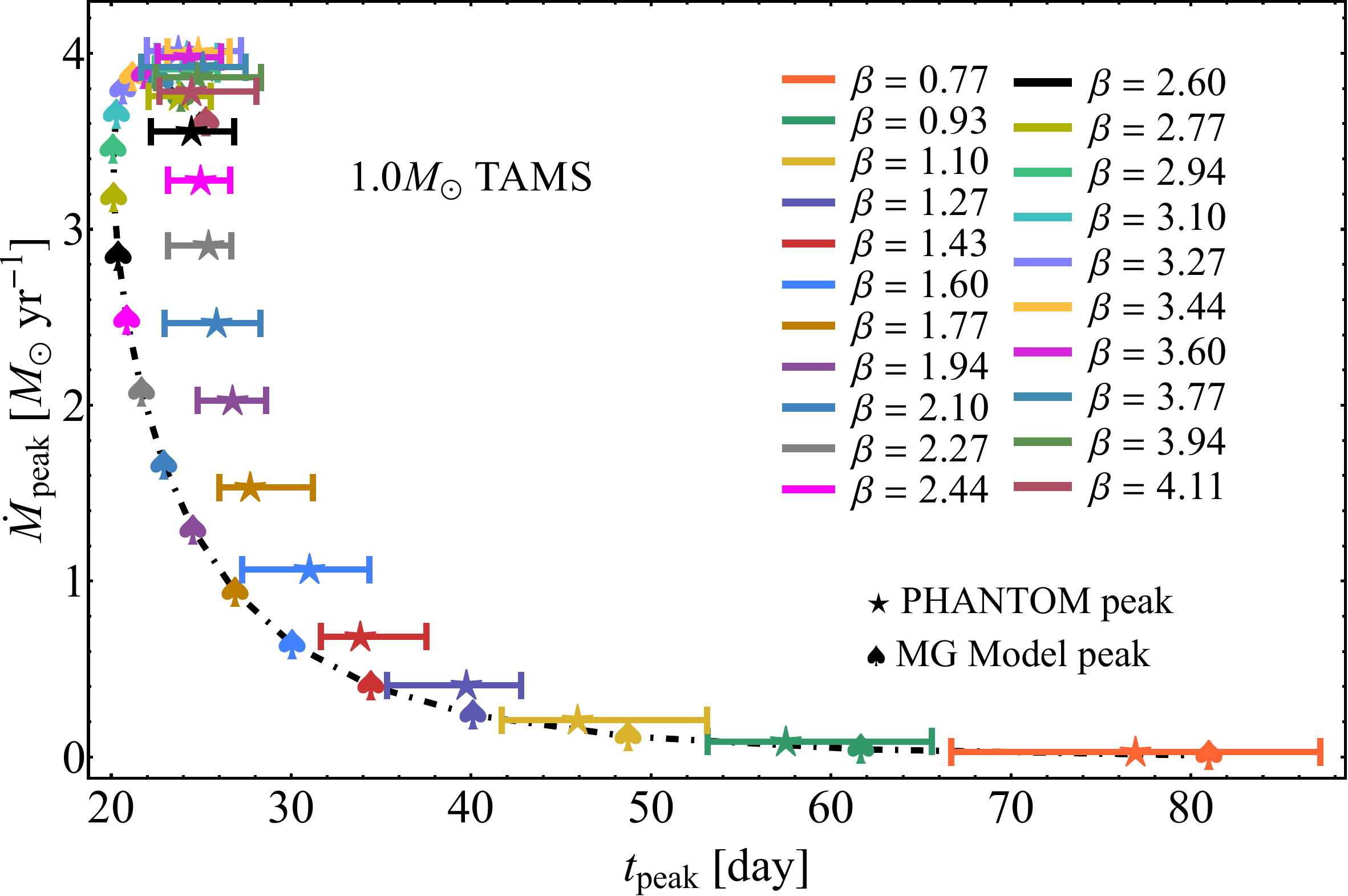}
    \includegraphics[width=0.52\textwidth]{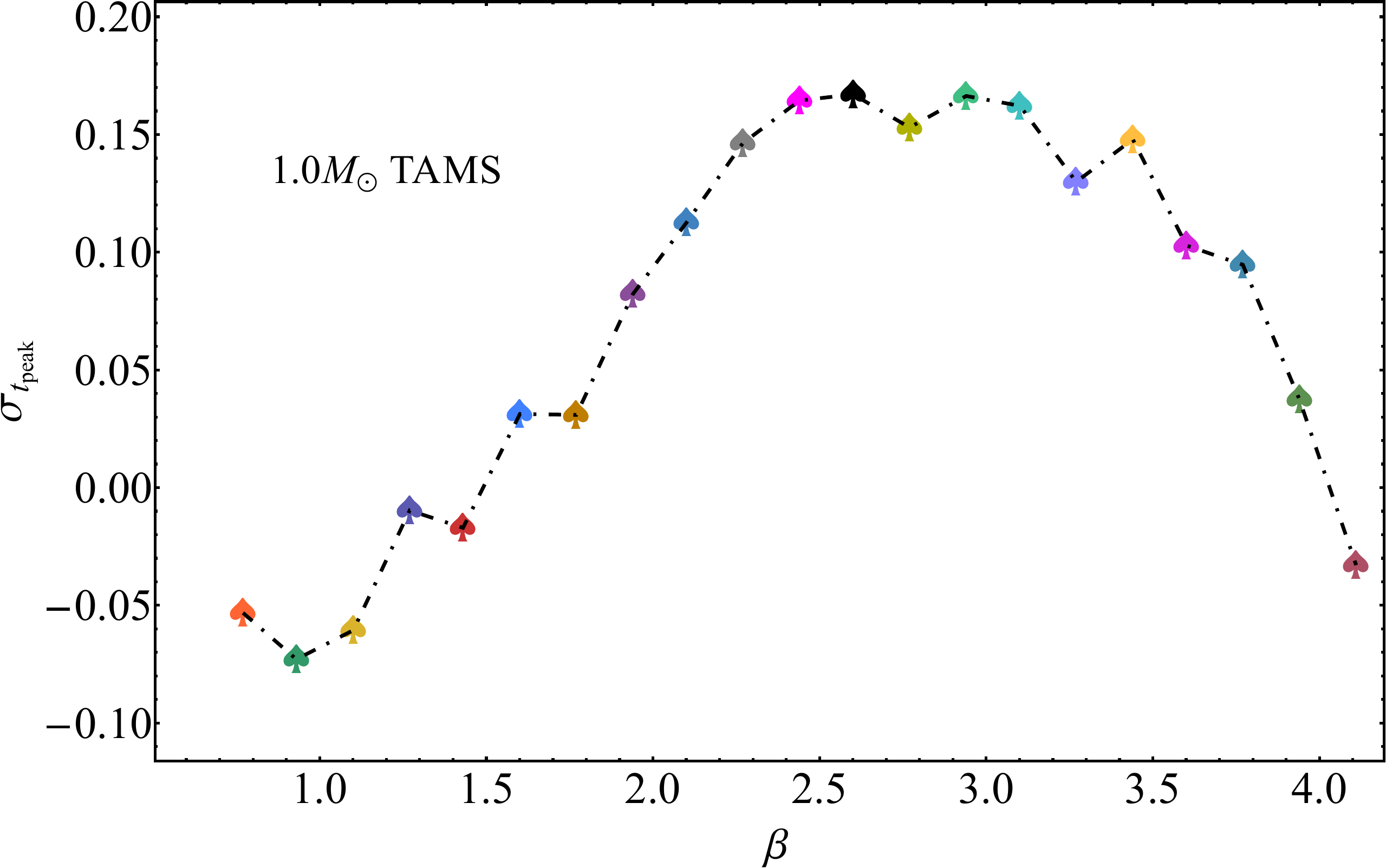}
    \caption{{\bf Left: }A comparison of the peak timescale $t_{\rm peak}$ and peak fallback rate $\dot{M}_{\rm peak}$ as measured from the {\sc phantom} simulations (depicted with star symbols), and as predicted by the MG model (depicted with spade symbols). The colors represent the penetration factor of the orbit, $\beta$ as shown in the legend. With an increase in $\beta$, the peak timescale $t_{\rm peak}$ shifts to earlier times, and the peak fallback rate $\dot{M}_{\rm peak}$ increases. The error bars show the duration for which $\dot{M}>95\%\dot{M}_{\rm peak}$. {\bf Right: }the relative error in $t_{\rm peak}$, defined as $\sigma_{\rm t_{\rm peak}} \equiv (t_{\rm peak,hydro}-t_{\rm peak,MG})/t_{\rm peak,hydro}$, as a function of $\beta$. The top, middle and bottom panel shows results for a $1.0M_\odot$ star at its ZAMS, MAMS and TAMS stages respectively.} \label{fig:solar-mass-peak-fbrs}
\end{figure*}
\begin{figure*}
    \includegraphics[width=0.5\textwidth]{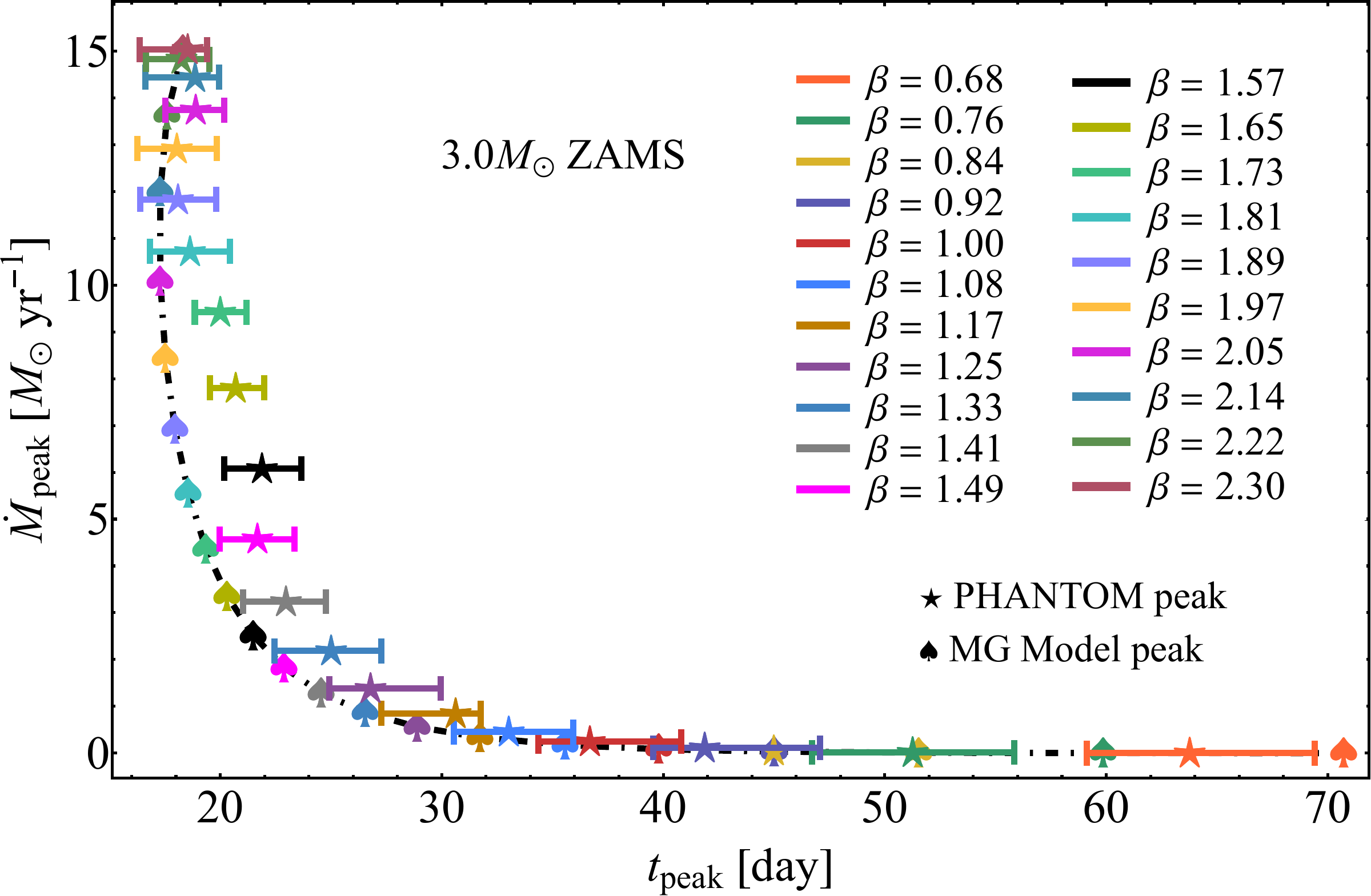} 
    \includegraphics[width=0.515\textwidth]{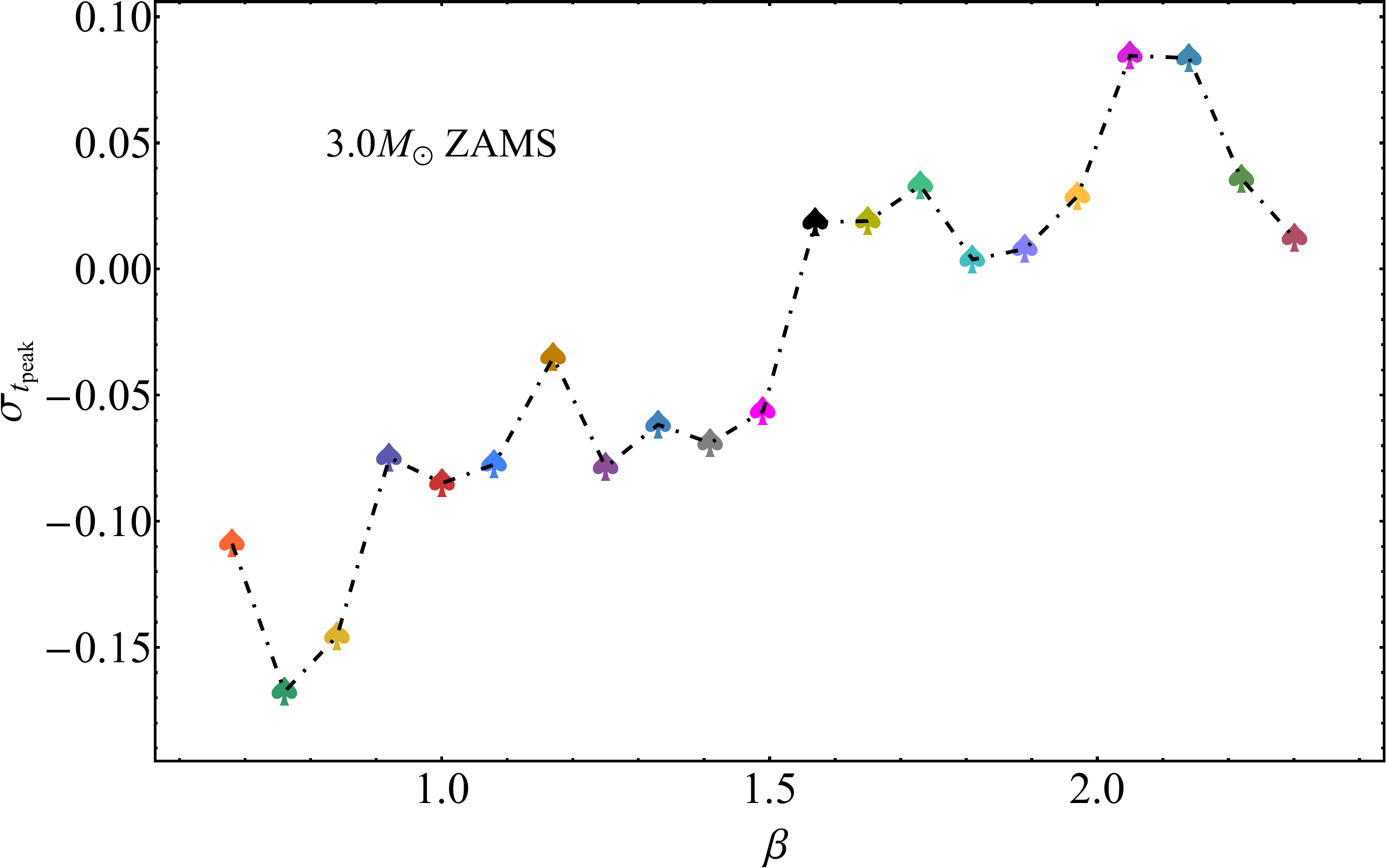}\\
    \includegraphics[width=0.5\textwidth]{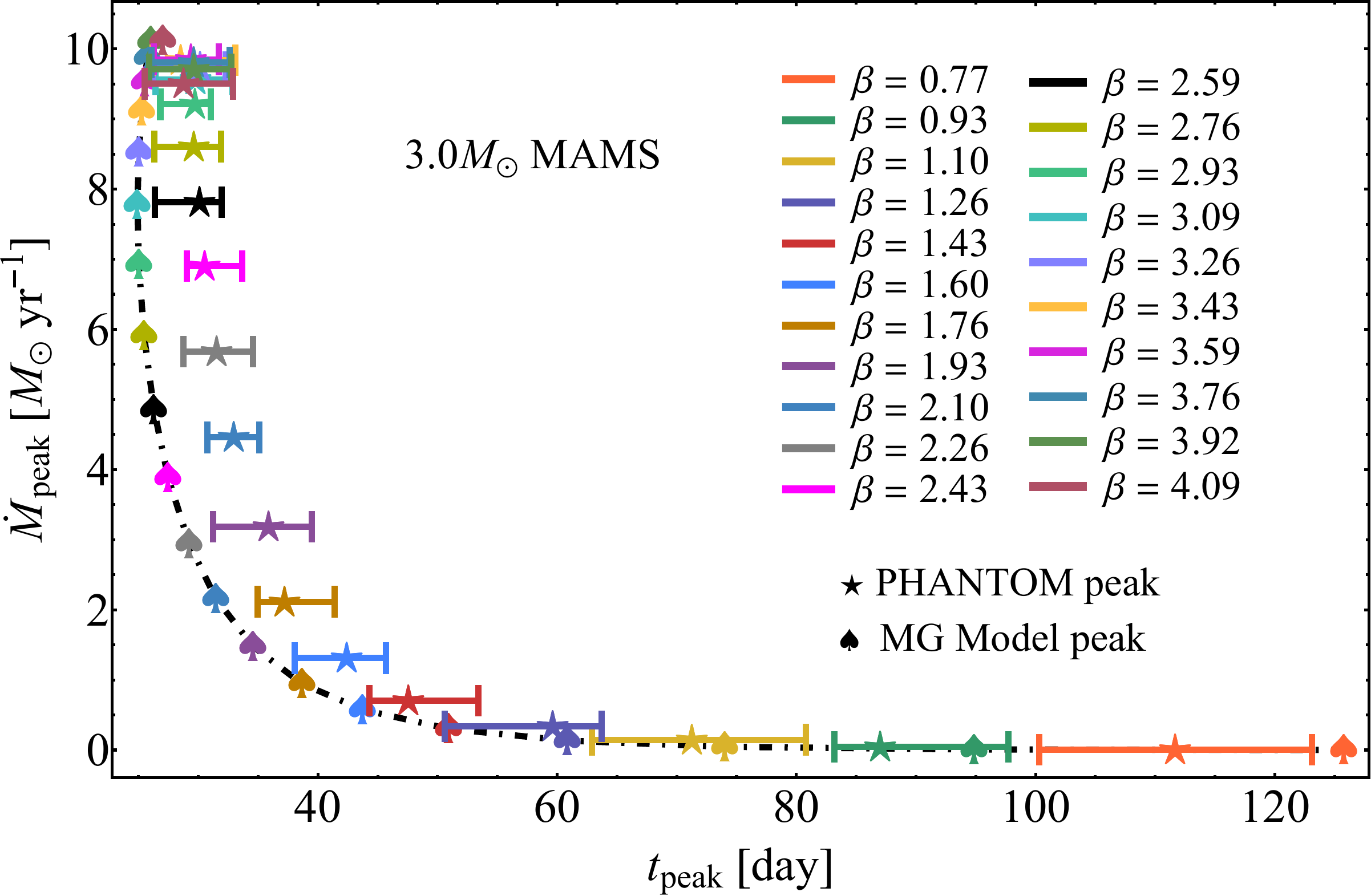} 
    \includegraphics[width=0.515\textwidth]{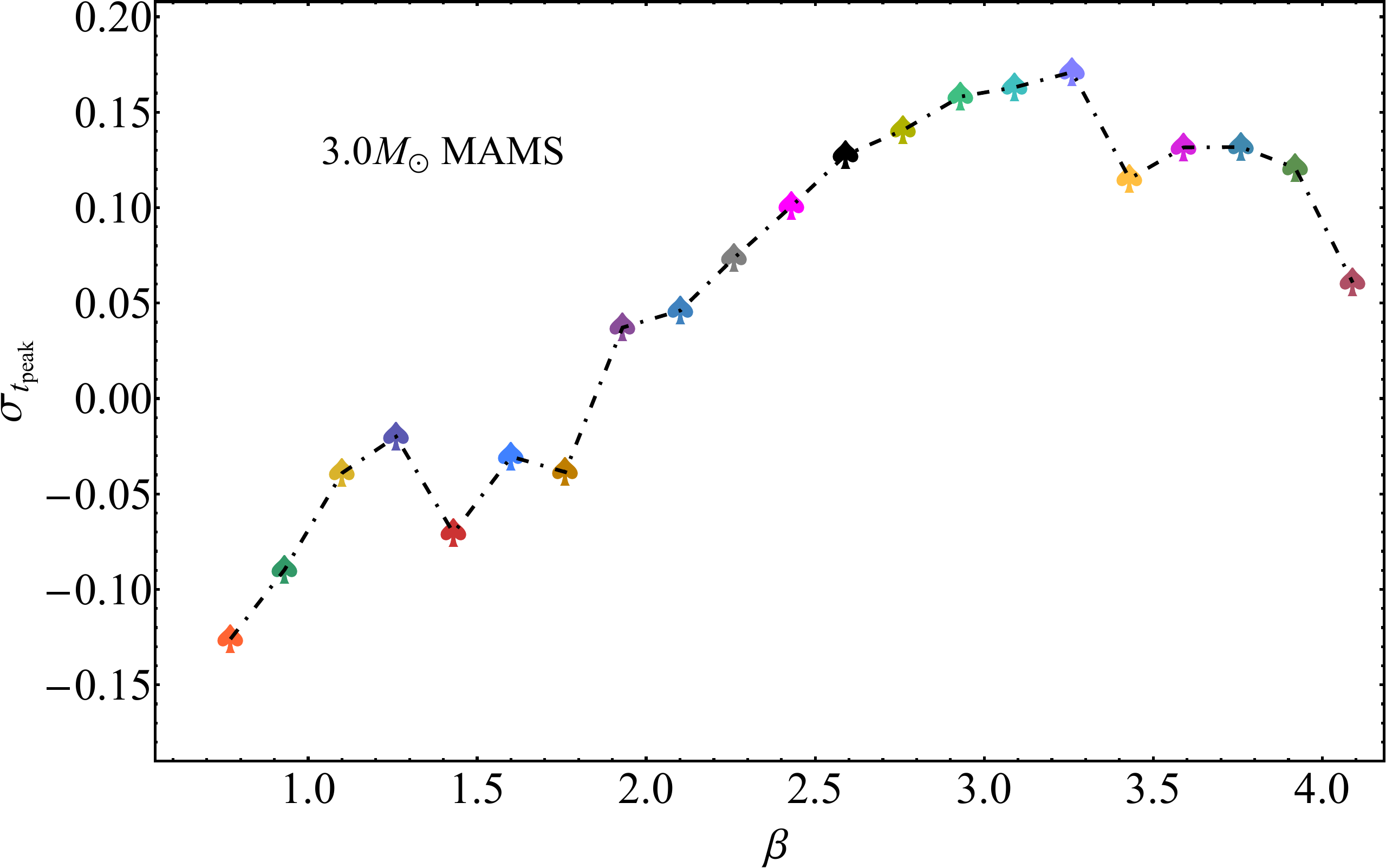}\\
    \includegraphics[width=0.5\textwidth]{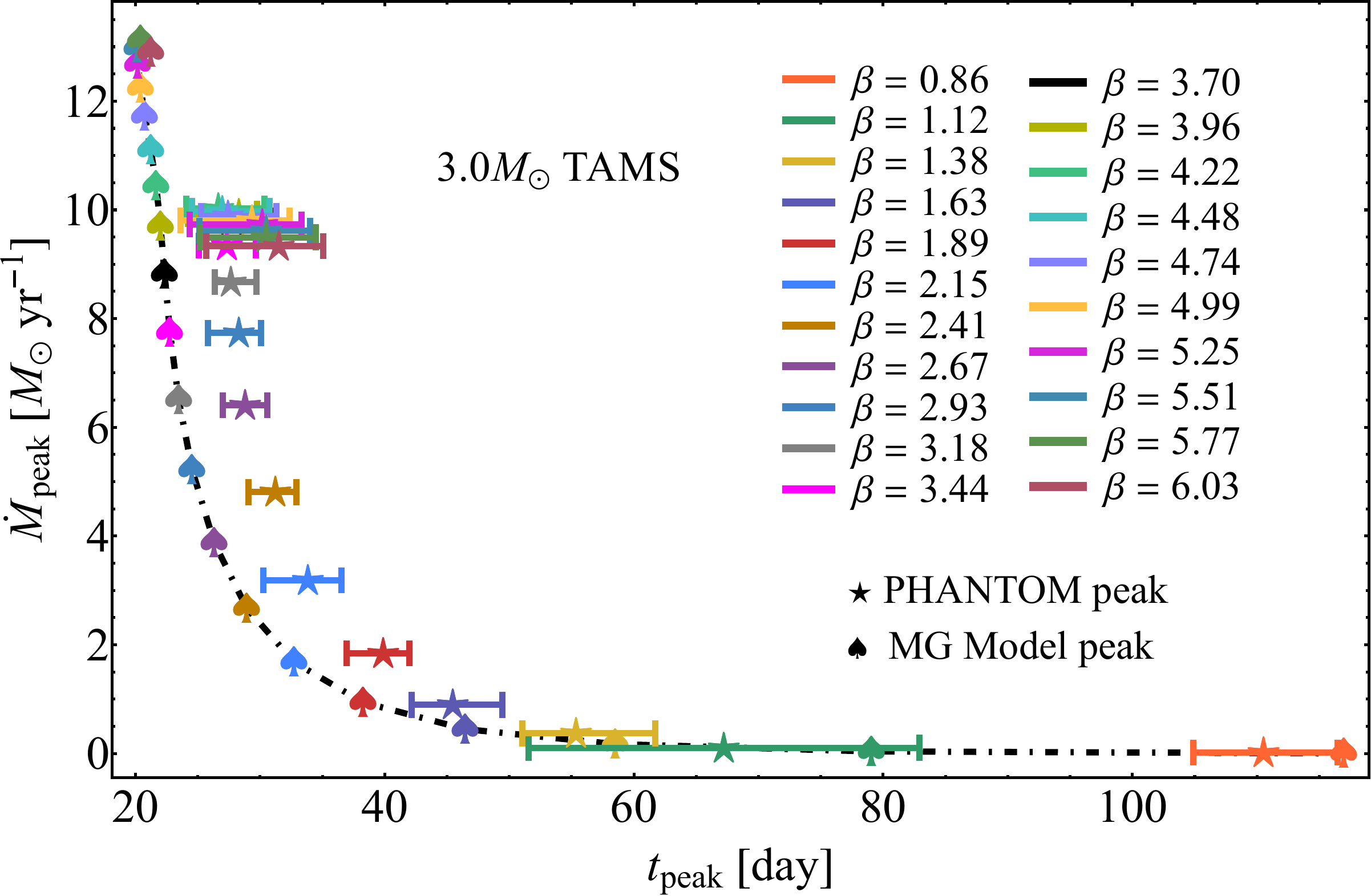}
    \includegraphics[width=0.515\textwidth]{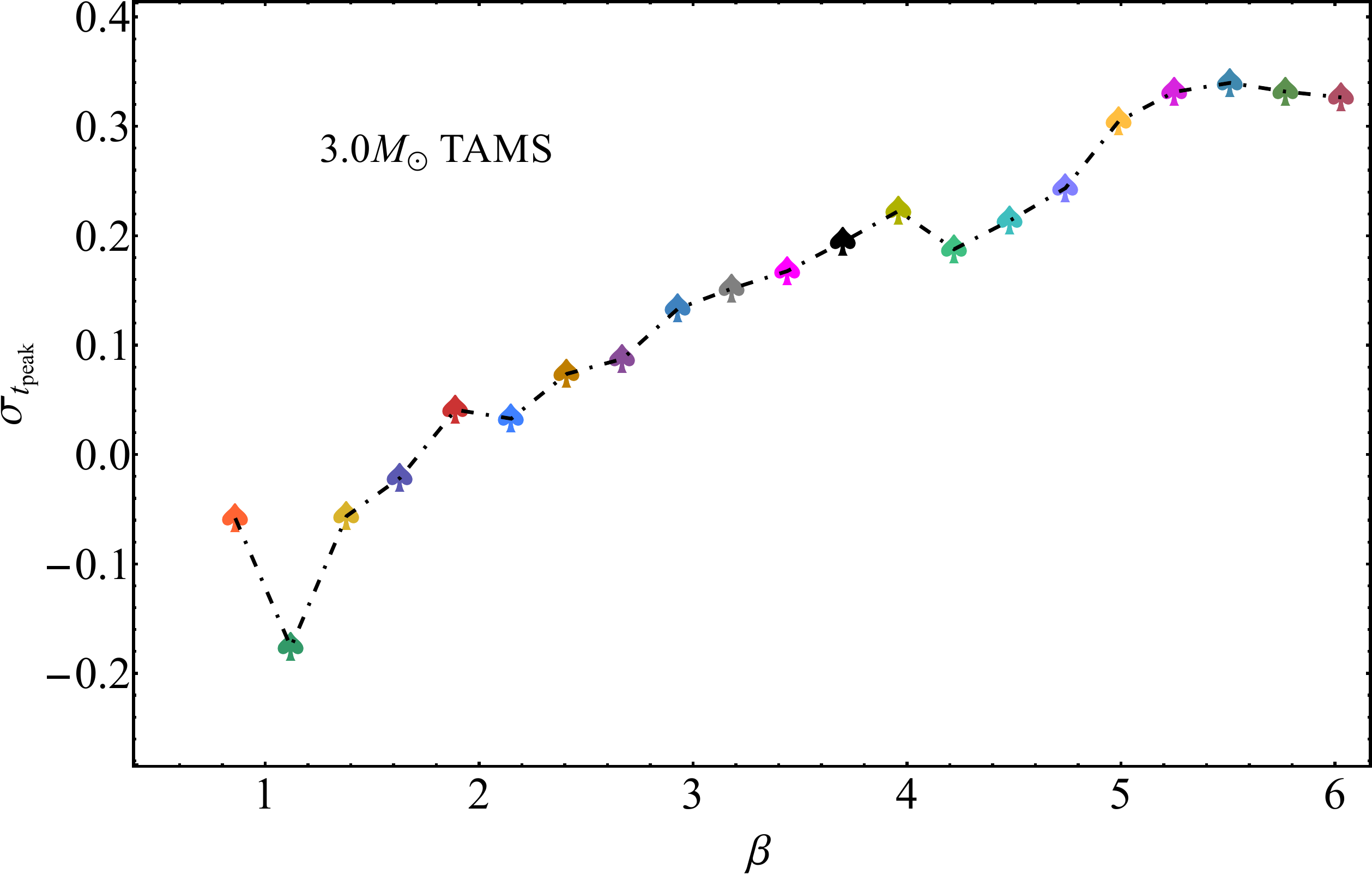}
    \caption{Same as Figure~\ref{fig:solar-mass-peak-fbrs}, but for the $3M_\odot$ star at its ZAMS (top panel), MAMS (middle panel) and TAMS (bottom panel) stages. As seen in the left panel of the figure, for the ZAMS star, $t_{\rm peak}$ decreases and $\dot{M}_{\rm peak}$ increases with an increase in $\beta$. This trend is reversed for $\beta\gtrsim3.76$ ($\beta\gtrsim4.22$) in the {\sc phantom} simulations for the MAMS (TAMS) star, with longer peak timescales and lower peak fallback rates as $\beta$ approaches $\beta_{\rm c}$.} \label{fig:3msun-peak-fbrs}
\end{figure*}
\subsubsection{Sun-like stars}
A $1M_\odot$ star at its ZAMS is well approximated by a $4/3-$polytrope. Figure~\ref{fig:solar-mass-peak-fbrs} shows the peak fallback times $t_{\rm peak}$ and rates $\dot{M}_{\rm peak}$ (left) for a $1M_\odot$ star at its ZAMS (top), MAMS (middle) and TAMS (bottom), for $\beta$'s in the range $\beta\in[0.6,\beta_{\rm c}]$, along with the relative error in $t_{\rm peak}$ (right). The $t_{\rm peak}$ predictions are all in good agreement with those obtained from the {\sc phantom} simulations, with the majority lying within the error bars associated with the numerical results. The peak fallback times for lower $\beta$'s are substantially longer than those obtained for low-mass stars ($M_\star \lesssim 0.5M_\odot$), and increasingly so for the more evolved stars. The {\sc phantom} simulations performed at $\beta=\beta_{\rm c}$ as predicted by the MG model yield partial disruptions for the $1M_\odot$ MAMS and TAMS stars (the $1M_\odot$ MAMS star, however, is completely destroyed at $\beta_{\rm c}+1$), due to the reformation of a stellar core shortly after pericenter. We return to a discussion of core reformation and its implications for the fallback rate in the context of high-mass stars, described in the next subsection. 

\begin{figure*}
    \includegraphics[width=0.51\textwidth]{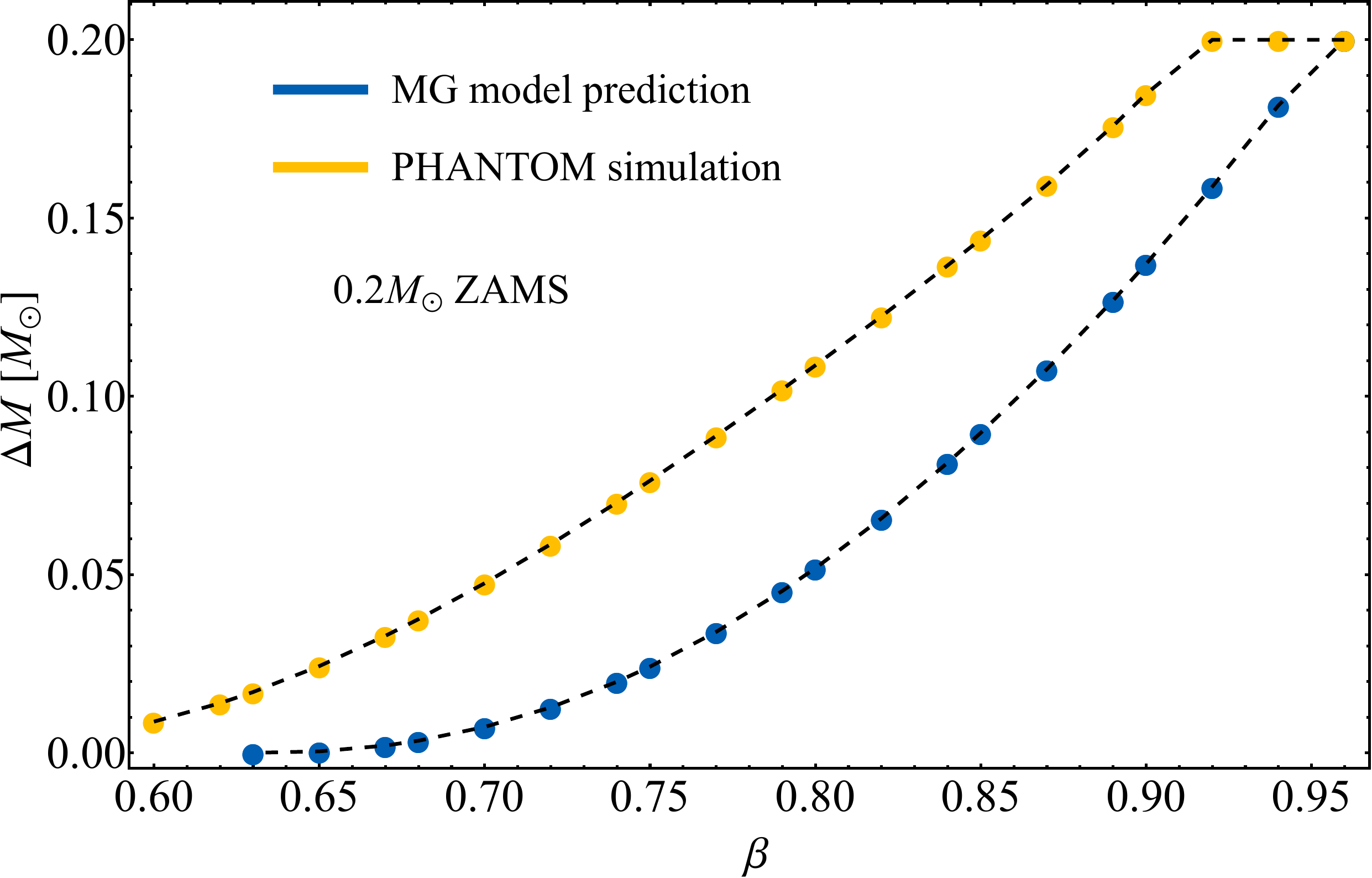}
    \includegraphics[width=0.51\textwidth]{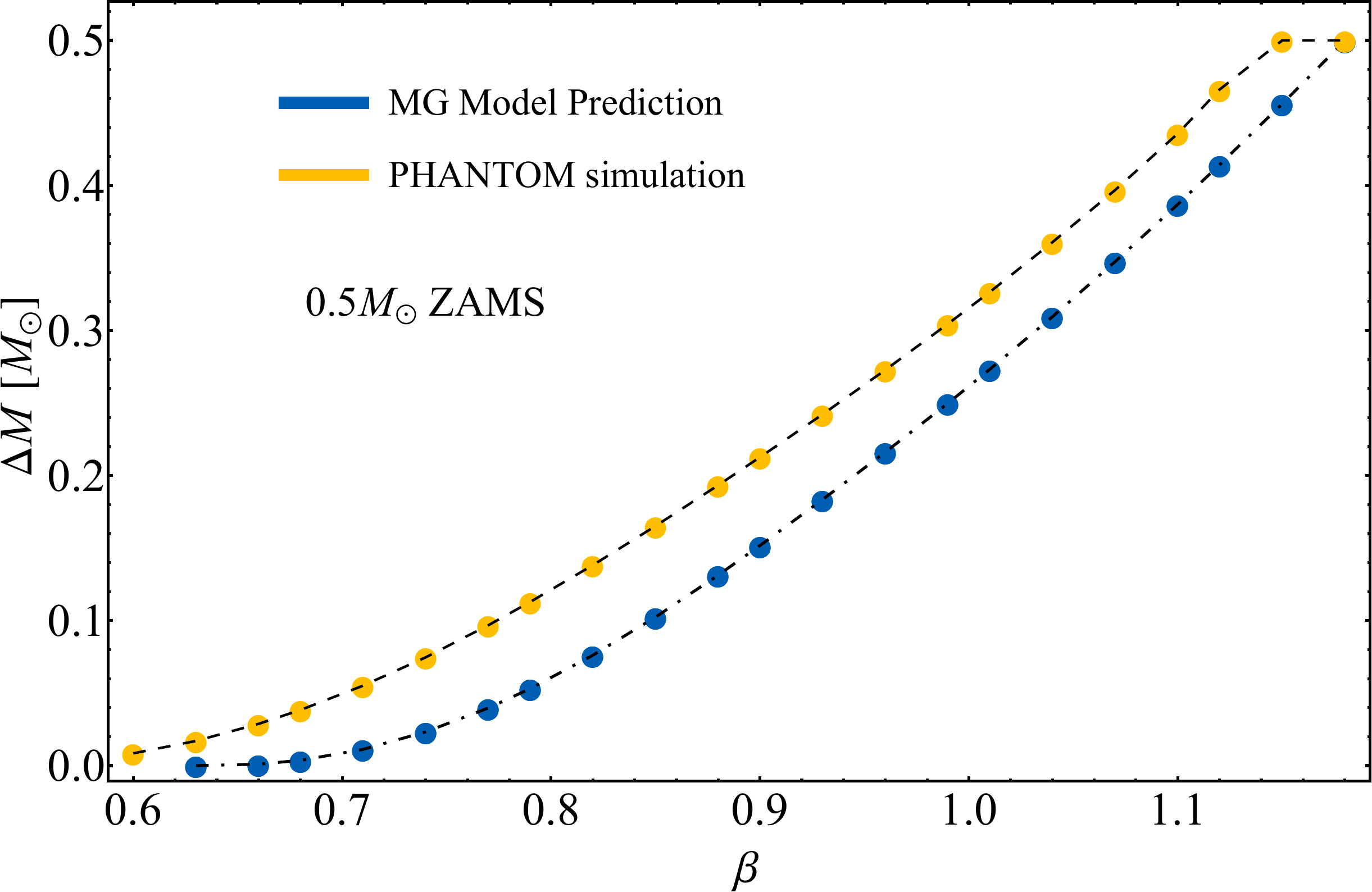}\\
    \includegraphics[width=0.51\textwidth]{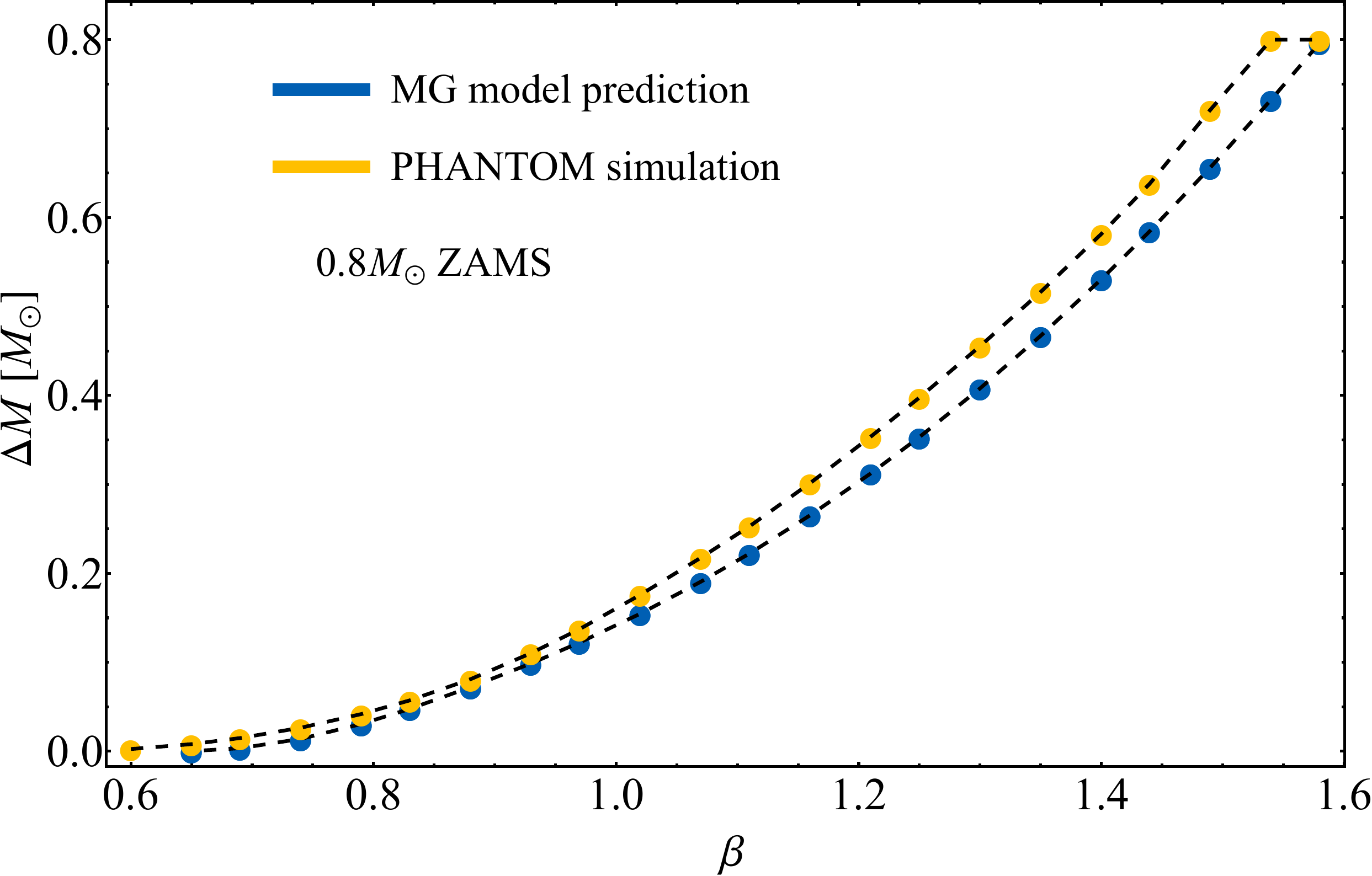}
    \includegraphics[width=0.51\textwidth]{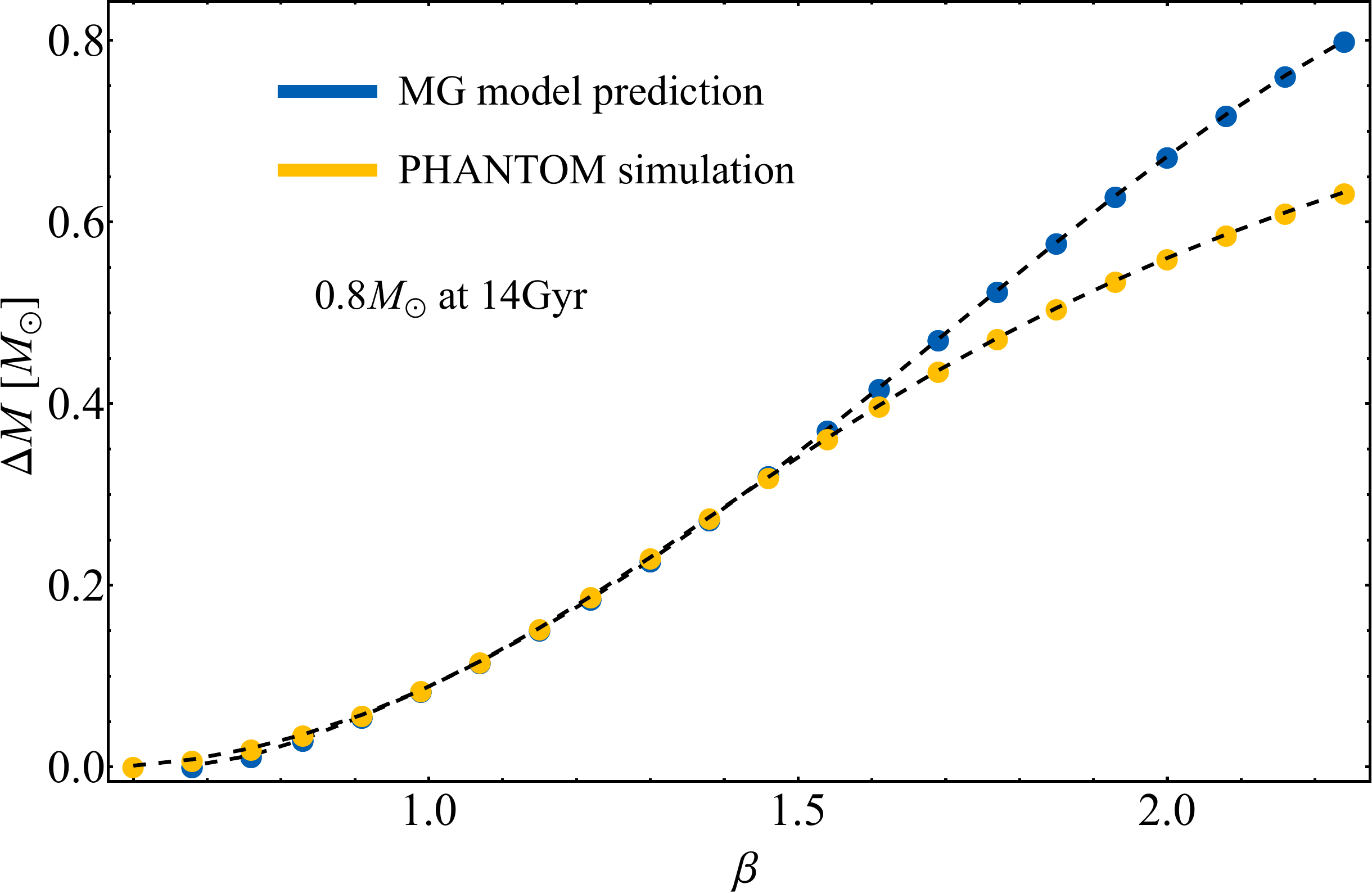}
    \caption{The amount of mass stripped, $\Delta M$, as a function of the penetration factor $\beta$,  for a $0.2M_\odot$ ZAMS star (top left), a $0.5 M_\odot$ ZAMS star (top right), a $0.8 M_\odot$ ZAMS star (bottom left), and a $0.8M_\odot$ star at $14$Gyr (botom right), with $\beta$ ranging from $0.6$ to $\beta_{\rm c}$ (critical $\beta$ for complete disruption). The blue data points depict the predictions of the MG model, and the yellow ones are as measured from the {\sc phantom} simulations.} \label{fig:low-mass-deltam}
\end{figure*}
\begin{figure}
    \includegraphics[width=0.495\textwidth]{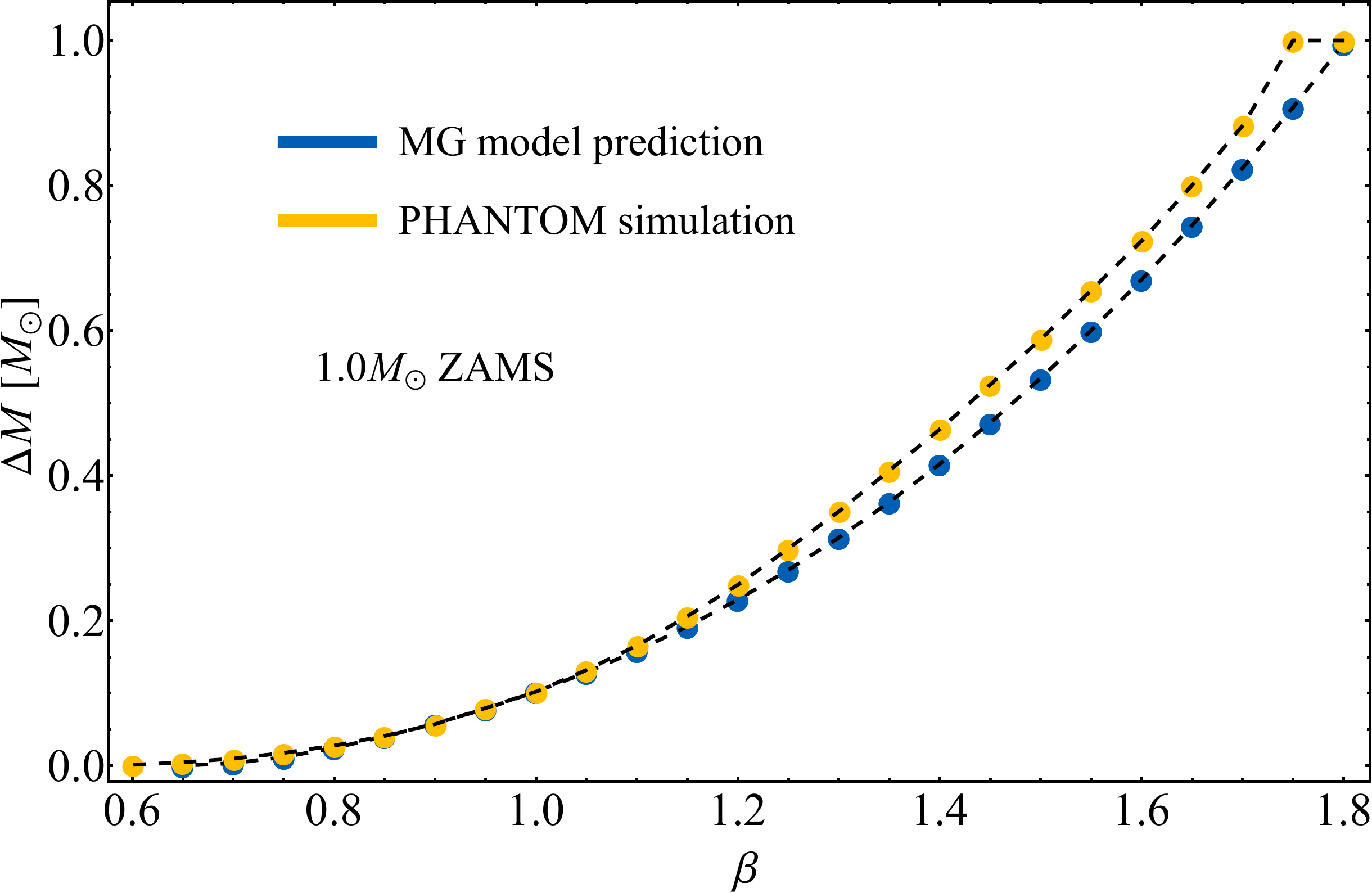} \\
    \includegraphics[width=0.495\textwidth]{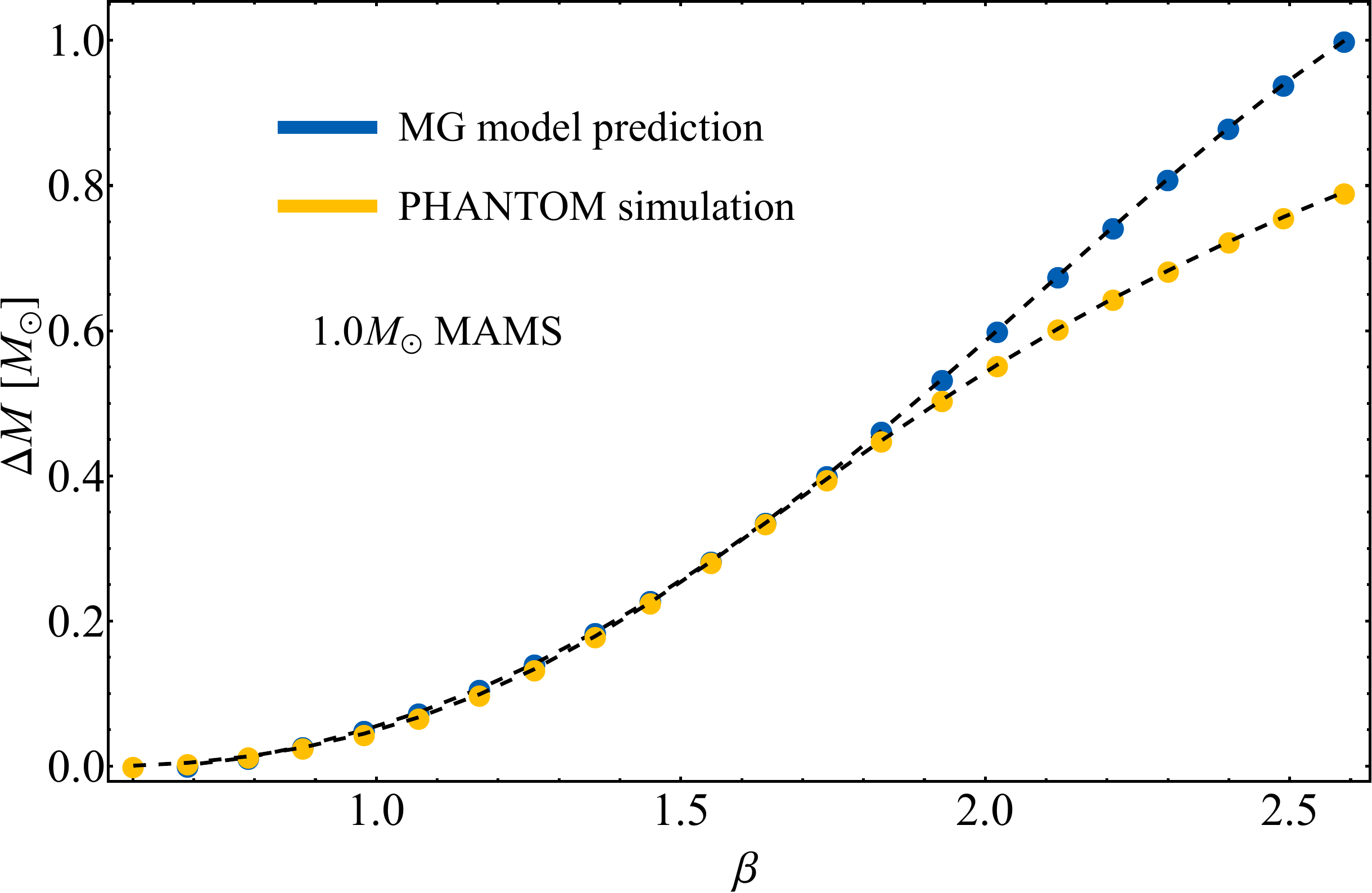} \\
    \includegraphics[width=0.495\textwidth]{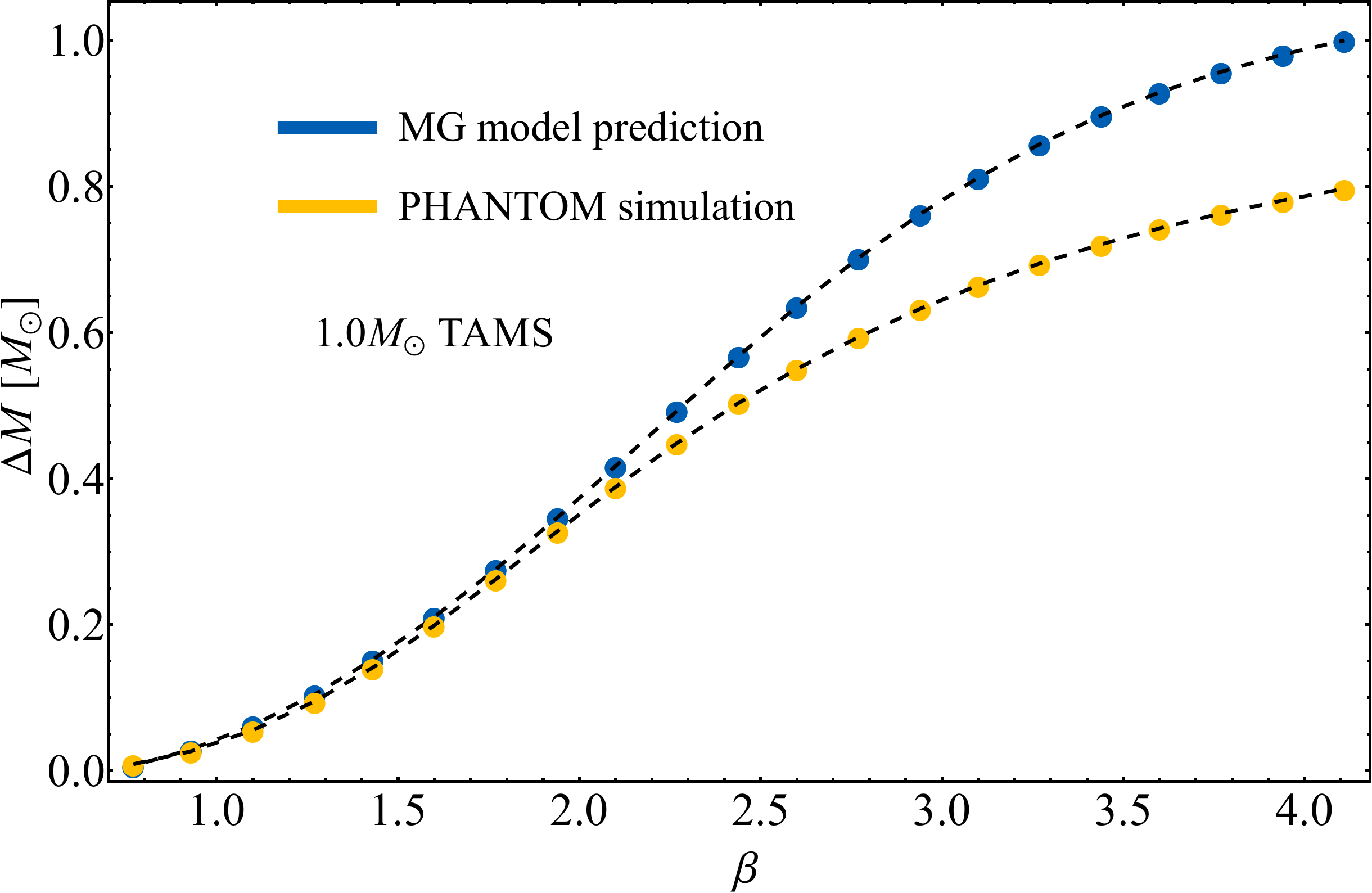}
    \caption{Same as Figure~\ref{fig:low-mass-deltam}, but for a $1.0M_\odot$ star at its ZAMS (top panel), MAMS (middle panel) and TAMS (bottom panel) stages. } \label{fig:solar-mass-deltam-beta}
\end{figure}
\begin{figure}
    \includegraphics[width=0.495\textwidth]{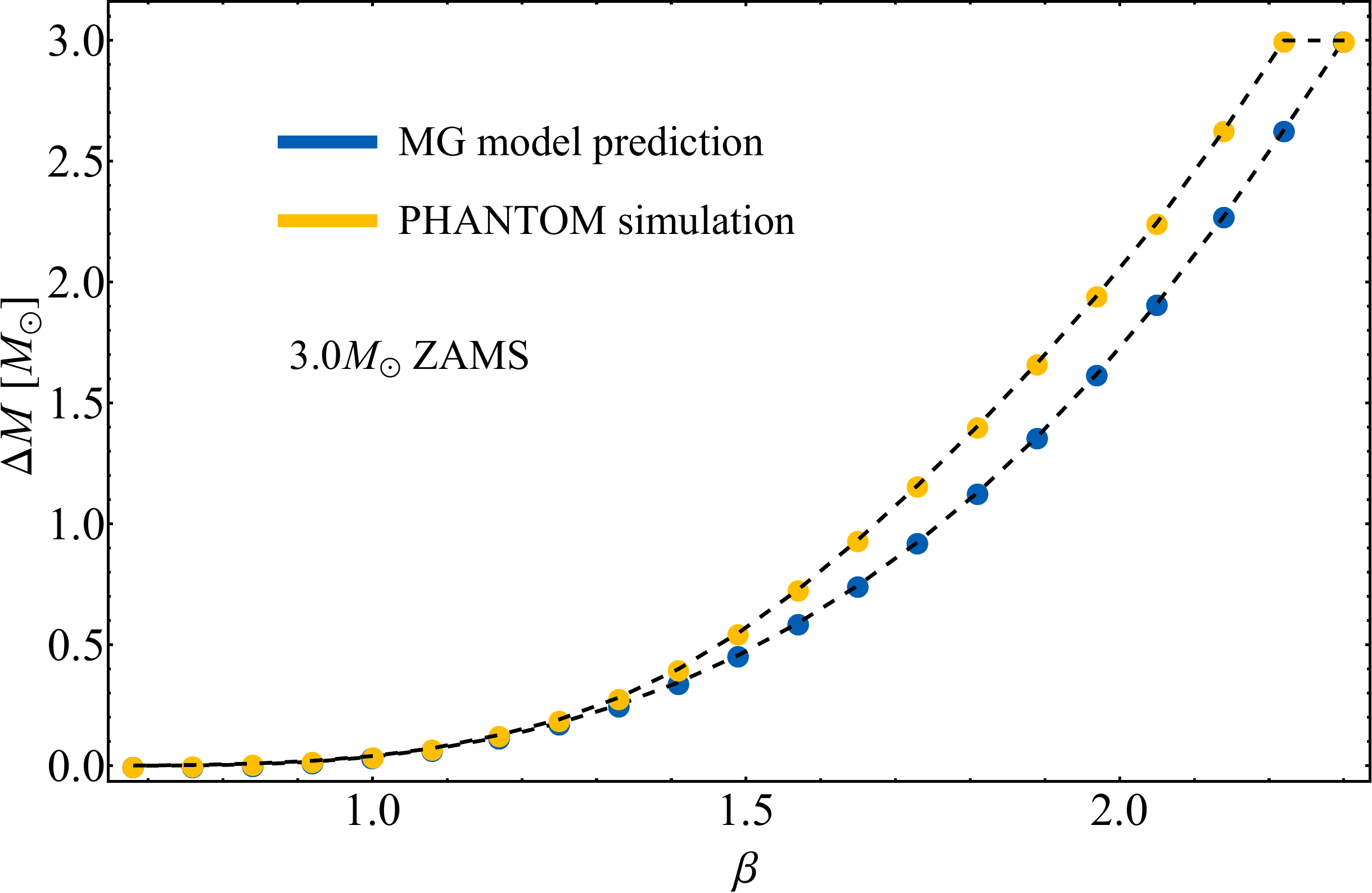} \\
    \includegraphics[width=0.495\textwidth]{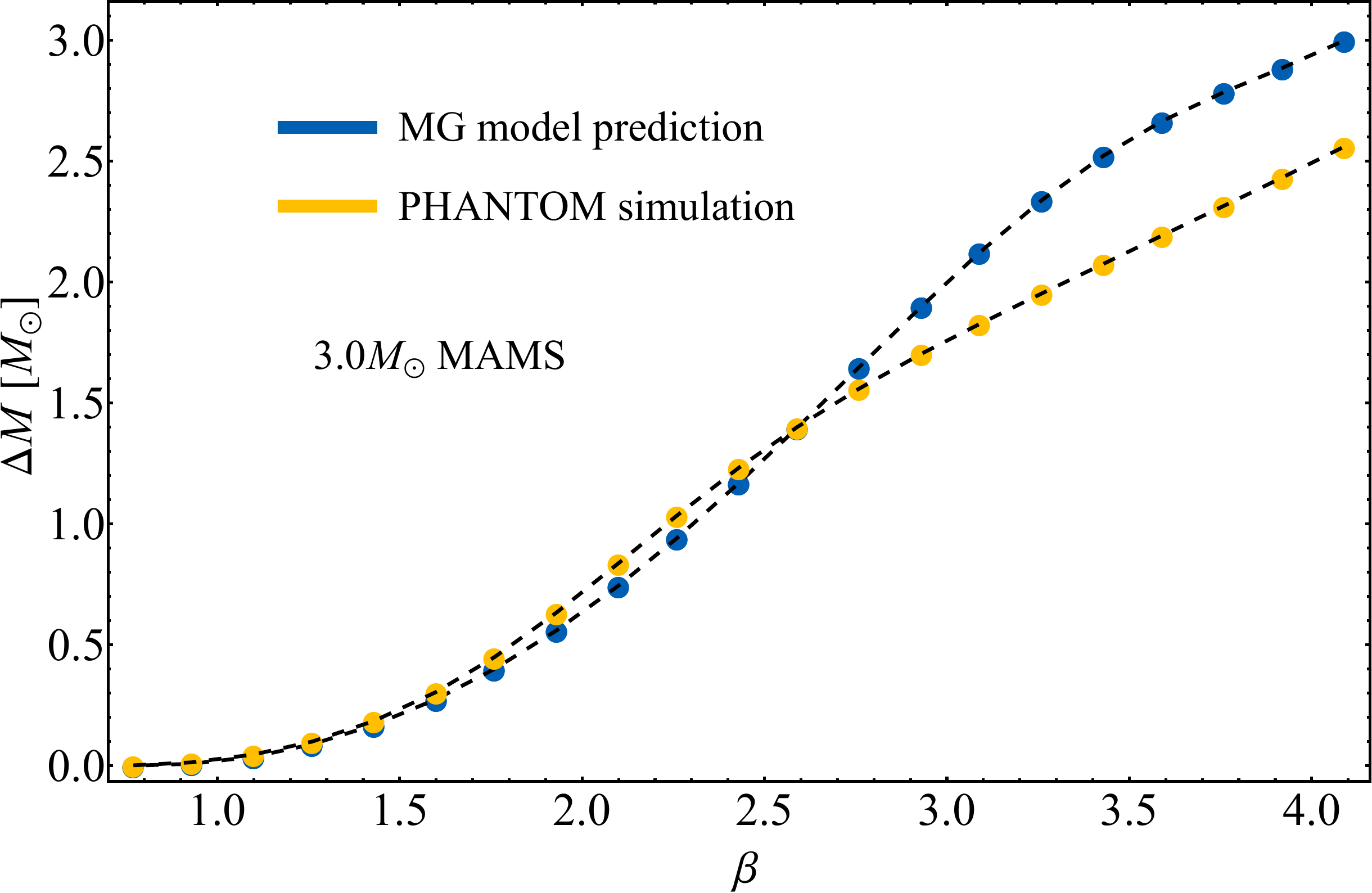} \\
    \includegraphics[width=0.495\textwidth]{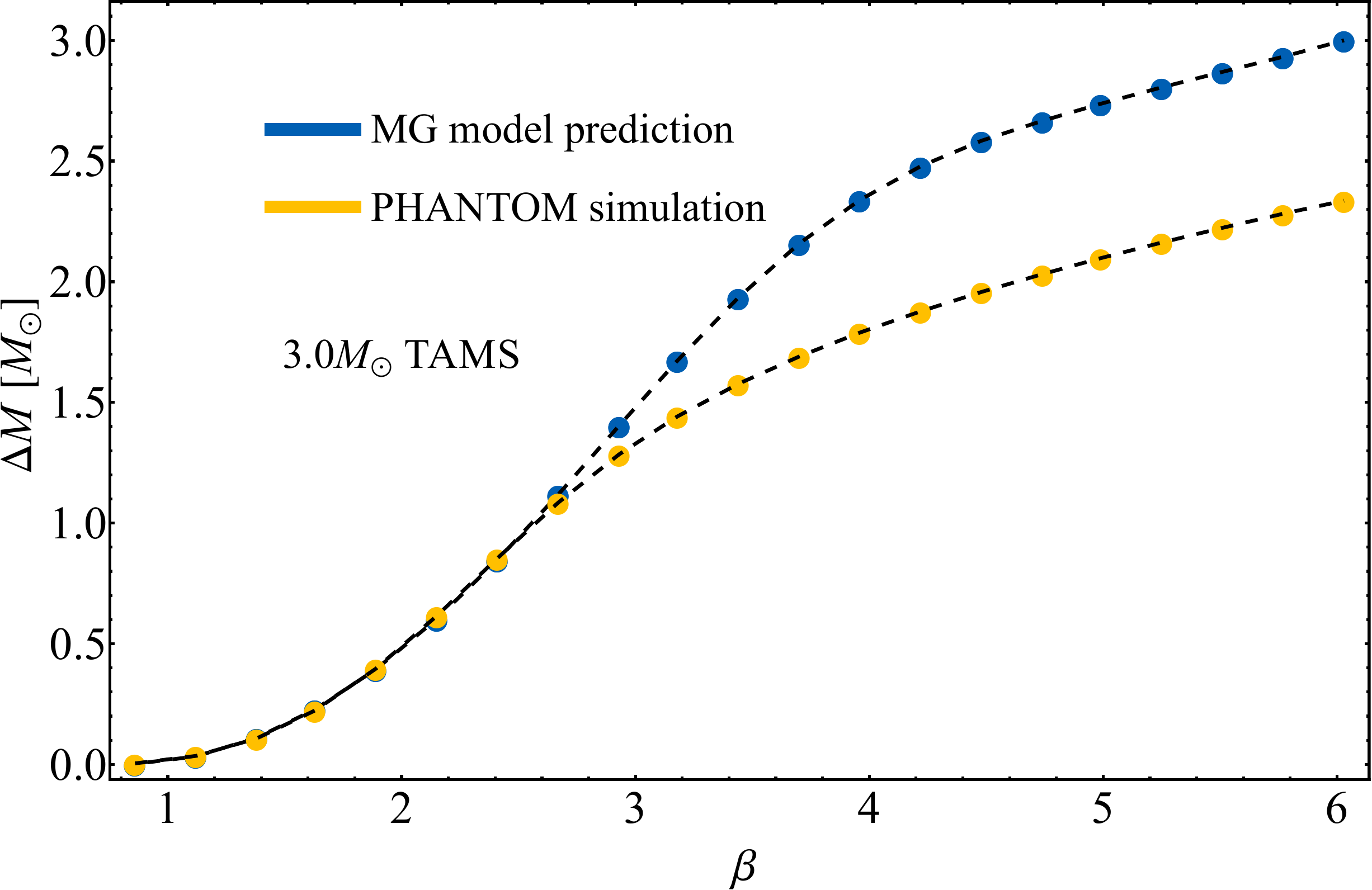}
    \caption{Same as Figure~\ref{fig:low-mass-deltam}, but for a $3.0M_\odot$ star at its ZAMS (top panel), MAMS (middle panel) and TAMS (bottom panel) stages.} \label{fig:3msun-deltam-beta}
\end{figure}
\subsubsection{High-mass stars}
\label{sec:high-mass}
High-mass stars ($M_\star \gtrsim 1.5 M_\odot$) have a more centrally concentrated structure compared to sun-like or low-mass and convective stars, and are in general not well approximated by a polytropic density profile. Moreover, as they evolve, high-mass stars develop a compact core and a diffuse radiative outer envelope, such that most of its mass is contained within the core of the star. Here we compare the MG model predictions for $(t_{\rm peak},\dot{M}_{\rm peak})$ with the {\sc phantom} simulations for a $3M_\odot$ star, as a representative example of a high-mass star. 

Figure~\ref{fig:3msun-peak-fbrs} shows the peak fallback rates and times (left panel) for a $3M_\odot$ star at its ZAMS, MAMS and TAMS stages, along with the relative error in $t_{\rm peak}$, $\sigma_{t_{\rm peak}}$ as a function of $\beta$ (right panel). For all three ages of the $3M_\odot$ star, the amount of mass stripped at $\beta=0.6$ is a negligible fraction of $M_\star$ ($\Delta M /M_\star \sim 10^{-5}$), and does not yield a measurable fallback rate at a resolution of $10^6$ particles, hence we do not show the results for $\beta=0.6$ here. For the ZAMS star, the peak timescale ranges between $\sim 18-65$ days, and the peak fallback rate spans $10^{-3}-15 M_\odot \mathrm{yr}^{-1}$ for the range of penetration factors shown in the top left panel of the figure. As the star evolves its radius increases while the mass remains the same, such that the surface gravitational field -- which largely characterizes the partial disruption radius (cf.~\citealt{coughlin22}) -- weakens. Mass is therefore peeled off at larger distances from the SMBH for a more highly evolved star, implying the debris has a smaller binding energy (to the SMBH) and a longer return time.  

For low-$\beta$ orbits, the predictions of the MG model agree to within $\sim1-2$ days in $t_{\rm peak}$ and to within a factor of $\sim2-3$ for $\dot{M}_{\rm peak}$. With an increase in $\beta$ (as the pericenter distance is reduced), the model predicts a decrease in $t_{\rm peak}$ and a monotonic increase in $\dot{M}_{\rm peak}$. This monotonicity is recovered for modest values of $\beta$ for all stars partially destroyed with {\sc phantom}, but eventually fails for the MAMS and TAMS stars, with a maximum in $\dot{M}_{\rm peak}$ being reached at a $\beta$ of 3.76 (4.22) for the MAMS (TAMS) star (the predicted $\beta_{\rm c}$ at which the star is destroyed being 4.09 and 6.03 for the MAMS and TAMS stars respectively). Beyond this point, the peak of the fallback rate from the {\sc phantom} simulations exhibits a slight increase in $t_{\rm peak}$ and decrease in $\dot{M}_{\rm peak}$. 

As discussed in Section 2 of \cite{fancher25}, the tidal compression of the star, which can raise the central density in high-$\beta$ encounters, could result in an effective core radius $R_{\rm c}$ that is smaller than that calculated from the original stellar density profile (which the MG model uses). A smaller $R_{\rm c}$ would yield a longer $t_{\rm peak}$ (using Equation 11 of \citealp{coughlin22}) and a smaller $\dot{M}_{\rm peak}$, which agrees with the trend seen in the {\sc phantom} simulations. 

\subsection{Critical \texorpdfstring{$\beta$}{Lg} for complete disruption }
\label{sec:critical-beta}
The top-left, top-right, and bottom-left panels of Figure~\ref{fig:low-mass-deltam} show the amount of mass stripped as a function of $\beta$ for a $0.2 M_\odot$ ZAMS star, $0.5 M_\odot$ ZAMS star, and $0.8$ ZAMS star, respectively. For such low-mass stars, the MG model underpredicts the mass lost at a given $\beta$, but the discrepancy between the prediction and the simulations decreases as the mass of the star increases. For each of these stars, $\beta_{\rm c}$ is always slightly larger than the value recovered from the hydrodynamical simulations.

The bottom-right panel of this figure shows $\Delta M(\beta)$ for a $0.8 M_\odot$ star at 14Gyr. For this star the prediction of the MG model agrees nearly exactly with the {\sc phantom} simulations for $\beta \lesssim 1.5$, while for $\beta \gtrsim 1.5$, the model overpredicts the amount of mass stripped. Thus, unlike the ZAMS stars, the model $\beta_{\rm c}$ is smaller than that of the {\sc phantom} simulations, such that a $0.2 M_{\odot}$ core remains when the model predicts a complete disruption. This qualitative difference between the ZAMS star and the more evolved star at $14$ Gyr -- that the model overpredicts $\beta_{\rm c}$ for the former and underpredicts $\beta_{\rm c}$ for the latter -- strongly suggests that there is a distinct, underlying physical origin for each of these outcomes (see the next section for additional discussion). Figures~\ref{fig:solar-mass-deltam-beta} and \ref{fig:3msun-deltam-beta} are analogous to Figure \ref{fig:low-mass-deltam}, but for a $1M_\odot$ star and a $3M_\odot$ star at their ZAMS (top panel), MAMS (middle panel), and TAMS (bottom panel) stages. The behavior is similar to that recovered for the low-mass stars: the ZAMS model predictions are in good agreement with the numerical results, while the MG model overpredicts the amount of mass lost for MAMS and TAMS stars above a critical $\beta < \beta_{\rm c}$.

\section{Discussion and Conclusions}
\label{sec:discussion}
\subsection{Physical differences between the model and the simulations}
The Maximum Gravity model (originally developed in \citealp{coughlin22}) calculates the critical pericenter distance for complete disruption of a star by equating the maximum gravitational field of the unperturbed stellar configuration to the tidal field of the black hole. We extended this model to partial TDEs (wherein the strength of the tidal encounter is quantified by the penetration factor $\beta$), and equate the tidal field at a given pericenter distance to the self-gravity of the mass enclosed within some spherical radius $R$, which yields the amount of mass stripped, $\Delta M$, for a given value of $\beta$, and the peak of the fallback rate (where $\Delta M$ is normalized such that it equals $M_\star$ when the tidal field exceeds the maximum self-gravitational field within the star, which is the criterion for complete disruption following \citealp{coughlin22}). The associated peak timescale $t_{\rm peak}$ is calculated, using arguments similar to Equation 11 in \cite{coughlin22}, as the peak fallback time associated with the radius $R$, that solves Equation~\eqref{eq:critical_rtc} for the particular value of $\beta$. As discussed in Section~\ref{sec:hydro}, for high $\beta$ encounters the stellar structure is significantly distorted and the gravitational field of the core is amplified near pericenter due to strong tidal compression, which results in a smaller amount of mass stripped in the numerical simulations for high-mass and evolved stars (which have a large $\beta_{\rm c}$ for complete disruption) as compared to the predictions of the MG model. Additionally, since the amplified gravitational field implies a smaller effective core radius $R_{\rm c}$ compared to that of the unperturbed stellar structure, our model underpredicts $t_{\rm peak}$ and overpredicts $\dot{M}_{\rm peak}$ as $\beta$ approaches $\beta_{\rm c}$ for high-mass and evolved stars (as shown in the bottom panel of Figure~\ref{fig:3msun-peak-fbrs}).

Low-mass and fully convective stars expand in response to small amounts of mass lost from near their surface \citep{ray87,mcmillan87,gu04,linial24,bandopadhyay25,yao25}. This increases their susceptibility to further mass loss. Thus, the expansion of a low-mass star in response to the small fractional mass loss at a given $\beta$ leads to an increased amount of mass being stripped in the numerical simulations, which is not captured in our analytical model. Consequently, the estimates of $\Delta M$ for low-mass stars are underpredicted by our model for small values of $\beta$, as seen in the top panel of Figure~\ref{fig:low-mass-deltam}. This has an impact on the predictions for $\dot{M}_{\rm peak}$, which are systematically smaller than the numerical values for the low-mass ($M_\star \lesssim 0.6 M_\odot$) ZAMS stars. 

Additionally, this model ignores the continued gravitational influence of the surviving core: following its removal through tides, the orbital dynamics of a given mass shell is determined solely by its initial location within the star and, correspondingly, its gravitational binding energy with respect to the black hole.  Contrarily, the late-time fallback rate in a partial TDE is established exclusively by the gravitational field of the core, and approaches $\propto t^{-9/4}$ (vs.~the standard $t^{-5/3}$) because the material originates asymptotically close to its Hill sphere \citep{coughlin19}. Therefore, a corollary of the agreement between the model predictions for $t_{\rm peak}$ and the numerically obtained peak timescales (over a wide range of stellar masses and ages and orbital parameters, as shown in Figures \ref{fig:low-mass-peak-fbrs}-\ref{fig:3msun-peak-fbrs}) is that the core does not significantly impact the dynamics of the debris that returns at early times, particularly the fluid elements that establish $t_{\rm peak}$. This is consistent with the results of \cite{guillochon13}, who found that the continued gravitational influence of the core is primarily responsible for determining the late-time behavior of the fallback rate (see Figure 10 of their paper, which shows that the long term evolution of the energy distribution $dM/d\epsilon$ in the presence of the core leads to significant differences in the late time fallback, while the behavior around the peak is established within a few dynamical times of the star). 

Finally, the self-gravity of the debris stream, which plays a dominant role in determining the shape of the fallback rate \citep{fancher23}, is ignored in our model. This results in a discrepancy between the predicted value of $\dot{M}_{\rm peak}$ and the one obtained numerically (from the {\sc phantom} simulations). For ZAMS stars with $M_\star \gtrsim 0.8 M_\odot$, the prediction for the peak fallback rate at low $\beta$ values is discrepant with the numerically obtained value by a factor of $\sim2-3$, despite the very good agreement between the two as concerns $\Delta M$ and $t_{\rm peak}$ values. This can be attributed to the neglect of the self-gravitating nature of the stream in our model. Our results show that these physical differences between our simplified model for a partial TDE and the numerical simulations lead to non-negligible errors in $\dot{M}_{\rm peak}$, particularly for low values of $\beta$. However, and in spite of these differences, the peak timescales -- which can be used to place constraints on the SMBH properties \citep{guillochon13,mockler19,bandopadhyay24} -- are accurate to within $\sim50\%$ of the numerical results for any star and pericenter distance.

\subsection{\texorpdfstring{$t_{\rm peak}$}{Lg} distribution, and implications for TDE observations}
For the range of stellar masses considered here ($0.2 \leqslant M_\star/M_\odot \leqslant5.0$), our model predicts that the peak timescale for the partial disruption by a $10^6 M_\odot$ can range from tens to hundreds of days, which is corroborated by the results of the numerical simulations described in Section~\ref{sec:hydro}. Since the peak timescale relates to the mass of the SMBH as $t_{\rm peak} \propto M_\bullet^{1/2}$, we can use these results to predict the $t_{\rm peak}$ distribution of observed TDE candidates. 

The dependence of the $t_{\rm peak}$ distribution on stellar parameters can be incorporated by considering a truncated Kroupa initial mass function (IMF; \citealp{kroupa01,mangeshwaran15,stone16}), which yields a good approximation for an early-type galaxy. However, observational evidence suggests a possible bias in favor of E+A galaxies as host galaxies of observed TDE candidates \citep{arcavi14,french16}. Due to recent or ongoing star formation, E+A galaxies would have an over-representation of high-mass stars, and are likely better represented by a standard Kroupa IMF (as suggested by, e.g., \citealp{zhang18,toyouchi22}). Mass segregation can also modify the rate at which stars diffuse into the loss cone, as the two-body relaxation (which is the standard mechanism through which stars diffuse into the loss cone; \citealp{bahcall76,frank76,lightman77,cohn78,magorrian99,merritt13,stone16}) rate is generally dominated by the heaviest stellar species, possibly leading to an over-representation of high-mass stars among TDE candidates in E+A galaxies.

Since $t_{\rm peak}$ varies with the penetration factor $\beta$, the observed $t_{\rm peak}$ distribution would also depend on the distribution function of the penetration factor, $f_{\beta}(\beta)$. \cite{coughlin22b} derived the distribution function of $\beta$ for non-spinning (Schwarzschild) as well as spinning (Kerr) black holes in the pinhole regime 
(see Equation 12 of \citealp{coughlin22b} for the functional form of $f_{\beta}(\beta)$). The distribution of $t_{\rm peak}$ can then be obtained from the joint probability distribution of $f(M_\star,R_\star), f_\beta(\beta)$, and the black hole mass (and spin) distribution functions, and is simply proportional to the product of these if these parameters are assumed to be independent. With the increasing rate of TDE detections, this distribution can be compared to the $t_{\rm peak}$ distribution of observed events to, in principle, place constraints on the distribution of SMBH masses and spins. 

A caveat associated with $t_{\rm peak}$ obtained from the hydrodynamical simulations (or the MG model) is that it is measured relative to the time at which the star reaches pericenter, $t_{\rm peri}$. However, there is a finite delay between $t_{\rm peri}$ and the time of first light, $t_{\rm ret}$, from a TDE, which can be modeled in two possible ways, as follows. The time of first light, $t_{\rm ret}$ roughly corresponds to the return time of the most bound debris, which can be constrained as a function of $\beta$ using an analytical approach (e.g., \citealp{coughlin22} used scaling arguments similar to the frozen-in approximation to obtain a lower bound on $t_{\rm ret}$; see Equation 10 of their paper for details), which we will investigate in future work. Alternatively, one could determine $t_{\rm ret}$ from a TDE lightcurve by choosing an appropriate prior and treating it as an additional fitting parameter (as was done in, e.g., \citealp{mockler19}).

\subsection{Long Duration Transients}
\begin{figure}
    \advance\leftskip-1cm
    \includegraphics[width=0.59\textwidth,height=6.5cm]{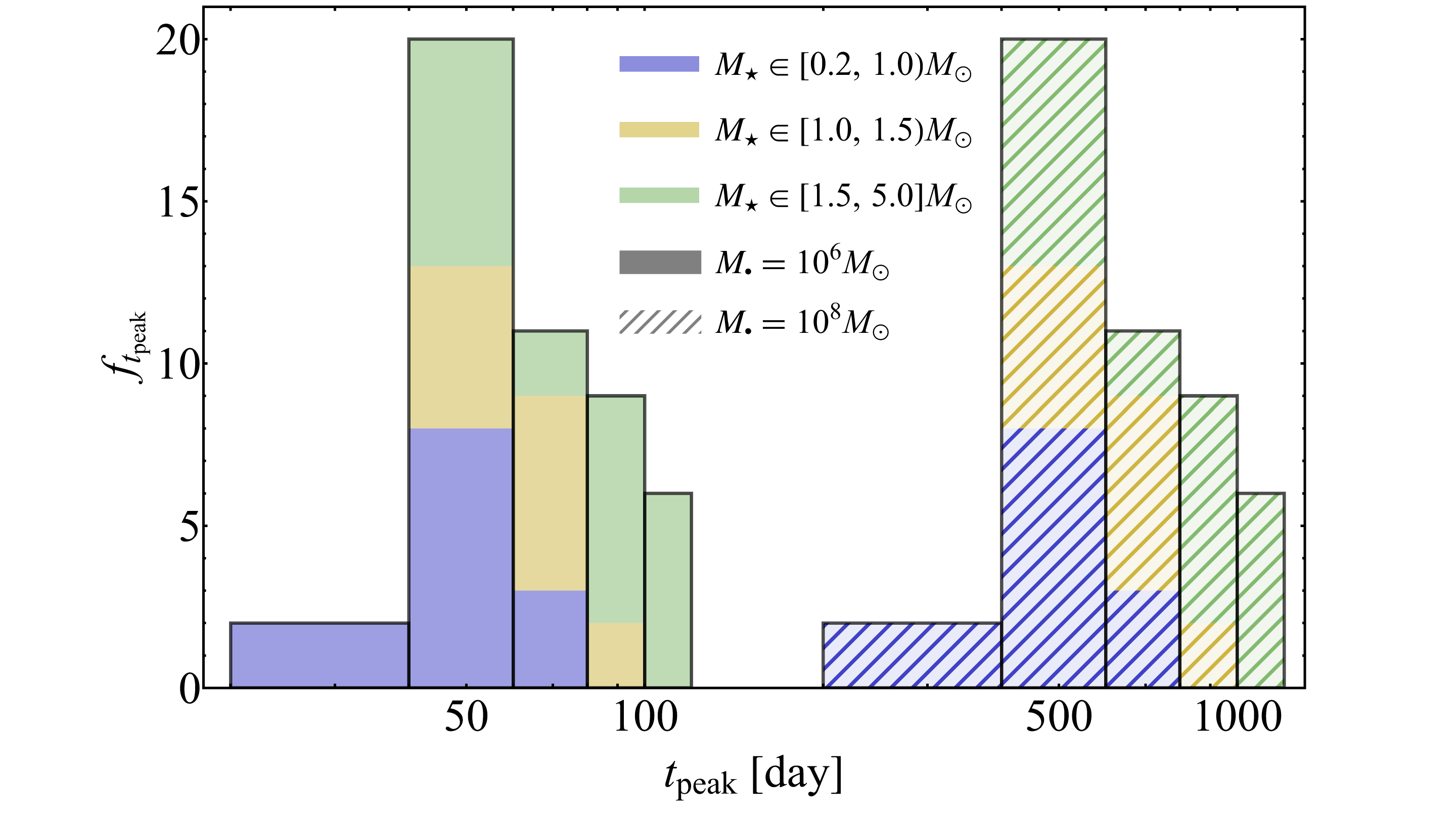}
    \caption{Histogram of $t_{\rm peak}(\beta_{\rm min})$ values for the range of stellar masses and ages considered in this work. Solid rectangles represent disruptions caused by a $10^6M_\odot$ SMBH, and hatched rectangles represent disruptions caused by a $10^8M_\odot$ SMBH. The colors indicate stellar masses (grouped into finite intervals), as shown in the legend. The height of each shaded region corresponds to the number of stars in a given mass interval (as indicated by its color), which have $t_{\rm peak}$ in the range demarcated by the edges of the bin. The total height of each histogram bar represents the total number of stars having $t_{\rm peak}$ within the given range.} \label{fig:tpeak-hist}
\end{figure}

The frozen-in model predicts that the characteristic fallback timescale of a TDE fallback rate is an increasing function of stellar mass. This behavior has been invoked by, e.g., \cite{lin17}, who suggested that the long-duration transient in GSN 069, which has a peak timescale of $\sim1000$ days and a lightcurve that decays over $\gtrsim11$ years, was due to the complete disruption of a high-mass star. \cite{bandopadhyay24} showed that the predictions for the peak timescale obtained using the frozen-in model are highly discrepant (the frozen-in model overpredicts $t_{\rm peak}$ by $\gtrsim2$ orders of magnitude for $M_\star\gtrsim 2M_\odot$) with numerical simulations of the complete disruption of high-mass stars, and the peak timescale for the complete disruption of any star can be approximated as $t_{\rm peak} \approx 30 \times (M_\bullet/10^6M_\odot)^{1/2} \textrm{ days}$, where $M_\bullet$ is the SMBH mass. We would therefore require $M_\bullet \gtrsim 10^8 M_\odot$ to give rise to a transient that reaches its peak in over $1000$ days. For low-mass and solar-like stars, the direct capture radius for a $10^8M_\odot$ SMBH is larger than the tidal radius, thus rendering the tidal disruption of such a star by a $10^8M_\odot$ SMBH unobservable. However, this is not the case for high-mass stars, for which the tidal radius ($r_{\rm t} \propto R_\star$ and $R_\star \sim R_\odot(M_\star/M_\odot)^{0.8}$ for main-sequence stars) can lie outside the direct capture radius for a $10^8 M_\odot$ SMBH. 

As seen in Figure~\ref{fig:3msun-peak-fbrs}, the fallback rate for grazing encounters of high-mass stars with a $10^6M_\odot$ SMBH can peak on timescales $t_{\rm peak}\gtrsim100$ days. Since the peak timescale of a TDE fallback rate scales with the SMBH mass as $t_{\rm peak} \propto M_\bullet^{1/2}$, we can determine the the region of parameter space that can yield $t_{\rm peak}\gtrsim1000$ days. Figure \ref{fig:tpeak-hist} shows a histogram of the peak timescales $t_{\rm peak}$ from the lowest-$\beta$ encounters that we simulated and that yield a measurable $t_{\rm peak}$ (i.e., the longest $t_{\rm peak}$ for a given stellar mass and age). 
The solid histogram shows $t_{\rm peak}$ values for $M_\bullet = 10^6M_\odot$ (directly measured from the simulations described in Section \ref{sec:hydro}, and as shown in Figures \ref{fig:low-mass-peak-fbrs}-\ref{fig:3msun-peak-fbrs}), whereas the histogram with a hatched shading assumes $M_\bullet = 10^8M_\odot$ (for which we scale the $M_\bullet = 10^6M_\odot$ results by a factor of $M_{\bullet,6}^{1/2}=10$). As indicated in the legend, each monochromatic block corresponds to stars in a given mass bin, such that (e.g.)~the blue rectangle with its base extending from $20-40$ days shows that $2$ stars with $M_\star<1M_\odot$ yield $20\leqslant t_{\rm peak}/\textrm{day} \leqslant40$ for $\beta=\beta_{\rm min}$. Since $t_{\rm peak}$ decreases with an increase in $\beta$ (grazing encounters yield the longest $t_{\rm peak}$ values), the peak timescales shown in this figure are the longest obtainable for a given stellar mass and age, and SMBH mass. This figure thus demonstrates that $>1000$ day timescales are obtained exclusively for high-mass stars being partially disrupted by high-mass SMBHs. This implies that the partial disruption of a high-mass star ($M_\star \gtrsim 2-3 M_\odot$) by a high-mass SMBH ($M_\bullet \gtrsim 10^{7.5} M_\odot$) can explain long-duration transients, such as the X-ray transient in GSN 069 \citep{miniutti19,miniutti23,miniutti23b}, or ``Scarie Barbie,'' a highly energetic ($L_{\rm max} =10^{45} \rm erg\, s^{-1}$) optical transient that faded over $\gtrsim 1000$ days \citep{subrayan23}. This is consistent with the SMBH mass constraint derived for ``Scarie Barbie'' in terms of the Eddington limit that would permit a luminosity $L_{\rm max} =10^{45} \rm erg\, s^{-1}$~\citep{subrayan23}. However, the inferred SMBH mass for GSN 069 based on the galactic $M-\sigma$ relation is $\sim 10^5 M_\odot$ \citep{miniutti19}, which our model suggests is too small to generate a $1000$ day peak timescale from the partial disruption of any star.

The long-duration transient associated with GSN 069, which was shown to be consistent with the partial tidal disruption of a main sequence star in \cite{miniutti23,miniutti23b} also exhibits quasi periodic eruptions (QPEs) on $\sim$hr timescales, which have been explained as the partial stripping of a white dwarf by an undermassive SMBH~\citep{king20}. However, as shown above, the SMBH mass that would be required to generate $t_{\rm peak}>1000$ days for the associated TDE is $M_\bullet\gtrsim10^{7.5}M_\odot$, which is significantly above the Hills mass for a white dwarf, implying that an orbiting white dwarf would be directly captured by the SMBH before being tidally destroyed. The discrepancy between the estimated black hole mass for GSN 069 and that required to generate a partial TDE having $t_{\rm peak}\gtrsim1000$ days thus implies that either the long duration flare cannot be explained by the partial disruption of a main-sequence star, or the long duration flare is due to the partial disruption of a high-mass star on a grazing orbit around a high-mass SMBH, in which case the QPEs cannot arise from the partial stripping of a white dwarf.

\section{Summary}
\label{sec:summary}
Using an approach based on the maximum gravity model (originally formulated in \citealt{coughlin22}), we developed an analytical prescription for the peak of the fallback rate ($t_{\rm peak},\dot{M}_{\rm peak}$) and the amount of mass stripped $\Delta M$ as a function of stellar mass and age, SMBH mass, and the penetration factor $\beta$ in a TDE. We performed hydrodynamical simulations of the partial disruption of $\sim 60$ main-sequence stars evolved using {\sc mesa}, for a range of penetration factors, and compared the numerically obtained fallback rates and amounts of mass stripped to the predictions of the model. The peak timescales predicted by the model are accurate to within $\sim50\%$ of the numerical simulations for any given star and penetration factor (in most cases the errors are $\lesssim10\%$, the highest errors being associated with high-mass and evolved stars with $\beta$ approaching $\beta_{\rm c}$; see Figure \ref{fig:low-mass-tpeak-error} and the right panels of Figures \ref{fig:solar-mass-peak-fbrs} and \ref{fig:3msun-peak-fbrs}). The critical penetration factor for complete disruption $\beta_{\rm c}$ predicted by the MG model also agrees with the results of the numerical simulations to within $\pm1$ for less-evolved (ZAMS and MAMS) stars, while $\beta_{\rm c}$ is underpredicted for TAMS stars. 

\begin{figure}
    \advance\leftskip-0.5cm
    \includegraphics[width=0.52\textwidth]{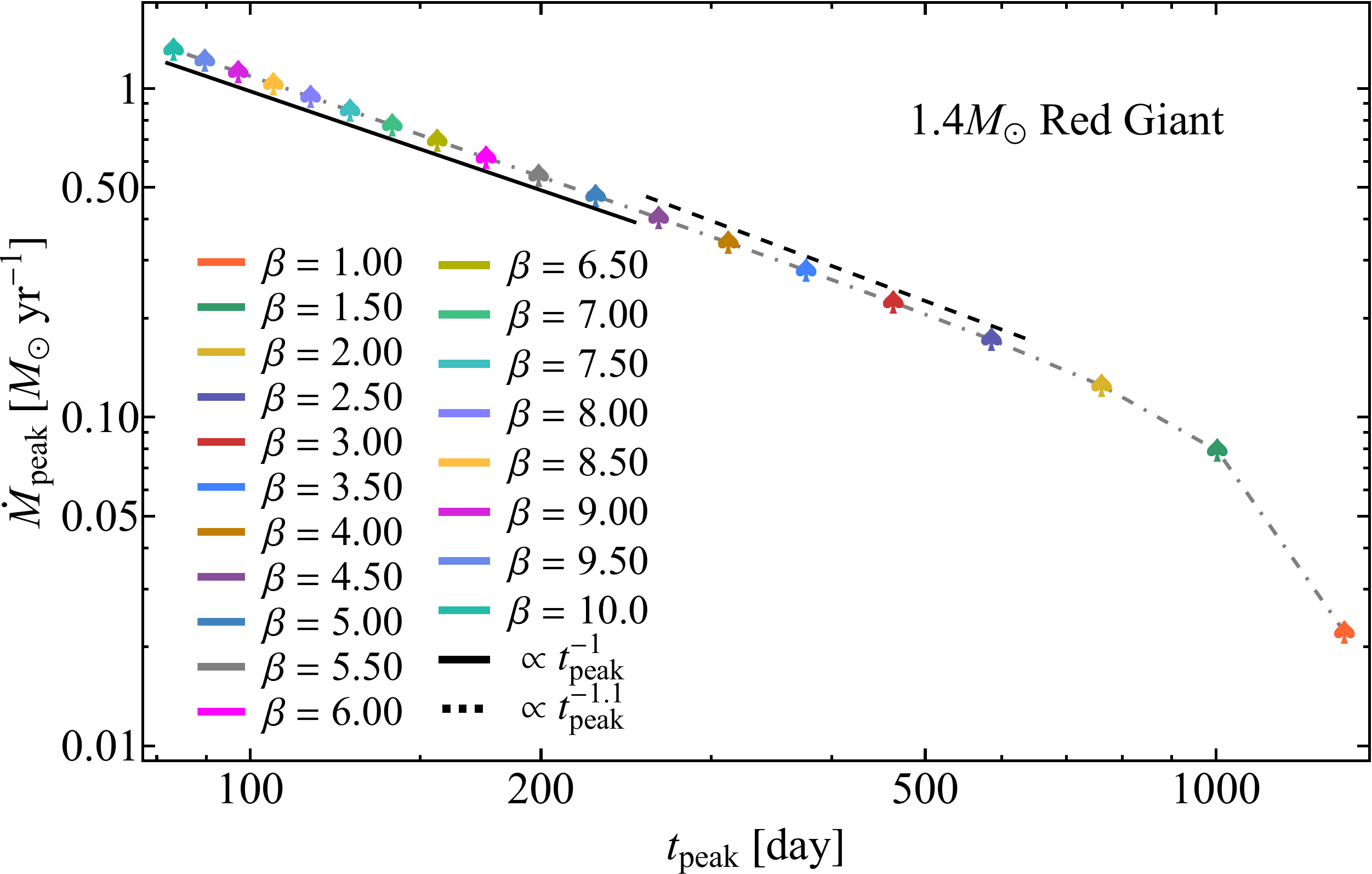}
    \caption{ The peak fallback times and rates for a {\sc mesa} evolved star with a ZAMS mass of $1.4M_\odot$, in its red giant phase. The black solid and dashed lines demonstrate that $\dot{M}_{\rm peak} \propto t_{\rm peak}^{-1.1}$ for low $\beta$ values, while the scaling becomes slightly shallower ($\dot{M}_{\rm peak} \propto t_{\rm peak}^{-1}$) for larger values of $\beta$.} \label{fig:RGB-peak-fbr}
\end{figure}

The hydrodynamical simulations presented in Section~\ref{sec:hydro} focus on main-sequence stars, but the model can be applied more broadly, e.g., to the disruption of giant stars by SMBHs \citep{macleod12} or to white dwarfs disrupted by intermediate mass black holes \citep{garain24,garain25,oconnor25}. The latter generalization has been discussed in \citet{oconnor25} and is in good agreement with the numerical simulations in \citet{garain24}. As concerns partial disruptions of giant stars, Figure~\ref{fig:RGB-peak-fbr} shows the peak fallback rates and times for a $1.4M_\odot$ (ZAMS mass) red giant star (RG I model in \citealp{macleod12}) for a range of $\beta$ values. The $(t_{\rm peak},\dot{M}_{\rm peak})$ prediction is in good agreement with the Figure 8 of \citep{macleod12}, which shows the fallback rate for $\beta=1.5$. As seen in the figure, $\dot{M}_{\rm peak}$ exhibits a power-law scaling in $t_{\rm peak}$, $\propto t_{\rm peak}^{-1.1}$ for low values of $\beta$, while the power-law becomes slightly shallower, $ \dot{M}_{\rm peak}\propto t_{\rm peak}^{-1}$, for larger $\beta$. The model in its current form likely breaks down when the tidal approximation ceases to be valid, e.g., in order-unity mass ratio systems in which the size of the star is not small compared to the radius at which it is destroyed. This may occur in systems such as the disruption of main-sequence stars by stellar-mass black holes (e.g., \citealt{kremer22}); to model these systems would require the inclusion higher order terms in the tidal field and to account for the reflex motion of the disrupter.

Our results demonstrate that the partial disruption of main-sequence stars with $0.2\leqslant M_\star/M_\odot \leqslant 5.0$ by a $10^6M_\odot$ SMBH can yield transients that peak on a timescale ranging from $\sim20-120$ days, and with peak luminosities spanning $\sim2-3$ orders of magnitude. Grazing encounters of high-mass and evolved stars $\Delta M \lesssim 1\% M_\star$ with high-mass SMBHs $M_\bullet \sim 10^8M_\odot$ can yield longer peak timescales $t_{\rm peak}\gtrsim1000$ days that are associated with some observed sources, e.g. GSN 069 and the ``Scarie Barbie'' \citep{miniutti23,subrayan23}. However, since the peak luminosities are inversely correlated with the peak timescale ($\dot{M}_{\rm peak}$ increases monotonically with $\beta$ until a star is completely destroyed, or a core is reformed, as discussed in Section \ref{sec:hydro} and shown in Figures \ref{fig:low-mass-peak-fbrs}-\ref{fig:3msun-peak-fbrs}), such encounters would yield lower peak luminosities compared to higher $\beta$ ones for a given star.

Given the agreement between the model predictions and the hydrodynamical simulations presented in the work, we can treat the predictions of the MG model as a first step in the direction of developing an analytical prescription for modeling the evolution of the lightcurve given any set of TDE parameters, such as stellar masses and ages, $\beta, M_\bullet$, etc.~(though clearly the fallback rate can be modified by relativistic effects, which we will investigate in a future work). 
To leading order, the fallback rate should closely track the lightcurve for at least the first few years of a TDE \citep{cannizzo90,lodato11}. However, over longer timescales, the lightcurve evolution would be governed by additional physics beyond the scope of the present work. For example, the observed luminosity of a TDE would depend on the efficiency of conversion of mass to energy, $\epsilon$, which can, in general, be a function of time. In cases where the fallback time $t_{\rm fb}$ becomes comparable to the viscous time $t_{\rm visc}$, viscous delays between the rate at which the material is supplied to the pericenter and the rate at which it viscously accretes can lead to significant differences between the fallback rate and the observed lightcurve. However, viscous delays have been shown to be insignificant in a large sample of observed optical/UV TDE candidates (e.g., \citealt{gezari15,mockler19,nicholl22}). The evolution of the accretion disk in a standard TDE environment, and its effect on the observed lightcurve, has been investigated in, e.g., \cite{guo25}. 

Barring these caveats, the MG model can be implemented as parameter estimation tool for TDEs under a Bayesian paradigm. In Appendix \ref{sec:appendix}, we describe an approach to analytically model the fallback rate for any given star and $\beta$, using the Pad{\'e} approximant described in Section \ref{sec:hydro}. While this approach is limited in its ability to model the exact shape of the fallback rate, it does not rely on any external input from hydrodynamical simulations and efficiently combines the analytical predictions for the peak of the fallback rate and its late-time scaling. Supporting data for this manuscript is archived on Zenodo \citep{bandopadhyay_2025_17822028}, and available on GitHub\footnote{\href{https://github.com/AnanyaBandopadhyay/mg-model-partials}{https://github.com/AnanyaBandopadhyay/mg-model-partials}}.

\section*{Acknowledgements}
We thank the anonymous referee for useful comments and suggestions that improved the manuscript. A.B.~acknowledges support from NASA through the FINESST program, grant 80NSSC24K1548. E.R.C.~acknowledges support from NASA through the Astrophysics Theory Program, grant 80NSSC24K0897. C.J.N.~acknowledges support from the Leverhulme Trust (grant No. RPG-2021-380). 
\clearpage
\appendix
\section{Analytical Fits to the Fallback Rates}
\label{sec:appendix}
\begin{figure}
    \advance\leftskip-0.7cm
    \includegraphics[width=0.51\textwidth]{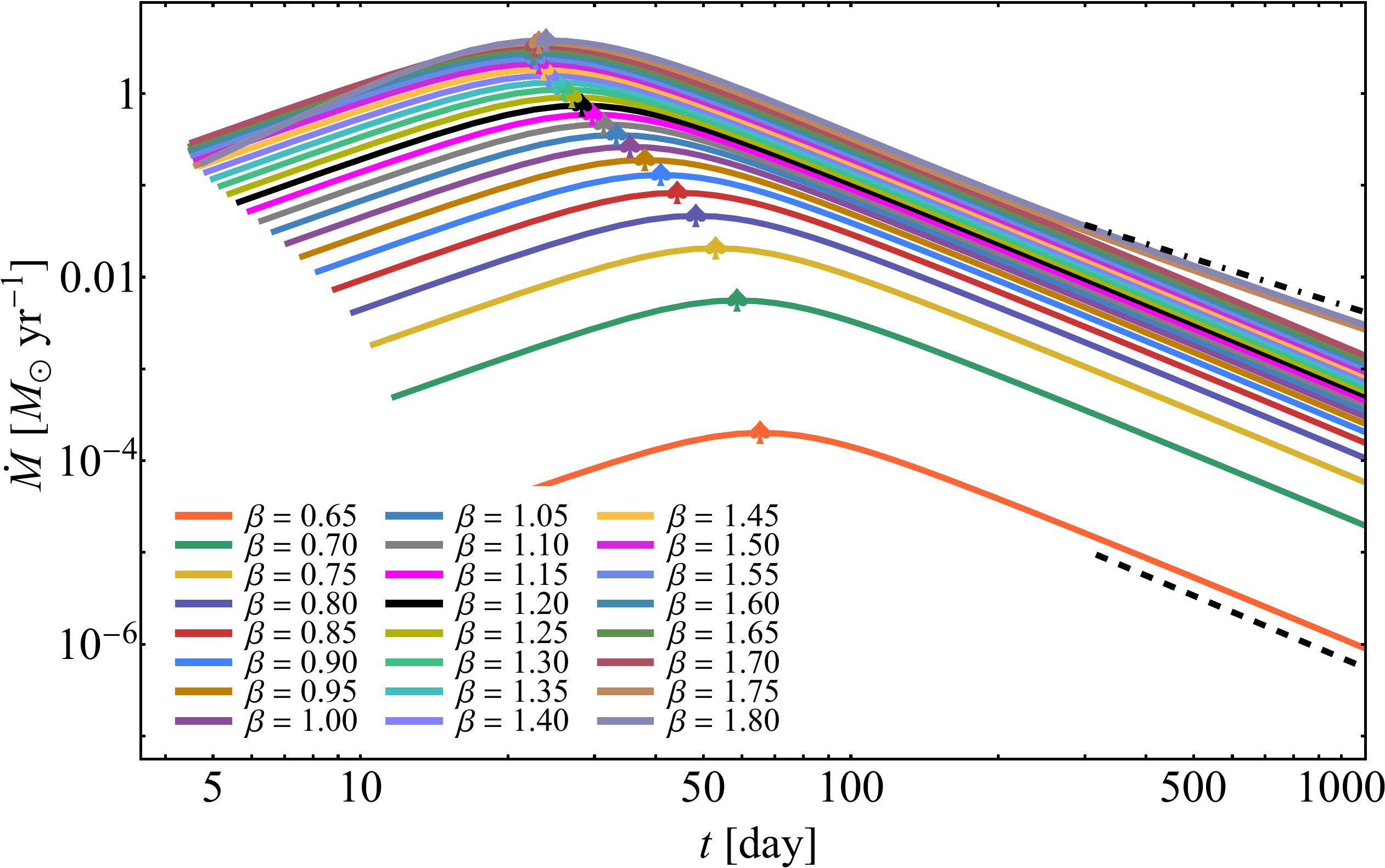}
    \caption{The analytical fits to the fallback rate obtained using Equation~\eqref{eq:fbr-pade-approx} for the $1M_\odot$ ZAMS star. The black dashed (dot-dashed) line represents a $\propto t^{-9/4}$ ($\propto t^{-5/3}$) power-law, which is the expected late-time scaling for partial (complete) disruptions.} \label{fig:ZAMS1-approx-fbrs}
\end{figure}

Here we use the predicted peak of the fallback rate, in combination with the Pad{\'e} approximant based approach for fitting numerically calculated fallback rates described in \cite{nixon21}, to obtain an anaytical expression for the fallback rate for any given star and $\beta$. In Section~\ref{sec:hydro}, we used the approach described in \cite{nixon21} to fit the numerically obtained fallback rates to the expression given by Equation~\eqref{eq:pade-approx}, using $N_{\rm max}=10$ to obtain an accurate estimate of the peak. As demonstrated in Figures~\ref{fig:low-mass-peak-fbrs}-\ref{fig:3msun-peak-fbrs}, the peak values obtained from the numerical simulations are reasonably well-approximated by the analytical prediction for the peak, given by Equation~\eqref{eq:peak-fbr}. We can use the prediction for the peak fallback rate $(t_{\rm peak},\dot{M}_{\rm peak})$ as one data point that lies on the fallback rate for a given star and pericenter distance. We also know that the late time scaling of the fallback rate is $\propto t^{-9/4}$ for partials, and $\propto t^{-5/3}$ for complete disruptions. Furthermore, the total mass accreted onto the SMBH is $\sim \Delta M/2$, where $\Delta M$ is the mass stripped at a given pericenter distance (given by Equation~\eqref{eq:deltaM}), which therefore yields the additional constraint
\begin{equation}
    \Delta M /2 = \int \limits_0^\infty \dot{M}_{\rm fit} (\tilde{t}) \mathrm{d} \tilde{t}. \label{eq:mass-int-dim}
\end{equation} 
Using this information (i.e., the predicted peak, the late time scaling $\propto t^{n_\infty}$ -- where $n_{\infty}=-9/4$ for partial disruptions and $n_{\infty}=-5/3$ for a complete disruptions -- and the total accreted mass), we can fit a Pad{\'e} approximant with 2 additional parameters to model the fallback rate, and choose the functional form
\begin{equation}
    \dot{M}_{\rm fit} = \frac{\Delta M}{4 t_{\rm peak}}\frac{a\tilde{t}^m}{1+ b \tilde{t}^{m-n_\infty}}\frac{1+ c_1 \tilde{t}+\tilde{t}^{2}}{1+\tilde{t}^{2}}, \label{eq:fbr-pade-approx}
\end{equation}
where we have fixed the value of $m=2$. Since the fallback rate peaks at $\tilde{t} =1$, and the magnitude of the peak (as given by Equation~\ref{eq:peak-fbr}) is $\dot{M}_{\rm peak} = \Delta M / 4 t_{\rm peak}$, this yields the following relation between $a,b$ and $c_1$,
\begin{equation}
     \frac{a(c_1+2)}{2(1+b)} = 1. \label{eq:peak}
\end{equation}
\begin{figure*}
    \includegraphics[width=0.51\textwidth]{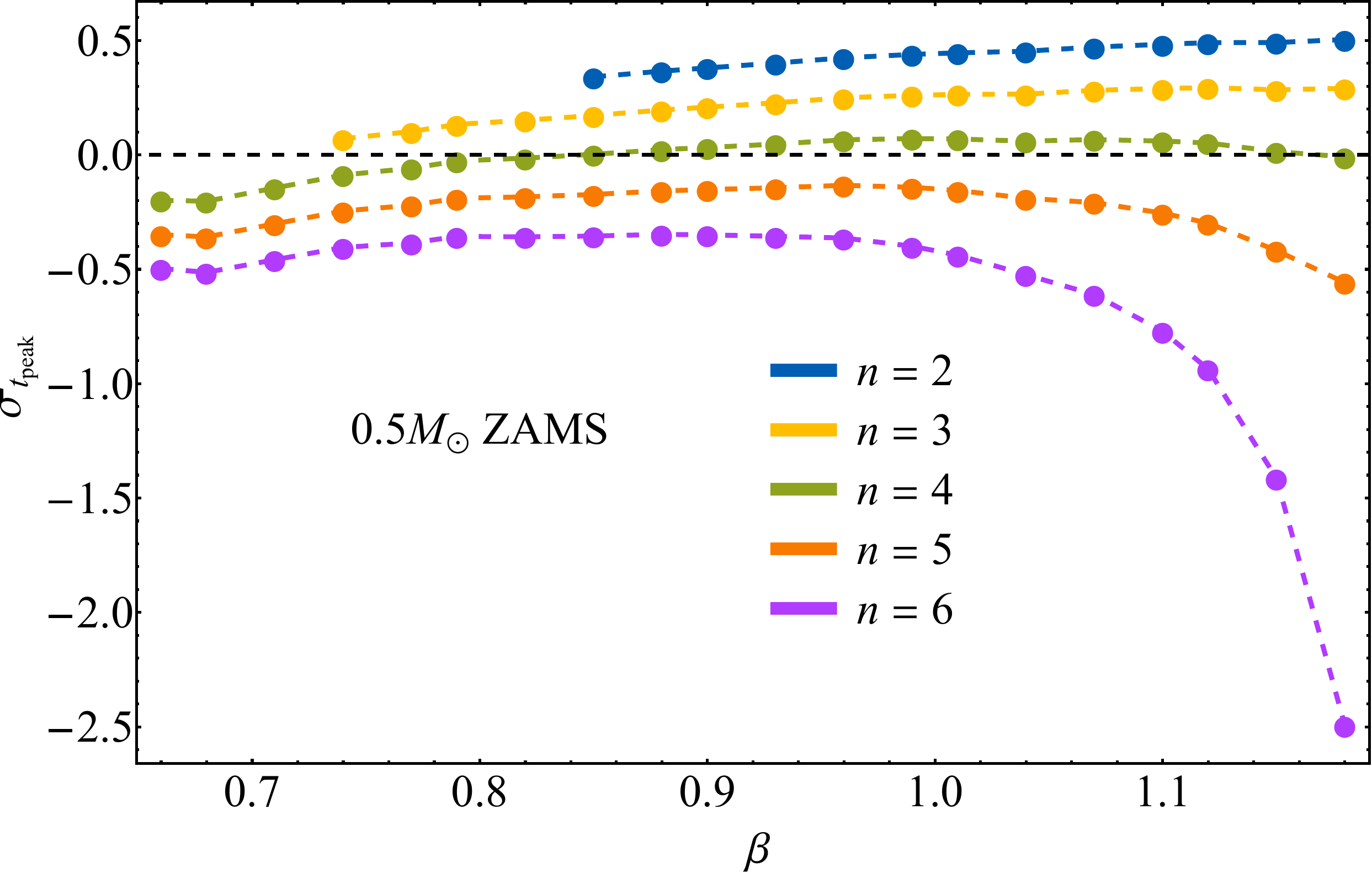}
    \includegraphics[width=0.51\textwidth]{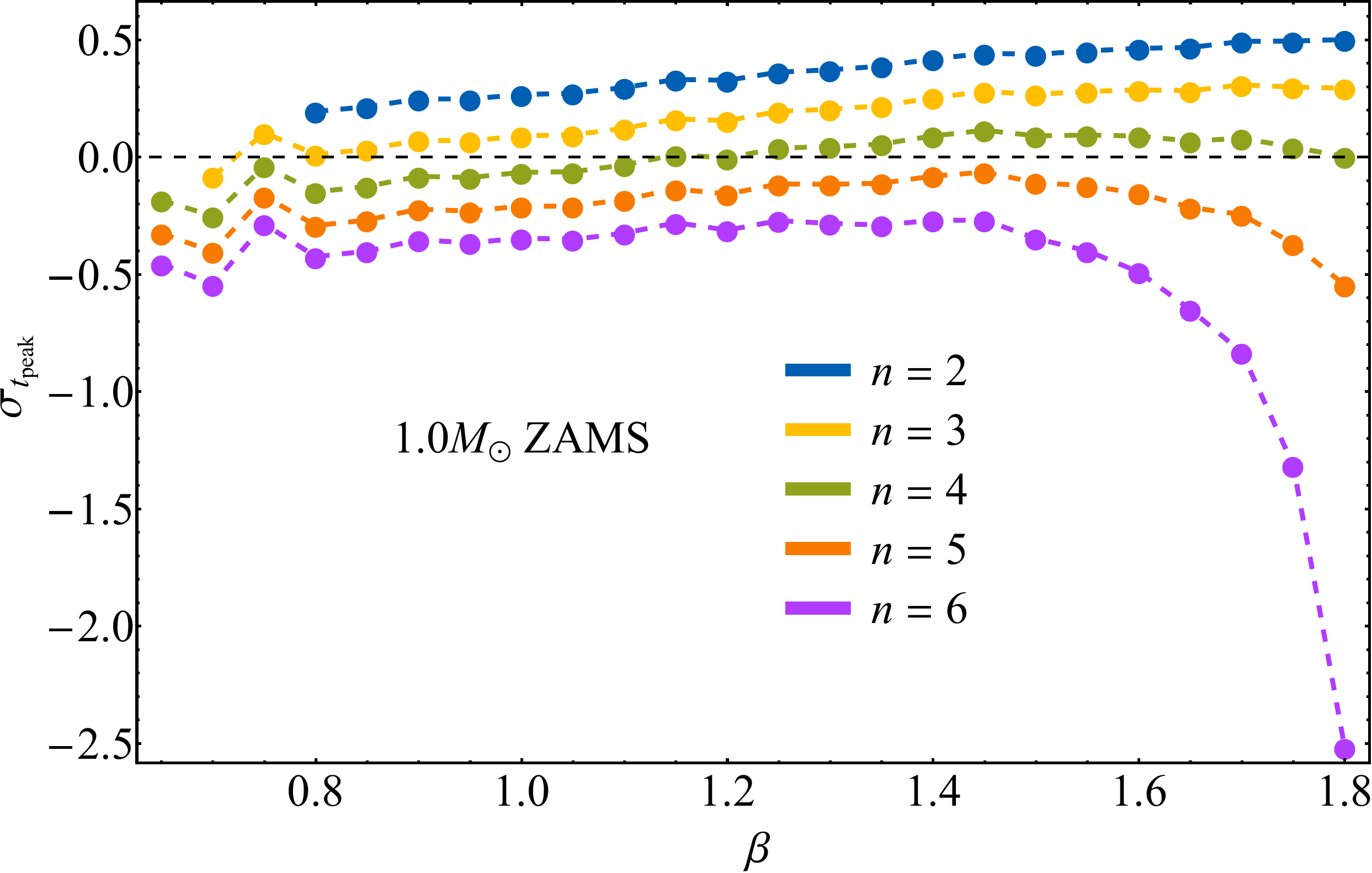}\\
    \includegraphics[width=0.51\textwidth]{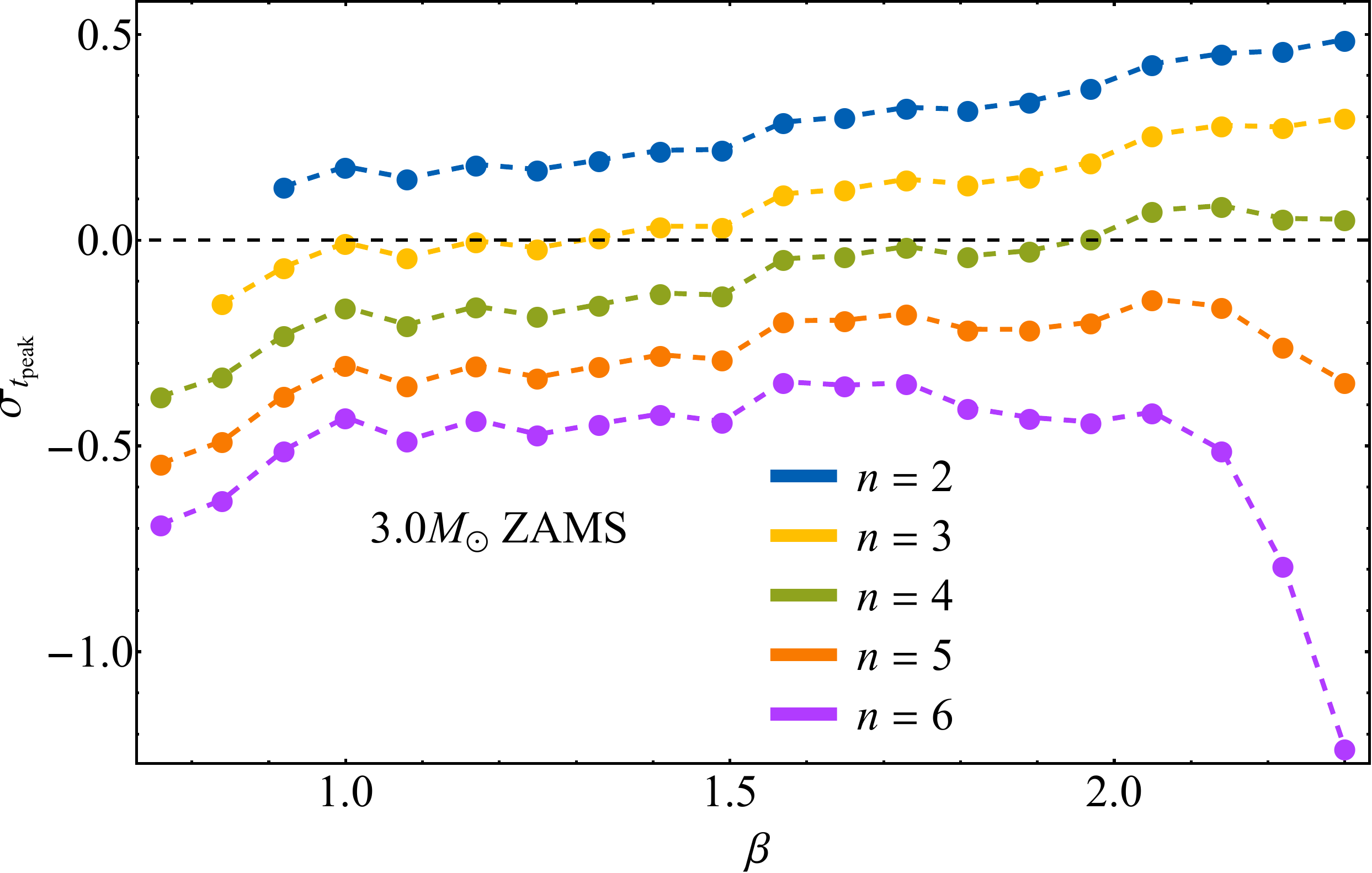}
    \includegraphics[width=0.51\textwidth]{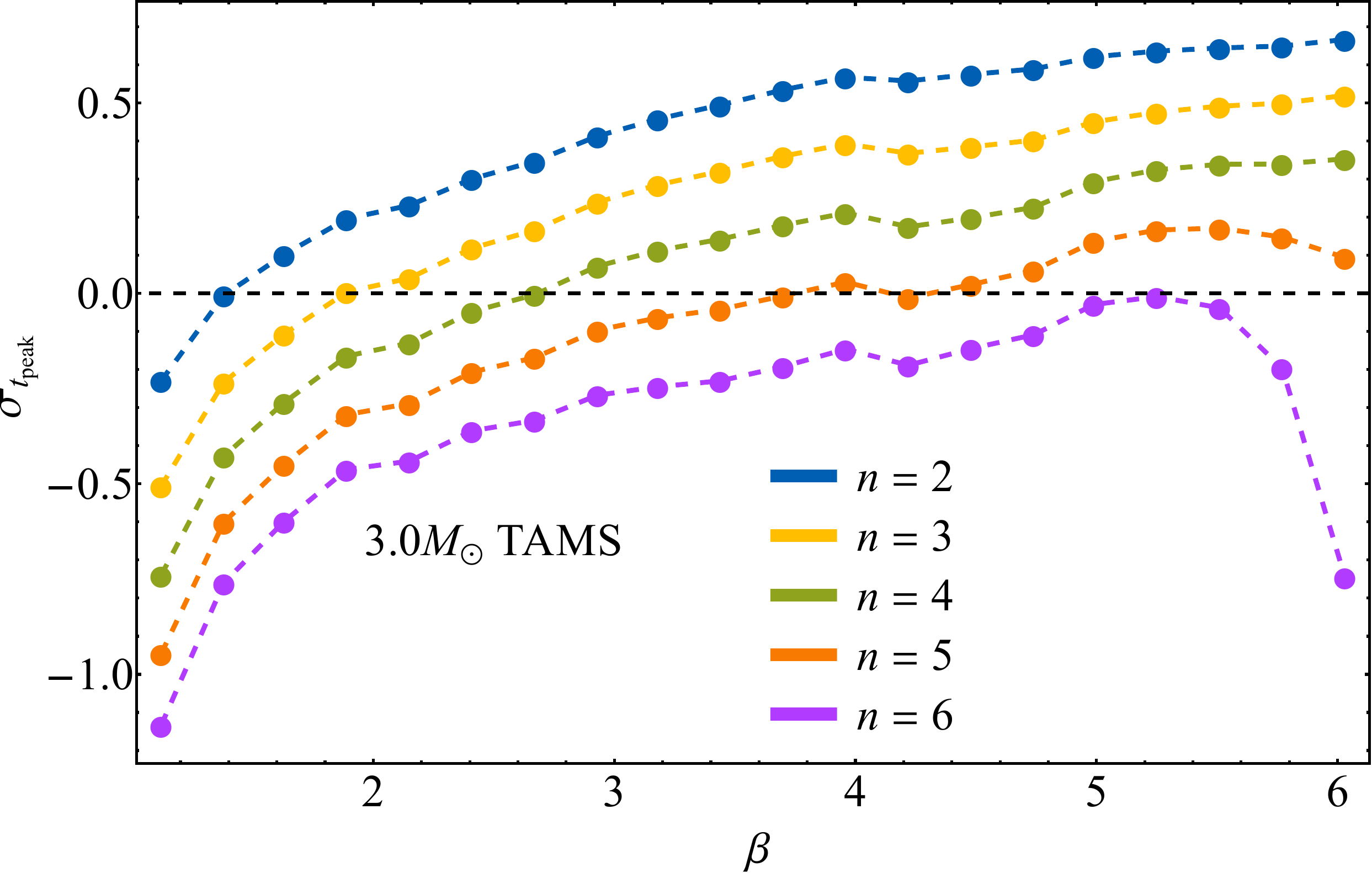}
    \caption{ The error in $t_{\rm peak}$, $\sigma_{t_{\rm peak}} \equiv (t_{\rm peak, hydro}-t_{\rm peak, MG})/t_{\rm peak, hydro}$ as a function of $\beta$ for a $0.5M_\odot$ ZAMS (top left), $1.0M_\odot$ ZAMS (top right), $3.0M_\odot$ ZAMS (bottom left), and $3.0M_\odot$ TAMS (bottom right) star. The different colors represent the order unity parameter $n$ in the approximation for the tidal field, i.e., $f_{\rm t} = n G M_\bullet R/r^3$. The absolute error, $|\sigma_{t_{\rm peak}}|$, is minimized for $n \sim 3-4$ for low-mass and younger stars, while a slightly higher value $n \sim 4-5$ is favored for high-mass and evolved stars.} \label{fig:tidal-field}
\end{figure*}
Also, since $\tilde{t} = 1$ is a maximum of the fallback rate, this yields, for its derivative
\begin{equation}
    \left. \frac{d\dot{M}_{\rm fit}}{d\tilde{t}} \right|_{\tilde{t}=1}=0.
\end{equation}
Setting the derivative to zero yields the following relation between the parameters $a,b$ and $c_1$,
\begin{equation}
    \frac{a (c_1+2) (m+bn_{\infty})}{(1+b)^2} = 0. \label{eq:derivative}
\end{equation}
Since $a=0$ and $c_1=-2$ both yield $\dot{M}_{\rm peak} = 0$, the above equation constrains $b=-m/n_{\infty}$. Thus, for $m=2$, $b=8/9$ for partial TDEs, and $b=6/5$ for complete disruptions. Finally, substituting the functional form for $\dot{M}_{\rm fit}$ given by Equation~\eqref{eq:fbr-pade-approx}, the mass integral in Equation~\eqref{eq:mass-int-dim} yields,
\begin{equation}
     \int \limits_0^{\infty} {\frac{a\tilde{t}^m}{1+ b \tilde{t}^{m-n_\infty}}\frac{1+ c_1 \tilde{t}+\tilde{t}^{2}}{1+\tilde{t}^{2}} \mathrm{d} \tilde{t}} = 2. \label{eq:mass-int-dim-less}
\end{equation}
Since $b$ is fixed by Equation~\eqref{eq:derivative}, Equations~\eqref{eq:peak} and \eqref{eq:mass-int-dim-less} can be solved simultaneously to constrain $a$ and $c_1$. The two constraints are satisfied for $a=2.44,c_1=-9/20$ for partial TDEs, and $a=0.51,c_1=6.68$ for complete disruptions. Figure~\ref{fig:ZAMS1-approx-fbrs} shows the analytical fits to the fallback rate for a range of $\beta$ values for the $1 M_\odot$ ZAMS star. Since the star is completely destroyed at $\beta=1.75$ in the {\sc phantom} simulation (the MG model predicts $\beta_{\rm c} \approx 1.80 $ for this star, as seen in Figure~\ref{fig:solar-mass-deltam-beta}), we treat the $\beta=1.75$ and $\beta=1.80$ cases as complete disruptions (i.e., use $n_\infty = -5/3$ in Equation~\ref{eq:fbr-pade-approx}) and the remaining ones as partial TDEs ($n_\infty = -9/4$).

\section{Tidal Field Strength}
\label{sec:tidal_field}
The tidal radius of a star of mass $M_\star$ and radius $R_\star$ being disrupted by an SMBH of mass $M_\bullet$, defined as $r_{\rm t} \equiv R_\star (M_\bullet/M_\star)^{1/3}$, is obtained by equating the tidal field $f_{\rm t} = GM_\bullet R_\star/r^3$ (where $r$ is the distance of the stellar COM from the SMBH) to the surface gravity of the star, $G M_\star /R_\star^2$. \cite{coughlin22} argued that the strength of the tidal field across a sphere of radius $R$ should be augmented by a factor of $4$ relative to the canonical expression above, in order to account for the velocity divergence induced across the diameter of the fluid sphere. In this work, we adopt the same normalization for the tidal field (i.e., preserving the factor of $4$ in its definition) to arrive at a straightforward generalization of the model for partial TDEs. As demonstrated in Figures~\ref{fig:low-mass-peak-fbrs}-\ref{fig:3msun-deltam-beta}, this yields good agreement in the predicted values of $t_{\rm peak}, \dot{M}_{\rm peak}, \beta_{\rm c}$ across the range of stars considered in this work.

Here we assess the accuracy of this factor of $4$ by replacing it with a free parameter $n$, and compare the TDE parameters thereby predicted with the results of our hydrodynamical simulations. We thus let the tidal field across a sphere of radius $R$ whose COM is at a distance $r$ from the SMBH be
\begin{equation}
    \label{eq:tidal-field-n}
    f_{\rm t} = \frac{n G M_\bullet R}{r^3},
\end{equation}
where $n$ is an unspecified parameter that should physically be of the order unity. With this expression and for a range of $n$, we can use the approach described in Section~\ref{sec:analytical-model} to solve for the radius $R$ within the star at which the tidal field for a given value of $\beta$ equals the self-gravity due to the mass contained within $R$, and subsequently solve for $\Delta M, t_{\rm peak}, \textrm{ and } \dot{M}_{\rm peak}$. 

Figure~\ref{fig:tidal-field} shows the error in $t_{\rm peak}$, $\sigma_{t_{\rm peak}}$ as a function of $\beta$ (with $\beta \in (0.6,\beta_{\rm c}]$, where $\beta_{\rm c}$ is the model prediction using $n=4$) for four different stars, for $n$ ranging from $2-6$. For the $0.5M_\odot, 1M_\odot$ and $3M_\odot$ ZAMS stars (shown in the top left, top right and bottom left panels of the figure), the absolute error $|\sigma_{t_{\rm peak}}|$ is minimized for $n\sim3-4$, whereas for the more evolved $3M_\odot$ TAMS star (shown in the bottom right panel), $|\sigma_{t_{\rm peak}}|$ is minimized for $n\sim4-5$. Thus, while $n=3$ or $n=5$ yields marginally better agreement for a subset of the stars and a restricted range in $\beta$, the results obtained using $n=4$ (for $t_{\rm peak}$ as well as $\beta_{\rm c}$ and $\dot{M}_{\rm peak}$, as shown in Section~\ref{sec:hydro}) are in good agreement with the results of the hydrodynamical simulations over a wide range of stellar masses, ages and orbital pericenter distances.

\end{document}